\documentclass[preprint2]{aastex62}
\usepackage{float, bm, graphicx, amsmath, morefloats}
\def\ra#1#2#3{#1$^{\rm h}$#2$^{\rm m}$#3$^{\rm s}$}
\def\dec#1#2#3{$#1^\circ#2'#3''$}

\graphicspath{{./}{figures/}}
\received{October 6, 2019}
\revised{June 25, 2020}
\accepted{July 31, 2020}
\submitjournal{ApJ}

\shorttitle{First ZTF SLSNe}
\shortauthors{Lunnan et al.}

\begin{document}

\title{Four (Super)luminous Supernovae from the First Months of the ZTF Survey}

\correspondingauthor{Ragnhild Lunnan}
\email{ragnhild.lunnan@astro.su.se}

\author[0000-0001-9454-4639]{R.~Lunnan}
\affil{The Oskar Klein Centre \& Department of Astronomy, Stockholm University, AlbaNova, SE-106 91 Stockholm, Sweden}

\author[0000-0002-0786-7307]{Lin~Yan}
\affil{Caltech Optical Observatories, California Institute of Technology, Pasadena CA 91125, USA}

\author[0000-0001-8472-1996]{D.~A.~Perley}
\affil{Astrophysics Research Institute, Liverpool John Moores University, 146 Brownlow Hill, Liverpool L3 5RF, UK}

\author[0000-0001-6797-1889]{S.~Schulze}
\affil{Department of Particle Physics and Astrophysics, Weizmann Institute of Science, Rehovot 76100, Israel}

\author[0000-0002-5748-4558]{K.~Taggart}
\affil{Astrophysics Research Institute, Liverpool John Moores University, 146 Brownlow Hill, Liverpool L3 5RF, UK}

\author{A.~Gal-Yam}
\affil{Department of Particle Physics and Astrophysics, Weizmann Institute of Science, Rehovot 76100, Israel}

\author{C.~Fremling}
\affil{Division of Physics, Mathematics and Astronomy, California Institute of Technology, Pasadena, CA 91125, USA}

\author[0000-0001-6753-1488]{M.~T.~Soumagnac}
\affil{Department of Particle Physics and Astrophysics, Weizmann Institute of Science, Rehovot 76100, Israel}
\affil{Lawrence Berkeley National Laboratory, 1 Cyclotron Road, Berkeley, CA 94720, USA}

\author{E.~Ofek}
\affil{Department of Particle Physics and Astrophysics, Weizmann Institute of Science, Rehovot 76100, Israel}

\author{S.~M.~Adams}
\affil{Division of Physics, Mathematics and Astronomy, California Institute of Technology, Pasadena, CA 91125, USA}

\author{C.~Barbarino}
\affil{The Oskar Klein Centre \& Department of Astronomy, Stockholm University, AlbaNova, SE-106 91 Stockholm, Sweden}

\author[0000-0001-8018-5348]{E.~C.~Bellm}
\affiliation{DIRAC Institute, Department of Astronomy, University of Washington, 3910 15th Avenue NE, Seattle, WA 98195, USA}

\author{K.~De}
\affil{Division of Physics, Mathematics and Astronomy, California Institute of Technology, Pasadena, CA 91125, USA}

\author{C.~Fransson}
\affil{The Oskar Klein Centre \& Department of Astronomy, Stockholm University, AlbaNova, SE-106 91 Stockholm, Sweden}

\author[0000-0001-9676-730X]{S.~Frederick}
\affiliation{Department of Astronomy, University of Maryland, College Park, MD 20742, USA}

\author{V.~Z.~Golkhou} 
\affiliation{DIRAC Institute, Department of Astronomy, University of Washington, 3910 15th Avenue NE, Seattle, WA 98195, USA} 
\affiliation{The eScience Institute, University of Washington, Seattle, WA 98195, USA}

\author{M.~J.~Graham}
\affil{Division of Physics, Mathematics and Astronomy, California Institute of Technology, Pasadena, CA 91125, USA}

\author[0000-0002-0430-7793]{N.~Hallakoun} 
\affil{School of Physics and Astronomy, Tel-Aviv University, Tel-Aviv 6997801, Israel}

\author{A.~Y.~Q.~Ho}
\affil{Division of Physics, Mathematics and Astronomy, California Institute of Technology, Pasadena, CA 91125, USA}

\author{M.~M.~Kasliwal}
\affil{Division of Physics, Mathematics and Astronomy, California Institute of Technology, Pasadena, CA 91125, USA}

\author{S.~Kaspi}
\affil{School of Physics and Astronomy, Tel-Aviv University, Tel-Aviv 6997801, Israel}

\author{S.~R.~Kulkarni}
\affil{Division of Physics, Mathematics and Astronomy, California Institute of Technology, Pasadena, CA 91125, USA}

\author[0000-0003-2451-5482]{R.~R.~Laher}
\affil{IPAC, California Institute of Technology, 1200 E. California Blvd, Pasadena, CA 91125, USA}

\author[0000-0002-8532-9395]{F.~J.~Masci}
\affil{IPAC, California Institute of Technology, 1200 E. California Blvd, Pasadena, CA 91125, USA}
             
\author[0000-0002-6716-4179]{F.~Pozo Nu\~{n}ez} 
\affil{Haifa Research Center for Theoretical Physics and Astrophysics, University of Haifa, Haifa, Israel}
\affil{Astronomisches Institut, Ruhr--Universit\"at Bochum, Universit\"atsstra{\ss}e 150, 44801 Bochum, Germany}

\author[0000-0001-7648-4142]{B.~Rusholme}
\affiliation{IPAC, California Institute of Technology, 1200 E. California Blvd, Pasadena, CA 91125, USA}
  
\author[0000-0001-9171-5236]{R.~M.~Quimby}
\affiliation{Department of Astronomy/Mount Laguna Observatory, San Diego State University, 5500 Campanile Drive,
San Diego, CA 92812-1221, USA}

\author[0000-0003-4401-0430]{D.~L.~Shupe}
\affiliation{IPAC, California Institute of Technology, 1200 E. California Blvd, Pasadena, CA 91125, USA}
                        
\author{J.~Sollerman}
\affil{The Oskar Klein Centre \& Department of Astronomy, Stockholm University, AlbaNova, SE-106 91 Stockholm, Sweden}

\author{F.~Taddia}
\affil{The Oskar Klein Centre \& Department of Astronomy, Stockholm University, AlbaNova, SE-106 91 Stockholm, Sweden}

\author{J.~van Roestel}
\affil{Division of Physics, Mathematics and Astronomy, California Institute of Technology, Pasadena, CA 91125, USA}

\author{Y.~Yang}
\affil{Department of Particle Physics and Astrophysics, Weizmann Institute of Science, Rehovot 76100, Israel}

\author[0000-0001-6747-8509]{Yuhan Yao}
\affil{Division of Physics, Mathematics and Astronomy, California Institute of Technology, Pasadena, CA 91125, USA}

\begin{abstract}
We present photometry and spectroscopy of four hydrogen-poor luminous supernovae discovered during the two-month science commissioning and early operations of the Zwicky Transient Facility (ZTF) survey. Three of these objects, SN\,2018bym (ZTF18aapgrxo), SN\,2018avk (ZTF18aaisyyp) and SN\,2018bgv (ZTF18aavrmcg) resemble typical SLSN-I spectroscopically, while SN\,2018don (ZTF18aajqcue) may be an object similar to SN\,2007bi experiencing considerable host galaxy reddening, or an intrinsically long-lived, luminous and red SN~Ic.  We analyze the light curves, spectra, and host galaxy properties of these four objects and put them in context of the population of SLSN-I. SN\,2018bgv stands out as the fastest-rising SLSN-I observed to date, with a rest-frame $g$-band rise time of just 10 days from explosion to peak -- if it is powered by magnetar spin-down, the implied ejecta mass is only $\simeq 1~{\rm M}_{\odot}$. SN\,2018don also displays unusual properties -- in addition to its red colors and comparatively massive host galaxy, the light curve undergoes some of the strongest light curve undulations post-peak seen in a SLSN-I, which we speculate may be due to interaction with circumstellar material. We discuss the promises and challenges of finding SLSNe in large-scale surveys like ZTF given the observed diversity in the population.
\end{abstract}

\keywords{supernovae:general
, supernovae:individual (SN2018avk, SN2018don, SN2018bym, SN2018bgv) --- surveys}

\section{Introduction} \label{sec:intro}
In the past decade, wide-field transient surveys have greatly expanded the parameter space of known stellar explosions. One such intriguing new type of supernova is the ``superluminous'' supernovae (SLSNe), initially broadly defined as supernovae with peak absolute magnitudes $< -21$~mag (\citealt{gal12}; see \citealt{gal19} for a recent review). Now with more than 100 such objects reported (e.g., \citealt{gpk+17,gal19}), the observed diversity of the class is growing but trends are emerging. It has been shown that most SLSNe without hydrogen in their spectra (SLSN-I) form a distinct spectroscopic class, and can be separated from other stripped-envelope supernovae based on their spectra alone \citep{qdg+18}; the observed luminosity function of SLSN-I selected spectroscopically does extend fainter than $-21$~mag \citep{lcb+18,dgr+18,ass+19}. For SLSNe with hydrogen in their spectra (SLSN-II), most display intermediate-width Balmer lines, classifying them spectroscopically as SN~IIn. There are also examples of SLSN-II with broad Balmer lines, however \citep{ghg+09,mcp+09,isg+18}, as well as objects classified as SLSN-I based on their peak spectra that show signatures of interaction with H-rich material at late times \citep{yqo+15,ylp+17}, complicating the picture. 

One reason why SLSNe have garnered so much attention in the community is that the underlying physical mechanism behind their enormous luminosities is still not well-understood. One class of models, particularly for the SLSN-I, involves energy injection from the spin-down of a millisecond magnetar born in the explosion \citep{kb10,woo10}. This model has proved successful in explaining the light curves and spectra of a wide variety of SLSN-I \citep{isj+13,lww+17,ngb+18,dhw+12,des19}, including out to very late times \citep{nbb+18}. Direct evidence of a central engine, such as the predicted X-ray breakout at late times \citep{mvh+14} has largely remained elusive, however \citep{mcm+18,bcm+18}; one exception may be the recent detection of radio emission from the position of the SLSN-I PTF10hgi consistent with a young magnetar nebula \citep{ebm+19}. Alternative explanations include circumstellar interaction, in which large amounts of kinetic energy can be converted to radiation as the ejecta collides with and shocks a dense circumstellar medium (CSM) \citep{ci11,cwv12,mbt+13,sbn16,wcv+17}. This model is certainly relevant to the SLSN-II that show narrow emission lines indicating CSM interaction; while such spectroscopic signatures are not seen in SLSN-I, observations such as light curve undulations \citep[e.g.][]{nbs+16,vlg+17,bnb+18}, late-time H$\alpha$ emission \citep{yqo+15,ylp+17} and the recent direct detection of a CSM shell around a SLSN-I through a light echo \citep{lfv+18} at least shows that some SLSN-I progenitors experience significant mass loss close to explosion. Finally, some SLSN-I have been proposed to be powered by radioactive decay of $^{56}$Ni through the pair-instability explosion of a very massive star (helium core mass $M_{\rm He} \approx 65-130 {\rm M}_{\odot}$; \citealt{brs67,hw02,wbh07}). The large ejecta masses and therefore long implied diffusion times means this model is only applicable to the slowest-evolving SLSNe, however, and even for these objects it is debated \citep{gmo+09,nsj+13,lcb+16,jsh16,gbn+19}.

The SLSN discovery rate has accelerated in the past decade, much thanks to untargeted transients surveys like the Panoramic Survey Telescope and Rapid Response System (Pan-STARRS; \citealt{cmm+16}), the Palomar Transient Factory (PTF; \citealt{rkl+09}), the Asteroid Terrestrial-impact Last Alert System (ATLAS; \citealt{tdh+18}), the \textit{Gaia} Photometric
Science Alerts programme\footnote{\url{http://gsaweb.ast.cam.ac.uk/alerts/}} \citep{gaia16}, and the All-Sky Automated Survey for Supernovae (ASAS-SN; \citealt{spg+14}). 
The Zwicky Transient Facility (ZTF), with an upgrade to a 47 square degree camera (compared to 7.8 square degrees in its predecessor PTF) represents a new generation of transient surveys, and functions as an important intermediate step between PTF-scale surveys and the upcoming Large Synoptic Survey Telescope (LSST) in the 2020s. 

Here, we present the discovery, data, and analysis of the first SLSN-I from ZTF, discovered during the commissioning and science validation period (April-May 2018). The total number of SLSN-I found during this period was four; we focus our analysis on two particularly noteworthy events: SN\,2018bgv (ZTF18aavrmcg), which shows unusually rapid timescales for a SLSN-I; and SN\,2018don (ZTF18aajqcue), a heavily reddened event that shows unusual light curve variation post-peak. This paper is structured as follows. Section~\ref{sec:data} details the discoveries, ZTF photometry and follow-up data. We analyze the supernova properties and place them in context of the population of SLSNe in Sections~\ref{sec:lightcurves} and \ref{sec:spec}, and do the same with their host galaxies in Section~\ref{sec:hosts}. We discuss our results in the context of diversity within the population of (super)luminous supernovae, and prospects and implications for studying SLSNe with large surveys such as ZTF in Section~\ref{sec:disc}, and summarize our findings in Section~\ref{sec:conc}. Throughout this paper, we assume a flat $\Lambda$CDM cosmology with $\Omega_{\rm M} = 0.27$, $\Omega_{\Lambda} = 0.73$, and $H_0 = 70~{\rm km~s}^{-1}~{\rm Mpc}^{-1}$ \citep{ksd+11}.

\section{Data} 
\label{sec:data}

\subsection{ZTF Survey Overview}

The Zwicky Transient Facility \citep{bkg+19,gkb+19} is an optical time-domain survey utilizing a new, 47 square degree field of view camera \citep{dsb+16} on the Palomar 48-inch telescope. A public-private partnership, the time is divided between public surveys (40\%), surveys undertaken by the ZTF partnership (40\%), and Caltech surveys (20\%); an overview of the major surveys undertaken in Year 1 is given in \citet{bkb+19}. The data is processed real-time at IPAC \citep{mlr+19}, including a novel image differencing algorithm \citep{zog16} and machine learning based vetting of candidates \citep{mrw+19,dima2019}. Transient candidates are then distributed in alert packages using the Apache Avro format\footnote{\url{https://avro.apache.org/}} and distributed using the Apache Kafka streaming system\footnote{\url{https://kafka.apache.org/}} \citep{pbr+19}. 

The objects described in this paper were identified following alert filtering using the GROWTH marshal \citep{kcb+19}, first in a general ``science validation'' filter requiring multiple detections, positive subtractions, not coincident with a point source (using star-galaxy scores derived by \citealt{Tachibana2018} based on PS1 images), and not rejected as an artifact by the machine learning (though this threshold was set low, as the model was still being trained at this stage). Thus, these transients were not initially flagged as SLSN candidates, but rather identified as such as more photometry, spectroscopic follow-up or external information became available. Our goal here is not to present a carefully selected or complete sample, but rather to showcase the variety of luminous supernovae found in a short time period by ZTF, and discuss implications for SLSN selection strategies in large surveys. We focus our attention on the two most unusual objects, SN\,2018bgv and SN\,2018don, which could have indeed easily been missed by stricter filtering on ``typical'' SLSN properties such as long rise times, blue colors and/or faint host galaxies. The full first two year sample of ZTF SLSNe will be presented by Perley et al. (2020, in preparation) and Yan et al. (2020, in preparation).

\subsection{SLSNe found in early ZTF data}
In this section we list the details of the discovery and classification of the four SNe discussed in this paper. The coordinates, redshifts and Galactic extinction for each object are listed in Table~\ref{tab:slsn_list}. These objects were all detected by multiple surveys; we describe the survey first reporting the transient to the Transient Name Server\footnote{\url{}https://wis-tns.weizmann.ac.il/} (TNS) as the ``discovery'' regardless of when the first ZTF detection was. Since all the objects discussed here are from the beginning of the survey, the ZTF template images contain varying amounts of transient flux in most cases. Thus, the first ZTF alert issued is typically later than the first ZTF detection of the transient, as recovered in reprocessing the images. 

\subsubsection{ZTF18aaisyyp = SN\,2018avk}
SN\,2018avk was discovered by Gaia (as Gaia18ayq) on 2018 Apr 14 and reported to the TNS on 2018 Apr 16 \citep{2018avk}. The transient is clearly present in the ZTF reference images; running subtractions using a PS1 reference image as described in Section~\ref{sec:p48phot} we find the earliest ZTF detection to be 2018 Mar 25.32 at $g = 20.54 \pm 0.12~{\rm mag}$. Due to the presence of transient flux in the reference, the first ZTF alert was 2018 Apr 11.26 (as ZTF18aaisyyp). A spectrum taken with the Andalucia Faint Object Spectrograph and Camera (ALFOSC) on the 2.5m Nordic Optical Telescope (NOT) on 2018 May 04 gives a redshift of $z=0.132$ from narrow host galaxy lines, and features matching typical SN~Ic, consistent with the classification reported by \citet{ngb+18}. Combined with the light curve information, this indicates SN\,2018avk is a SLSN-I, with the first spectra taken slightly after peak.

\subsubsection{ZTF18aapgrxo = SN\,2018bym}
SN\,2018bym was discovered by ATLAS on 2018 May 10 (as ATLAS18ohj) and reported to the TNS \citep{2018bym,tdh+18}; the first ZTF detection (as ZTF18aapgrxo) was 2018 Apr 21.49, at $g = 20.94 \pm 0.14~{\rm mag}$. An initial spectrum taken with NOT+ALFOSC shows a blue continuum with shallow absorption features, consistent with the \ion{O}{2} absorptions typically seen in SLSN-I \citep{qkk+11,qdg+18} at a redshift $z \simeq 0.28$ (assuming an expansion velocity of $10,000~{\rm km~s}^{-1}$). Narrow emission lines in the classification spectrum reported by \citet{bgb+18} gives the host galaxy redshift $z=0.274$, which is also confirmed in our own follow-up spectroscopy.

\subsubsection{ZTF18aavrmcg = SN\,2018bgv}
SN\,2018bgv was discovered by Gaia (as Gaia18beg) on 2018 May 6 and reported to the TNS on 2018 May 8 \citep{2018bgv}. Again the ZTF reference image contains transient flux; running host subtraction with PS1 pre-explosion images we find that ZTF caught the full rise and peak of this transient, with the first detection on 2018 May 5.18 at $r = 19.73 \pm 0.06~{\rm mag}$. Again due to significant transient flux in the reference image, the first ZTF alert (as ZTF18aavrmcg) was not until 2018 May 22.17, at which point the transient had already been publicly classified as a SLSN-I at $z \simeq 0.08$ by \citet{dbs+18}. Narrow emission lines from the host galaxy seen in our subsequent spectra give a precise redshift of $z=0.0795$.

\subsubsection{ZTF18aajqcue = SN\,2018don}
SN\,2018don was first detected in ZTF data on 2018 Apr 14.32 as ZTF18aajqcue, though again it is also likely present in the reference images. A spectrum taken with the Double Beam Spectrograph (DBSP; \citealt{og82}) on the 200-in Hale telescope at Palomar Observatory (P200) gives the redshift $z=0.0734$ from narrow emission and absorption lines, and shows spectroscopic features similar to slowly evolving SLSNe such as SN\,2007bi \citep{gmo+09} and PS1-14bj \citep{lcb+16}. 
This long-lived transient was also discovered by Pan-STARRS (as PS18aqo) on 2018 Jun 16, and reported to the TNS by them on 2018 Jul 13, given the IAU name 2018don \citep{2018don}.

\subsection{Photometry} 
\label{sec:phot}

\subsubsection{P48 photometry}
\label{sec:p48phot}

The ZTF photometric pipeline and data products are described in \citet{mlr+19}. However, since the supernovae discussed here were discovered at the very beginning of the survey, the objects have varying degrees of supernova flux present in the reference images, meaning the pipeline photometry in the alert packages will report magnitudes that are too faint. Additionally, if an object falls on multiple fields/quadrants each of these will have a separate reference image which may in turn contain different amounts of supernova flux, resulting in considerable scatter in the default photometry -- this is the case for SN\,2018don (3 fields) and SN\,2018avk (2 fields).

\begin{figure*}
\centering
\begin{tabular}{cc}
\includegraphics[width=3.5in]{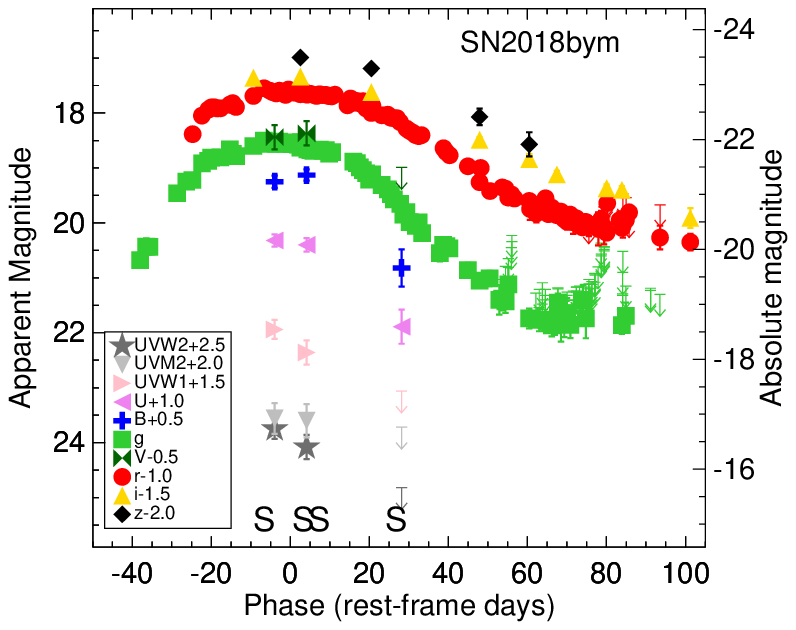} & \includegraphics[width=3.5in]{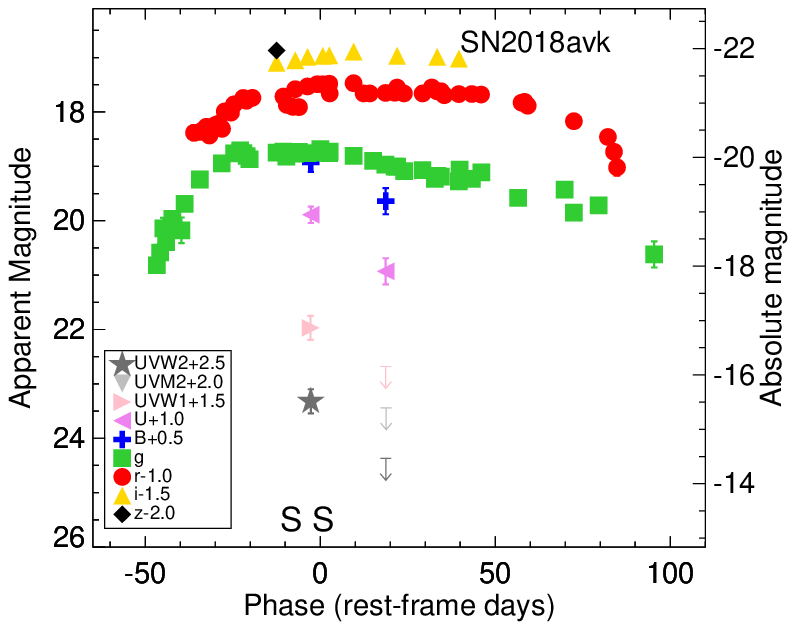} \\
\includegraphics[width=3.5in]{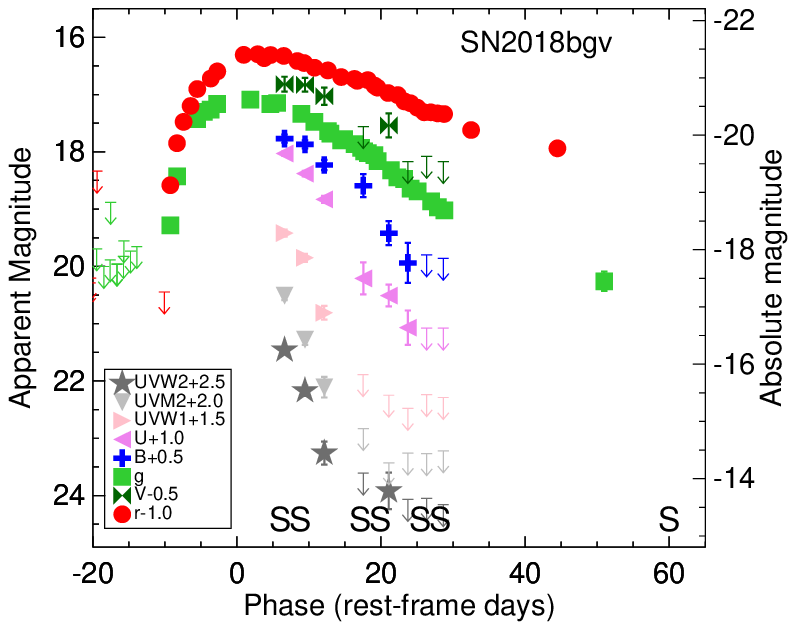} &
\includegraphics[width=3.5in]{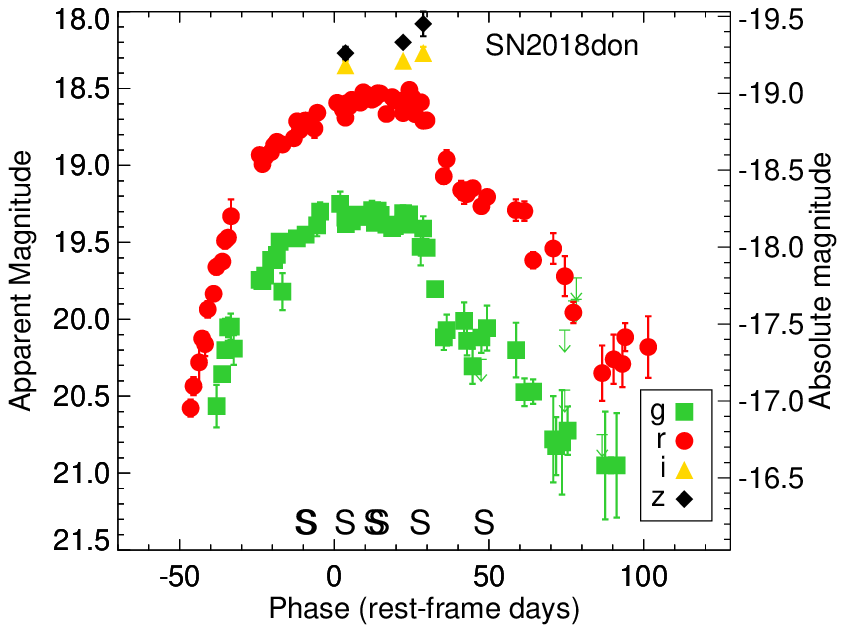} 
\end{tabular}
\caption{Light curves of the four SLSN-I. Filters are offset as indicated in the individual legends. Phase is relative to the (rest-frame) $g$-band peak. ZTF data has been binned by day for presentation purposes. Epochs of spectroscopy are marked with the letter `S' along the bottom axis. 
The absolute magnitude y-axis on the right side of each plot is calculated using an approximate, constant $K$-correction of $2.5\times\log(1+z)$. The difference to the full $K$-correction near peak is generally $< 0.1$~mag in the filters we have the spectroscopic coverage to compute; $K$-corrected $g$-band light curves are shown in Figure~\ref{fig:lc_comp_slsni}.
\label{fig:lc}}
\end{figure*}

For this reason, we reprocess the light curves using the ZTF forced photometry service \citep{mlr+19}, which performs forced PSF-fit photometry on the archived difference images. This allows us to adjust the flux baseline for each field/filter combination using data taken after the supernovae has faded below ZTF detectability (generally, we find this to be the case for the 2019 observing season). This will correct the photometry for any transient flux present in the template, but will not recover any of the epochs from the time period that went into making the template images used. For two objects, SN\,2018avk and SN\,2018bgv, a significant part of the light curve is contained in the images that went into building the reference. For these, we use FPipe \citep{fst+16} to redo the subtractions using Pan-STARRS1 (PS1; \citealt{cmm+16,fmc+16}) images as references. We find that the photometry from the two pipelines agree well up to a small zeropoint offset, and shift the FPipe photometry to be consistent with the IPAC forced photometry. All photometry from the first observing season, including that recovered from the reference-building images, is reported in Table~\ref{tab:phot}, and the light curves are shown in Figure~\ref{fig:lc}.

\subsubsection{LT photometry}
We acquired multi-filter observations of several of the SLSNe discussed in this work with the optical imager (IO:O) on the robotic Liverpool Telescope \citep[LT;][]{ssr+04} located at the Observatorio del Roque de los Muchachos on La Palma.  Observations were taken in the $g$, $r$, $i$, or $z$ bands.  We use reduced images provided by the basic IO:O pipeline, stacking using SWarp in cases where multiple exposures were taken on a given night. Digital image subtraction was performed versus PS1 reference imaging, again following the techniques of \citet{fst+16}.  PSF photometry was performed relative to PS1 photometric standards.

\subsubsection{Wise photometry}
In addition, we observed SN\,2018bgv in $g'$ and $r'$ band with the 0.71-m C28 Jay Baum Rich and the 1-m telescopes at the Wise Observatory in Israel. The data were reduced in a standard fashion, including bias correction and flat fielding and the calibration of the world-coordinate system, using the Matlab package for astronomy and astrophysics \citep{Ofek2014a}.

These data were obtained when the host contamination was minimal. Hence, we applied aperture photometry, using the tool presented in \citet{skl+18}\footnote{\href{https://github.com/steveschulze/Photometry}{https://github.com/steveschulze/Photometry}}, to extract the light curve. To measure the zeropoint of each image, we measured the brightness of several stars in the same way and compared their instrumental magnitudes against tabulated measurements in the SDSS DR9. To mitigate colour differences between Wise and ZTF filter system, we shifted the $g'$-band data by 0.05~mag and the $r'$-band data by 0.1~mag.

\subsubsection{Swift photometry}
\label{sec:uvot}
Three of the supernovae, SN\,2018bym, SN\,2018avk and SN\,2018bgv had UV data obtained with the UV/Optical Telescope \citep[UVOT;][]{rkm+05} aboard the \textit{Neil Gehrels Swift Observatory} \citep{gcg+04}. We retrieved the UVOT data from the NASA \textit{Swift} Data Archive\footnote{\url{https://heasarc.gsfc.nasa.gov/cgi-bin/W3Browse/swift.pl}}, and used the standard UVOT data analysis software distributed with {\tt HEASoft} version 6.19\footnote{\url{https://heasarc.nasa.gov/lheasoft/}}, along with the standard calibration data to process it. 

In each case, we combined individual integrations of a given epoch in each band using the command {\tt uvotimsum}. For SN\,2018avk and SN\,2018bym, we measured the flux in a 5\arcsec\ aperture using the tool {\tt uvotsource}. No host galaxy corrections were applied for these two sources, but the UVOT observations were obtained near peak; based on the host galaxy brightnesses (Section~\ref{sec:hosts}) we do not expect this data to be significantly affected by host contamination.

For SN\,2018bgv, we do see the UVOT fluxes leveling off in the later epochs suggesting host contamination. To correct for this, we use the data taken between June 22 2018 and November 10 2018 to build a host template. We then measure the flux of the transient in a 6\arcsec\ circular aperture, and quantify the host flux by applying the same aperture to the template images. We then numerically subtract the host flux from the transient flux to recover the transient photometry. All Swift photometry is reported in Table~\ref{tab:phot}.

\subsection{Spectroscopy}
\label{sec:spec_obs}
\begin{figure}
\centering
\includegraphics[width=3.5in]{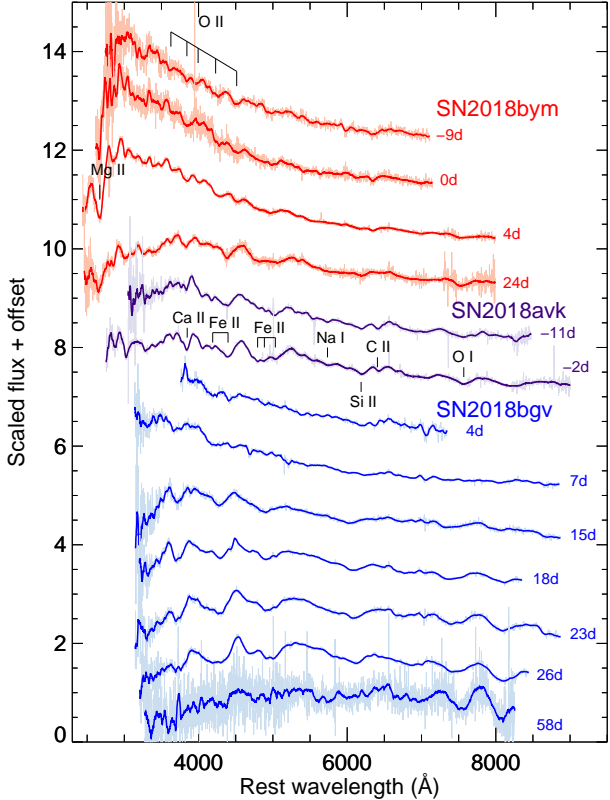}
\caption{Spectra of three of the SLSN-I: SN\,2018bym (red), SN\,2018avk (blue) and SN\,2018bgv (black). Spectra have been smoothed with a Savitzky-Golay filter; the unsmoothed spectra are shown in the background. Strong host galaxy emission lines have been clipped. As is typical of SLSN-I, pre-peak spectra are dominated by a blue continuum with shallow absorptions, while later-phase spectra resemble those of SN~Ic. Some of the strongest typical spectral features are marked. 
\label{fig:slsni_spec}}
\end{figure}

\begin{figure}
    \centering
    \includegraphics[width=3.5in]{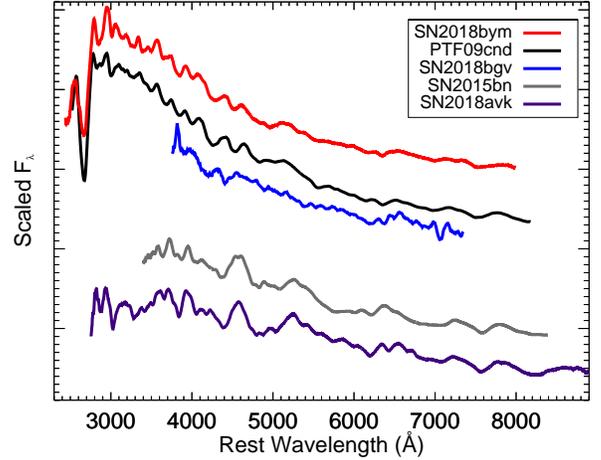}
    \caption{Comparisons of the near-peak spectra of the three SLSN-I shown in Figure~\ref{fig:slsni_spec}. The top three spectra show a comparison of SN\,2018bym and SN\,2018bgv with PTF09cnd \citep{qkk+11,qdg+18} -- note the characteristic blue continuum and the \ion{O}{2} ``W''-feature around 4500~\AA, typical of SLSN-I in the hot photospheric phase. SN\,2018avk is shown on the bottom compared to a spectrum of SN\,2015bn \citep{nbs+16}. The phase of the SN\,2015bn comparison spectrum (30 days post peak) was chosen for having a similar temperature to SN\,2018avk (which is considerably cooler than SN2018bym and SN2018bgv; see Section~\ref{sec:bbfits}); the features, which match up well, are typical of SLSN-I in the cool photosperic phase. }
    \label{fig:slsni_spec_comp}
\end{figure}

\begin{figure}
\centering
\includegraphics[width=3.5in]{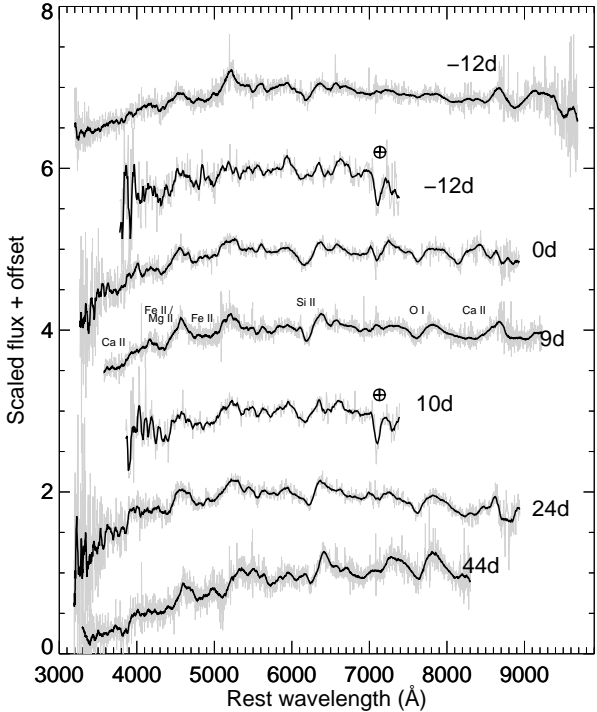}
\caption{Spectroscopic sequence of SN\,2018don. Unsmoothed spectra are shown in gray, while the spectra shown in black have been smoothed with a Savitzky-Golay filter. Areas strongly affected by telluric absorption are marked with a $\oplus$ symbol, and some of the strongest features are marked on the 9-day spectrum. Unlike the objects shown in Figure~\ref{fig:slsni_spec}, SN\,2018don shows a significantly redder spectrum and very slow spectroscopic evolution. 
\label{fig:18cue_spec}}
\end{figure}

\begin{figure}
\centering
\includegraphics[width=3.5in]{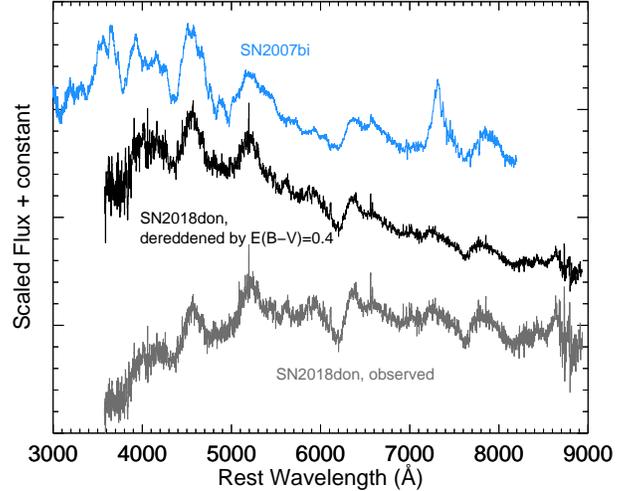}
\caption{Spectrum of SN\,2018don compared to that of SN\,2007bi. The bottom (gray) shows a stack of the near-peak SN\,2018don spectra, and the middle the same spectrum when dereddened by $E(B-V) = 0.4$~mag and assuming $R_{\rm V} = 3.1$. This spectrum is a good match to the first spectrum taken of SN\,2007bi, shown in blue, with the exception of the strong [\ion{Ca}{2}] emission seen from early times in SN\,2007bi. Spectra have been arbitrarily scaled and binned for display purposes.
\label{fig:07bi_vs_cue}}
\end{figure}

We obtained classification and follow-up spectra using the Double Beam Spectrograph (DBSP; \citealt{og82}) on the 200-in Hale telescope at Palomar Observatory; the Andalucia Faint Object Spectrograph and Camera (ALFOSC) on the 2.5m Nordic Optical Telescope (NOT); the SPectrograph for the Rapid Acquisition of Transients (SPRAT; \citealt{pia+14}) on the 2m Liverpool Telescope (LT); and the Low Resolution Imaging Spectrometer (LRIS; \citealt{occ+95}) on the 10m Keck~I telescope. Table~\ref{tab:spec} lists the details of the spectroscopic observations, and the spectra, along with classification comparisons, are displayed in Figures~\ref{fig:slsni_spec} - \ref{fig:07bi_vs_cue}.

Spectra were reduced using standard methods, including wavelength calibration against an arc lamp and using a spectrophotometric standard star for the flux calibration, using available pipelines where possible.
For SPRAT spectra we used the automatic reductions supplied in the archive; DBSP spectrum were reduced using a PyRAF-based pipeline \citep{Bellm2016}; LRIS spectra were reduced using LPipe \citep{Perley2019}.

\subsection{Host galaxy photometry}
\label{sec:host_phot}

To measure the brightness of the host galaxies from the rest-frame UV to the near-IR, we retrieved science-ready images from the archives of the Galaxy Evolution Explorer Data Release 7 \citep[GALEX;][]{galex}, Panoramic Survey Telescope And Rapid Response System DR1 \citep[PanSTARRS;][]{fmc+16}, Sloan Digital Sky Survey DR9 \citep[SDSS;][]{yaa+00}, Two Micron All-Sky Survey \citep[][]{2mass}, and and the unWISE \citep{Lang2014a} images from the NEOWISE Reactivation Year-3
\citep{Meisner2017a}. Furthermore, we augmented the data set of SN\,2018bym with deeper optical images from the Canada-France-Hawaii Telescope (CFHT, also science-ready) and complemented the data set of SN\,2018bgv with UV data obtained with \textit{Swift}/UVOT. The UVOT images were reduced as described in Section~\ref{sec:uvot}.

We used the software package {\tt LAMBDAR} (Lambda Adaptive Multi-Band Deblending Algorithm in R) \citep{Wright2016a}, which is based on a software package written by \citet{bmd+12}, to measure the brightness of each host. {\tt LAMBDAR} has three major input parameters: the point spread function (PSF), the zeropoint and source catalogue (coordinates and parameters of the elliptical apertures) of a given image. With this information in hand, {\tt LAMBDAR} can accurately position an aperture on the pixel grid, modify its shape by the PSF of the to-be-analysed image, and accurately measure the brightness while preserving the intrinsic colour.

We used the point-spread functions provided by the \textit{GALEX} in the \textit{GALEX} Technical Documentation\footnote{\href{http://www.galex.caltech.edu/wiki/Public:Documentation}{http://www.galex.caltech.edu/wiki/Public:Documentation}}. For unWISE images, we used the parameterisation of the PSF from \citet{Lang2014a} to generate templates for $W1$ and $W2$. For 2MASS, PS1 and SDSS images, we extracted the median full-width half-maximum of point sources in each image and assumed that the PSFs can be approximated by Gaussian profiles.

We used the tabulated zeropoints for \textit{GALEX}, PanSTARRS, SDSS and unWISE. Specifically, we set the zeropoint of the \textit{GALEX} FUV and NUV images to 18.82 and 20.08~mag(AB) \citep{mcb+07}, SDSS images to 22.5~mag(AB)\footnote{\href{https://www.sdss.org/dr12/algorithms/magnitudes/}{https://www.sdss.org/dr12/algorithms/magnitudes/}}, PS1 images to $25 + 2.5 \times \log ({\rm Exposure~time})$~mag(AB)\footnote{\href{https://outerspace.stsci.edu/display/PANSTARRS/PS1+FAQ+-+Frequently+asked+questions}{https://outerspace.stsci.edu/display/PANSTARRS/PS1+FAQ+-+Frequently+asked+questions}}, where the exposure time is given in seconds, and unWISE W1 and W2 images to 22.5~mag(Vega) \citep{Lang2014a}. To measure the zeropoint of the 2MASS images, we identified several tens of stars in a given field from the 2MASS Point source catalogue and compared the instrumental magnitudes to the tabulated magnitudes. The 2MASS and unWISE zeropoints were converted from the Vega to the AB system using the offsets reported in \citet{Cutri2013a} and \citet{br07}.

For SN\,2018avk, there is a faint galaxy visible $\sim 2.5\arcsec$ east of the transient position ($\sim 6$~kpc if at the same redshift). However, there is also diffuse emission at the location of the transient, and strong galaxy emission lines seen in the supernova spectra -- given that 6~kpc would be an unusually large offset for a SLSN-I (e.g., \citealt{lcb+15,skl+18}), we assume this diffuse emission is the true host galaxy. We perform photometry at the transient location in the same manner as with the other objects, using an unconvolved aperture radius of 0.75\arcsec.

Table~\ref{tab:host_phot} summarizes all host measurements.

\section{Light curve properties}
\label{sec:lightcurves}

\subsection{Peak luminosities and light curve timescales}
As we do not, in general, have multicolor data available, we do not attempt to construct bolometric light curves for the purposes of measuring light curve properties. Instead, we use the $g$-band (where available) as our baseline for computing light curve properties such as peak magnitudes and timescales, allowing for easy comparison with the PTF SLSN-I sample published in \citet{dgr+18}. We $K$-correct the light curves to rest-frame $g$-band using our available spectra; for SN\,2018bgv, SN\,2018avk and SN\,2018don we $K$-correct from observed $g$-band while for the slightly higher-redshift SN\,2018bym we calculate the cross-filter $K$-correction from observed $r$-band to rest-frame $g$-band. For light curve points before the earliest spectrum we use the first $K$-correction measured; for light curves point after the last spectrum we use the last spectrum measured, and for points in between we linearly interpolate between the two closest values. This approach has some drawbacks: as our spectroscopic coverage varies and we will only have a few spectra for some objects, we have to extrapolate the $K$-correction which introduces uncertainty. On the other hand, the alternative of using some well-measured SLSN (e.g. PTF12dam, as was used by \citealt{dgr+18}) will introduce uncertainties due to the potentially different color and spectroscopic evolution of PTF12dam compared to our objects. In practice, since the redshifts of our objects are all $z < 0.3$, the difference between these choices are generally $< 0.1$~mag.

To measure light curve timescales, we first smooth/interpolate the light curves using Gaussian Process regression. Following \citet{ipg+18} and \citet{ass+19}, we use the Python package {\tt george} \citep{afg+15} and an optimized Matern~3/2 kernel to perform the fits. We then measure the peak luminosity in the $g$-band, and the rise and decline timescales, defined as when the supernova flux is a factor of 2 below peak. The resulting peak times, peak luminosities, and timescales are reported in Table~\ref{tab:lc_prop}. 

\begin{figure*}
\centering
\begin{tabular}{cc}
\includegraphics[width=3.5in]{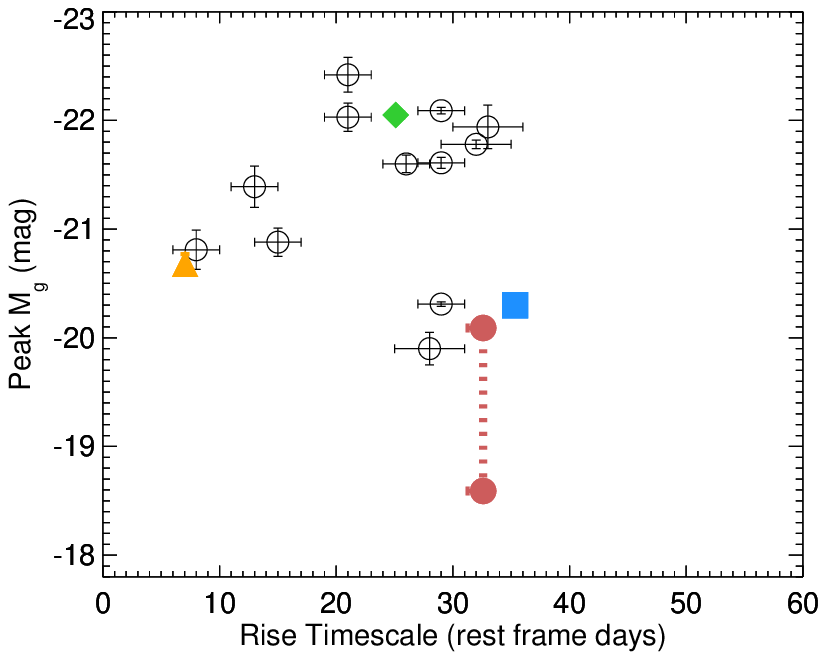} & \includegraphics[width=3.5in]{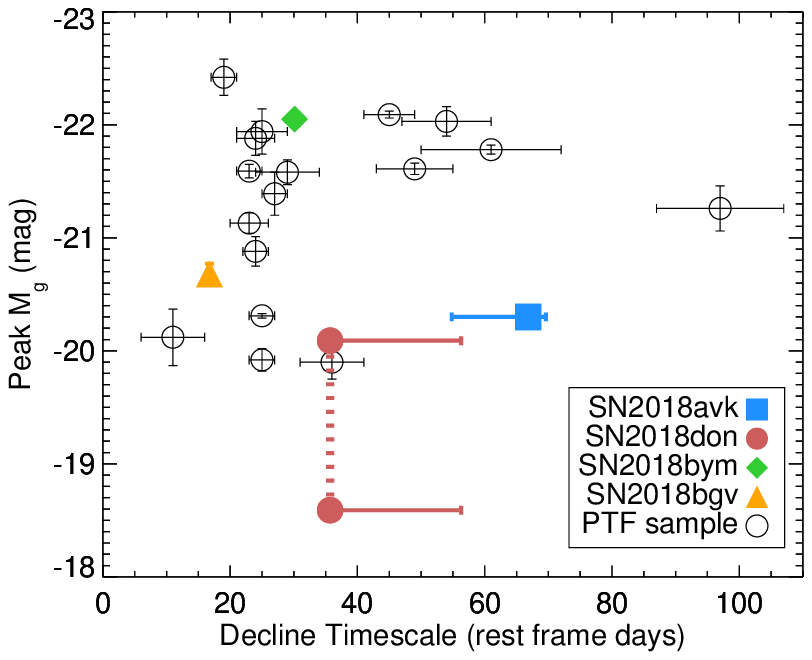}
\end{tabular}
\caption{Peak absolute $g$-band magnitude plotted against rise timescale (left) and decline timescale (right) for the H-poor SNe in our sample. Both timescales are defined as the time between peak and half flux. The smaller circle for SN\,2018don shows the peak $g$-band magnitude if correcting for host galaxy reddening of $E(B-V)=0.4$, corresponding to a 1.5~mag shift in $g$-band. The PTF sample from \citet{dgr+18} is shown as circles for comparison.
\label{fig:timescales}}
\end{figure*}

Figure~\ref{fig:timescales} shows the peak g-band luminosities plotted against rise- and decline timescales, respectively. SN\,2018bym and SN\,2018bgv both fall within the general cloud of measurements of the \citet{dgr+18} sample, though SN\,2018bgv has a faster rise than any of the objects measured. SN\,2018don is both significantly fainter and has a longer rise than most of the PTF objects - if the $g$-band suffers from $\gtrsim 1.5$~mag extinction, as suggested by spectral comparisons, it would be compatible with the PTF sample luminosity-wise. The steep drop in the light curve of this object affects the measured decline time -- had we chosen a different timescale measure (e.g. one magnitude below peak), we would instead have measured a decline timescale of 59~days, closer to that of SN\,2018avk. We also note that both SN\,2018don and SN\,2018avk are examples of relatively faint and long-lived SLSNe, a combination that is rare in the PTF sample.

Since we have coverage of the rising phase of the light curve in $g$-band for all four objects, we use this to also estimate the explosion dates. We do this by fitting a second-order polynomial to the rise of the observed $g$-band light curves in flux space, and extrapolating the fit to zero flux. (For SN\,2018don, where the observed $r$-band extends earlier, we use the observed $r$-band for this calculation instead, though note that we get consistent explosion dates within the error bars using either filter.) The estimated explosion dates for each object are also listed in Table~\ref{tab:lc_prop}. Note that the sharp, well-constrained rise of SN\,2018bgv caught in the ZTF data implies that this object had a total rise-time from explosion to peak of only 10 days. 

We also note that none of the SLSNe discussed here show evidence of a pre-peak light curve bump, which has been suggested to be an ubiquitous feature of SLSN-I (\citealt{ns16}; but see also \citealt{ass+19}). Our coverage of the earliest phases of the light curves is complicated by the issues with supernova flux in the template images, which result in either simply missing coverage of the earliest phases, or increased uncertainty where we had to use PS1 template images for subtraction in order to recover the early light curve. The full ZTF SLSN sample will be better equipped to address this question in more detail.

\subsection{Comparison to other SLSNe}

Figure~\ref{fig:timescales} gives some indication of how the SLSN-I discussed here compare to other SLSNe in terms of luminosity and timescale. A more detailed comparison can be done in the phase-space of luminosity, timescale and color, for example using the framework proposed by \citet{ipg+18} in their attempt to define SLSN-I statistically. We use our spectra to calculate $K$-corrections to the fiducial 400~nm and 520~nm bandpasses used in \citet{ipg+18}, and calculate the peak luminosity and decline rate over 30 days in the 400~nm filter, as well as the $400-520$ color at peak and at 30 days. We find that only SN\,2018bym falls within their statistical definition of the ``4-observables parameter space''\footnote{These four parameters are (1) the peak luminosity in the 400~nm filter, $M(400)_0$;
(2) the decline in magnitudes in the 400~nm filter over the 30\,days following peak brightness, $\Delta M(400)_{30}$;
(3) the $400-520$ color at peak, $M(400)_0 - M(520)_0$;
(4) and the $400-520$ color at +30 days, $M(400)_{30}- M(520)_{30}$}: SN\,2018avk is fainter than their sample given its slow decline and relatively blue color 30 days after peak, while SN\,2018bgv's color evolution is faster than their sample (i.e., redder than expected 30 days after peak given the very blue color at peak). SN\,2018don is, as expected, both fainter and redder than their sample, but this is not surprising if this object is suffering significant host extinction. We note that even when assuming host extinction according to $E(B-V)=0.4$ (which corresponds to 1.8~mag in the 400~nm band and 1.3~mag in the 520~nm band), this object is still too faint and red to be consistent with the ``4OPS'' parameter space.

\begin{figure*}
    \centering
    \begin{tabular}{cc}
    \includegraphics{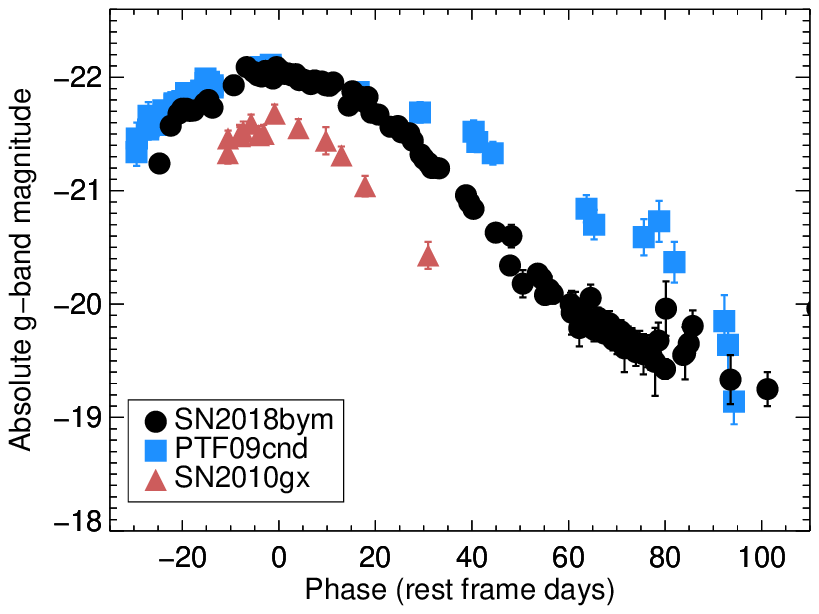} & \includegraphics{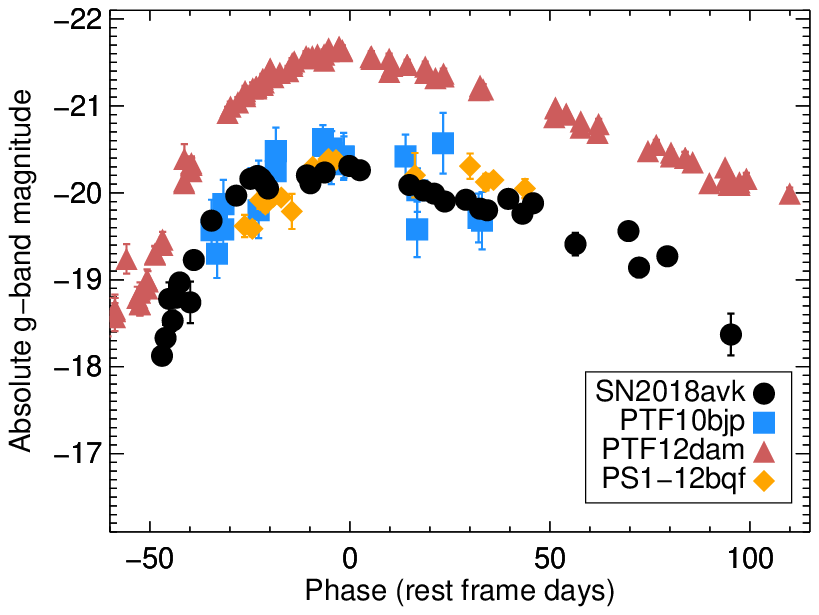} \\
    \includegraphics{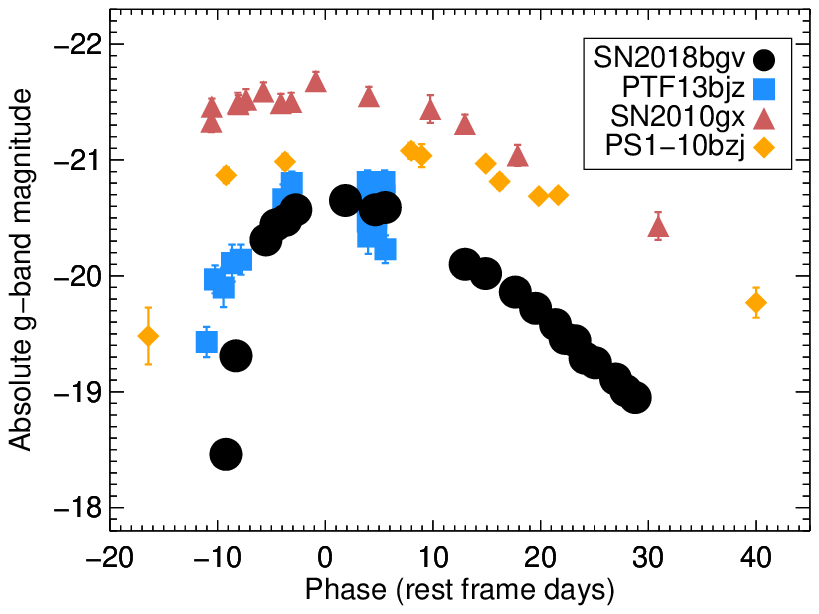} & \includegraphics{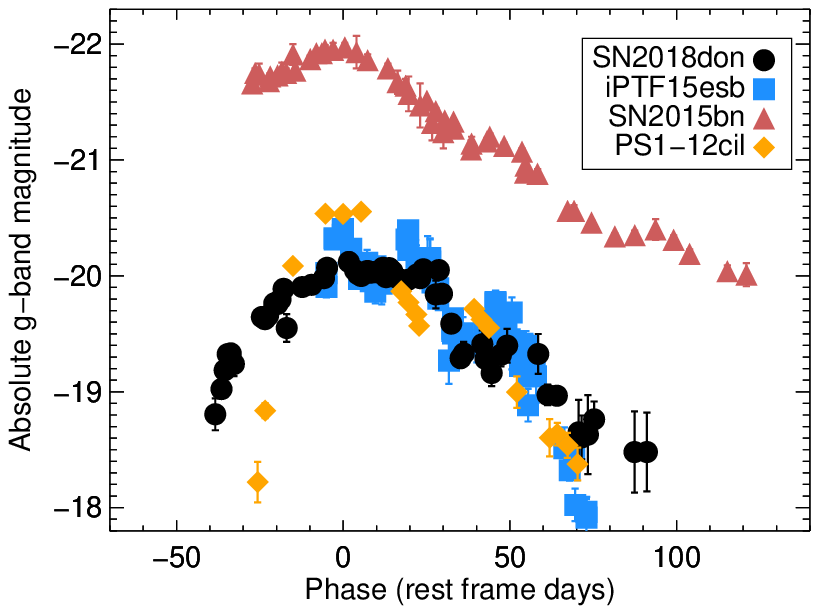}
    \end{tabular}
    \caption{Light curves of the four SLSNe compared to some similar SLSNe from the literature. Comparison data from \citet{dgr+18}, \citet{lcb+13,lcb+18}, \citet{nbs+16} and \citet{ylp+17}. ZTF data has been binned by day for presentation purposes. $K$-corrected data is used where available; in the case of slightly higher-redshift objects (PS1-12bqf, PS1-10bzj and PS1-12cil) the observed filter closest to rest-frame $g$-band is plotted with an approximate $K$-correction of $2.5\times \log(1+z)$. The light curve of SN\,2018don has been shifted upwards by 1.5~mag, as based on our estimate of the extinction from comparing the spectrum to SN\,2007bi (Figure~\ref{fig:07bi_vs_cue}).}
    \label{fig:lc_comp_slsni}
\end{figure*}

Figure~\ref{fig:lc_comp_slsni} shows these same four supernovae compared to some objects with similar light curves in the literature. SN\,2018bym has a decay slope similar to SN\,2010gx, but its peak luminosity and rise slope more resembles PTF09cnd. The timescales of SN\,2018avk are similar to the slowly-evolving PTF12dam, but the peak luminosity is quite a bit fainter, and the overall lightcurve resembles PTF10bjp and PS1-12bqf. For SN\,2018bgv, we do not find any clear equivalents in the literature -- the decay slope may be similar to fast-evolving objects like PS1-10bzj and SN\,2010gx, but the rise is faster and the peak luminosity fainter than either of these. The early light curve appears similar to PTF13bjz, but this object has very sparse data so we cannot ascertain exactly how similar these are.

SN\,2018don, with its steep post-peak drop and subsequent plateau, is shown in comparison to several other literature light curves with pronounced bumps. The most similar in terms of the post-peak slope and second bump is PS1-12cil, but this object had a much faster rise and narrower main peak, so that the rise and decay slopes are more consistent. Similarly, iPTF15esb displays a series of bumps, and is similar to SN\,2018don on the decline -- as the rise was not observed in this object, we cannot ascertain whether its overall light curve would have been more similar to PS1-12cil or SN\,2018don. Long-lived SLSNe such as  SN\,2015bn have also been shown to display post-peak bumps in their light curves, but with less contrast between the bump and an otherwise smooth and slow decline. 

\subsection{Color evolution}
\begin{figure}
\centering
\includegraphics[width=3.5in]{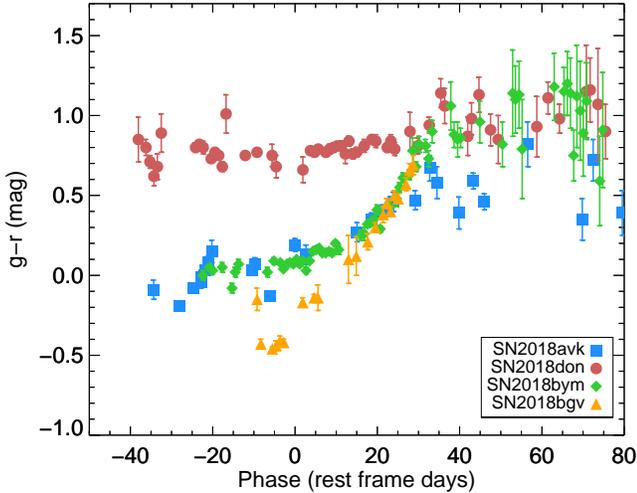}
\caption{Observed ZTF $g-r$ color of the four supernovae. Color is calculated at the time of the $g$-band observations (binned by day), and using Gaussian Process interpolation to get the $r$-band magnitude and uncertainties at the corresponding times. Two objects stand out: SN\,2018don, which shows a nearly constant color evolution and is much redder than the other objects; and SN\,2018bgv which evolves much faster than the other supernovae.
\label{fig:gr_color}}
\end{figure}

An advantage of the Gaussian Process interpolation is that it naturally outputs an estimate of the predicted magnitude and associated uncertainty between data points. We use these $g$-band magnitudes, together with the interpolated $r$-band magnitudes, to calculate the $g-r$ colors of our supernovae. The resulting color curves are shown in Fig.~\ref{fig:gr_color}. 

With the exception of SN\,2018don, all the supernovae show a color starting out blue and turning redder with time, consistent with an expanding and cooling photosphere. SN\,2018bgv stands out in both having the bluest initial colors, and in having the fastest color evolution by far. SN\,2018don is significantly redder than the remaining objects throughout its entire evolution. As previously noted, this object could be experiencing significant host galaxy reddening. Additionally, the color evolution of SN\,2018don is significantly flatter than the other objects, and as such resembles other slowly-evolving SLSNe such as PS1-14bj \citep{lcb+16}.

We also note that the color of SN\,2018don changes during the rapid decline and rebrightening, in particular that the decline in $g$-band is larger than in $r$-band. We discuss possible interpretations of this light curve feature and associated color change in Section~\ref{sec:18cue_disc}.

\subsection{Blackbody fits}
\label{sec:bbfits}
We use the code {\tt PhotoFit} \citep{sgg+19} to fit blackbody functions to epochs where we have at least three filters available (typically, epochs of either \textit{Swift}/UVOT or LT observations). As described in the appendix of \citet{sgg+19}, PhotoFit first interpolates the light curve in each filter onto the given epoch, then fits a Planck function at each epoch after correcting for extinction, redshift and filter transmission curves. Both of these fits are performed using Monte Carlo Markov Chain simulations using {\tt emcee} \citep{fhl+13}, allowing for uncertainties to be calculated at each epoch.

\begin{figure*}
    \centering
    \begin{tabular}{cc}
    \includegraphics[width=3.5in]{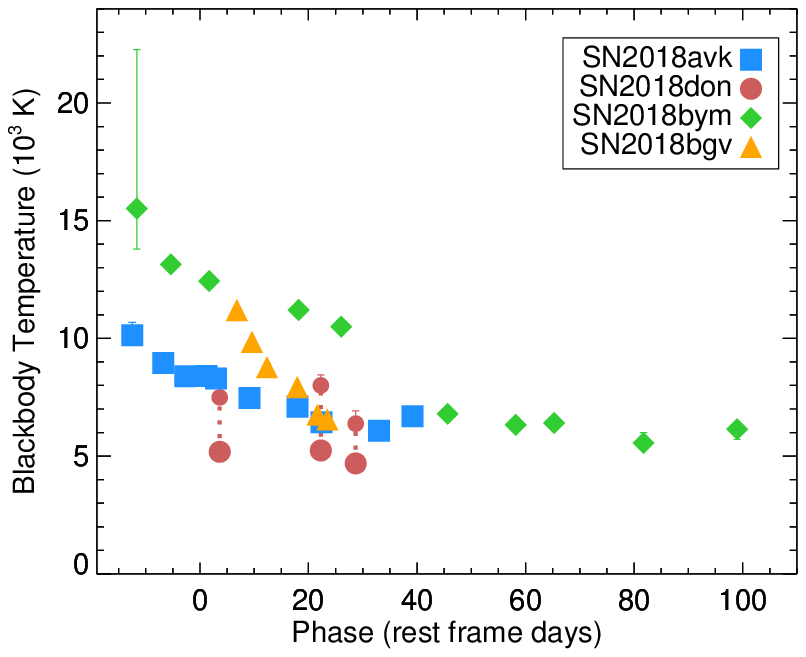} & \includegraphics[width=3.5in]{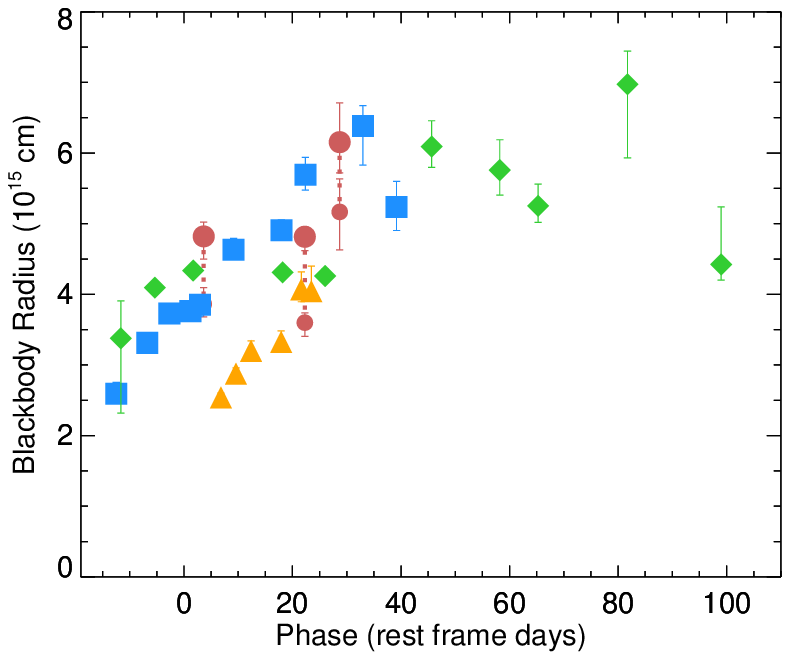}
    \end{tabular}
    \caption{Derived blackbody temperatures (left) and radii (right) from fitting a Planck function to photometry epochs with at least three filters available. The smaller points for SN\,2018don show the inferred values if assuming host galaxy extinction according to $E(B-V)=0.4$~mag.
    \label{fig:bbfits}}
\end{figure*}

The resulting best-fit blackbody temperatures and radii are shown in Figure~\ref{fig:bbfits}. SN\,2018bym and SN\,2018bgv show the hottest color temperatures, which is consistent with these two objects also displaying spectra with features corresponding to higher temperatures, such as \ion{O}{2}, over this time period. As with the color and overall light curve evolution, SN\,2018bgv shows the fastest temperature evolution. 
The two different values plotted for each epoch of SN\,2018don correspond to the results when including the potential host galaxy reddening of $E(B-V)=0.4$~mag versus not; as is seen with the colors, its temperature is close to flat over the time period we have multicolor photometry.) The derived blackbody radii are high ($>10^{15}$~cm), which is a natural consequence of the high luminosities. As expected, the blackbody radii are increasing with time up until and past peak in all the objects, as the ejecta expand and cool. We follow the evolution of SN\,2018bym for the longest; after $\sim 40$~days the temperature plateaus around 6000-7000~K and the blackbody radius starts to decrease, indicating that the photosphere is receding into the ejecta faster than the ejecta is expanding. A similar behaviour is seen in other SLSN-I followed to late times (e.g., \citealt{nbs+16}). 

\subsection{Magnetar model fits}
\label{sec:magfits}
The magnetar spin-down model (e.g., \citealt{kb10,woo10}) has been shown to be successful in reproducing the light curves of a wide variety of SLSN-I \citep{isj+13,dhw+12,msp+16,ngb17,dgr+18}. Fitting a magnetar model to our light curves thus offers both insight in the required parameters to explain the observed data, and another point of comparison with the observed population of SLSN-I. We use the Modular Open Source Fitter for Transients code ({\tt MOSFiT}; \citealt{gnv+18}), and fit a {\tt slsn} model (consisting of a magnetar engine, and an SED comprising a blackbody with some UV suppression). For speedup, we bin our light curves by day. Following the procedure outlined in \citet{ngb17}, we largely use the default priors, except we restrict the prior on the photospheric velocities to the range allowed by our spectroscopic measurements (Section~\ref{sec:velocity}).

\begin{figure*}
    \centering
    \begin{tabular}{cc}
    \includegraphics[width=3.4in]{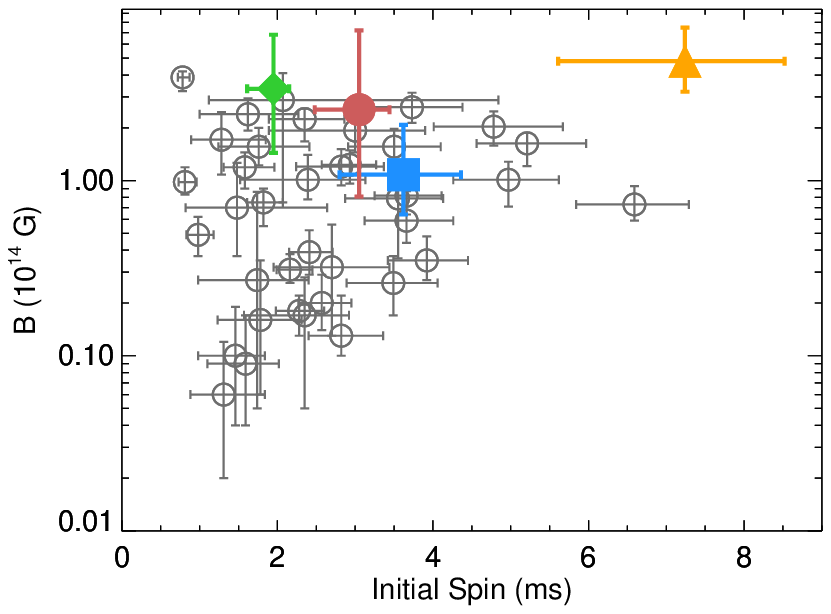} & \includegraphics[width=3.4in]{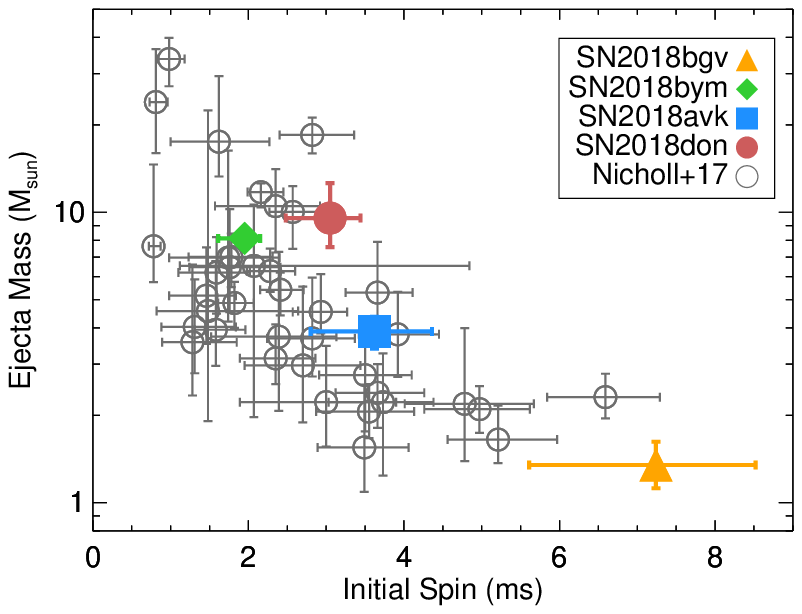}
    \end{tabular}
    \caption{Magnetic field (left) and ejected mass (right) vs. initial spin period derived from the {\tt MOSFiT} magnetar fits. The compilation from \citet{ngb17} is shown as gray circles. Note that the fit for SN\,2018don includes a host extinction of $A_{\rm V} = 2.26$~mag. SN\,2018bym, SN\,2018avk and SN\,2018don fall in the general locus of the literature compilation, while SN\,2018bgv is an outlier with lower ejecta mass and higher initial spin and magnetic field.}
    \label{fig:magfits}
\end{figure*}

The key parameters from these fits are summarized in Table~\ref{tab:magfits}. Figure~\ref{fig:magfits} show the main magnetar parameters (ejecta mass, magnetic field $B$, initial spin $P$) compared to the results from the compilation in \citet{ngb17}. We see that SN\,2018bym, SN\,2018avk and SN\,2018don all fall within the general locus of the SLSN-I population, indicating that if they are powered by magnetars, they would come from a similar source population as previously observed objects. The fit to SN\,2018don converged on a high host extinction of $A_{\rm V} = 2.26$~mag, while the derived host extinction for the other objects is negligible. This supports our interpretation that SN\,2018don is consistent with a SLSN-I undergoing significant host extinction. The parameters for SN\,2018bgv, with a lower ejecta mass, higher magnetic field, and high spin period, set it apart from the literature sample. We discuss potential powering mechanisms for this supernova in more detail in Section~\ref{sec:bgv_disc}. 

\section{Spectroscopic properties}
\label{sec:spec}

\subsection{Spectral Features and Comparisons}
The spectra of SN\,2018bym, SN\,2018bgv and SN\,2018avk all show features that are typical of SLSN-I (e.g., \citealt{qdg+18}). SN\,2018bym and SN\,2018bgv, in particular, show the blue continuum and weak absorption features typical of SLSN-I in the early, hot photospheric phase \citep{gal19}. The full set of \ion{O}{2} features, marked on Figure~\ref{fig:slsni_spec}, is clearly displayed in SN\,2018bym, which is a particularly good spectroscopic match to the prototypical event PTF09cnd \citep{qkk+11,qdg+18}, shown in Figure~\ref{fig:slsni_spec_comp}. The strongest feature in the \ion{O}{2} series, the ``W''-feature around 4500~\AA\,, is also clearly visible in the first spectrum of SN\,2018bgv (Figures~\ref{fig:slsni_spec}, \ref{fig:slsni_spec_comp}).

As the ejecta expand and cool, the spectra transition to the cool photospheric phase, showing features resembling SNe~Ic, with features from \ion{Ca}{2}, \ion{Fe}{2}, \ion{O}{1}, \ion{Si}{2} and \ion{C}{1}, among others. Both of our spectra of SN\,2018avk show features typical of this phase, despite being taken near peak. We note, however, that given the long rise of SN\,2018avk, this still translates to $>50$~days past explosion, and that the flat shape of the peak means that defining the peak epoch is tricky and could conceivably also be set $\sim 25$~days earlier. Regardless, the features seen in the spectrum are completely consistent with the cooler temperature of SN\,2018avk at this phase (Section~\ref{sec:bbfits}), and with SLSNe in the cool photospheric phase in general. Figure~\ref{fig:slsni_spec_comp} shows a comparison to SN\,2015bn at a comparable temperature (taken 30~days post peak), which is an excellent match.

The spectrum of SN\,2018don is significantly redder than the other objects, and shows features generally consistent with a Type I SN -- while spectral comparison codes like SNID \citep{snid} and Superfit \citep{hsp+05} finds matches to both SNe Ia and Ic, the light curve rules out a SN~Ia interpretation. Fig.~\ref{fig:07bi_vs_cue} shows a comparison of the near-peak spectrum of SN\,2018don compared to the first spectrum of SN\,2007bi (taken 54 days past peak; \citealt{gmo+09}). The spectral features match well, but the continuum of SN\,2007bi is significantly bluer -- the spectrum of SN\,2018don is a good match if dereddened by E(B-V) = 0.4~mag. If so, the implied extinction in $g$- and $r$ band is A$_g$=1.52~mag and A$_r$=1.09~mag, respectively, again assuming R$_V$=3.1. This suggests the peak luminosity of SN\,2018don was at least $-20$~mag in $r$-band, and supports the interpretation of this object as a slowly-evolving SLSN-I. We note that the emerging [\ion{Ca}{2}] feature seen in the SN\,2007bi spectrum is not seen in SN\,2018don. The early emergence of this nebular feature in SN\,2007bi while the spectrum is otherwise still displaying photospheric features is not well understood; it is not surprising that it is not seen in SN\,2018don given that the phase of this spectrum is significantly earlier, though.

Compared to the objects shown in Figure~\ref{fig:slsni_spec}, the spectroscopic evolution of SN\,2018don is slow, with the main features changing little over the nearly 60 days we observe it. The primary features are again typical of SN~Ic, and are marked on the +9d spectrum in Figure~\ref{fig:18cue_spec}. We do see a clear reddening of the continuum especially in the +44d spectrum -- this is partially due to the supernova itself turning redder as the $g$-band drops faster than the $r$-band at the light curve break around 30 days, and partially because the host galaxy continuum is starting to contribute significantly to the spectrum at these late epochs. Such a slow spectroscopic evolution has been seen in other SLSN-I of this subtype, for example the well-studied PS1-14bj \citep{lcb+16}, which also showed a spectrum similar to SN\,2007bi and which hardly showed any evolution over the $\sim 60$~days of spectroscopic observations in the photospheric phase.

\subsection{Velocities}
\label{sec:velocity}
Velocities in stripped-envelope supernovae are commonly measured from the absorption minimum of the \ion{Fe}{2}$\lambda$5169 feature, which has been suggested to be a good tracer of the photospheric velocity \citep{bbk+02}. We use this feature to measure the velocity of the H-poor SLSNe in our sample, following the method developed in \citet{mlb+16} for SNe Ic-BL. This method takes a SN~Ic template and applies a blueshift and broadening to match the \ion{Fe}{2}$\lambda$5169 and \ion{Fe}{2}$\lambda\lambda$4924,5018 features, which are often blended together in both Ic-BL and SLSN-I \citep{mlb+16,lmb+17}. A complication with applying this technique to SLSNe is that it is not necessarily obvious which phase SN~Ic template one should use, since SLSN-I past peak often resemble SN~Ic around or before peak (e.g., \citealt{psb+10}). In principle the template convolution method should be insensitive to which template is used (using a lower-velocity template would result in a larger blueshift measured but the same total velocity), though we find that in spectra of limited S/N there is still some scatter. Therefore, we try a range of template phases for each spectrum, and report the median velocity measured as our best estimate. Typically the template-to-template scatter is smaller than the uncertainties of each velocity measurement; in the cases where it is not we report the standard deviation from the measurements with different templates as the uncertainty. We note that the code did not converge for the final spectrum (+44~days) for SN\,2018don, and that we fit the higher S/N spectra from P200+DBSP rather than the LT+SPRAT spectra for this transient for the epochs where both were obtained the same night.

\begin{figure}
    \centering
    \includegraphics[width=3.5in]{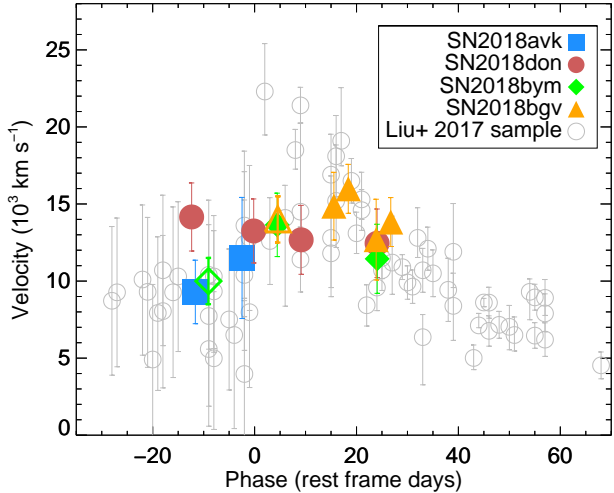}
    \caption{Velocities as measured from the \ion{Fe}{2} feature, using the method developed by \citet{mlb+16}. Data for other SLSNe, taken from \citet{lmb+17}, are shown as gray circles in the background. Also shown, as an open diamond and triangle, respectively, are \ion{O}{2} measurements from the earliest spectra of SN\,2018bym and SN\,2018bgv.}
    \label{fig:feii_vel}
\end{figure}

The velocities we measure are plotted in Figure~\ref{fig:feii_vel}, together with the SLSN-I sample measured by \citet{lmb+17} using the same method. They caution that in pre-peak measurements, there can be some contamination from \ion{Fe}{3} around the wavelength of \ion{Fe}{2}$\lambda$5169 in SLSN-I, and only consider measurements 10~days past peak as reliable. We see that the velocities measured for our objects, generally ranging from 10,000-15,000~km~s$^{-1}$, are in the typical range of SLSN-I as measured by \citet{lmb+17}.

An alternative to using the \ion{Fe}{2} lines at early times is to use the \ion{O}{2} lines instead. Figure~\ref{fig:feii_vel} also shows (as open symbols) the velocities inferred from the \ion{O}{2} lines in the first spectra of SN\,2018bym and SN\,2018bgv, using the wavelengths given in \citet{qdg+18} as the reference. The velocities from \ion{O}{2} are generally lower than those measured from \ion{Fe}{2} for these two supernovae, despite being at an earlier phase. This is consistent with the trend found in \citet{qdg+18} of the \ion{O}{2} lines both tracing lower velocities and declining more rapidly than the \ion{Fe}{2} lines, indicating that the \ion{O}{2} lines form in deeper levels of the ejecta.

\section{Host Galaxy Properties}
\label{sec:hosts}
Previous studies have shown that SLSN-I tend to preferentially happen in low-mass, metal-poor dwarf galaxies \citep{lcb+14,lcb+15,lsk+14,pqy+16,alp+16,csy+17,skl+18}. In this section, we fit spectral energy distributions (SEDs) to our compiled archive photometry of the host galaxies (Section~\ref{sec:host_phot}) in order to derive basic host properties and compare to the literature sample.

To fit SEDs, we use the stellar population synthesis code FAST \citep{kvl+09}, using the \citet{mar05} stellar population library, a Salpeter initial mass function (IMF), an exponentially declining star formation history and a Milky Way-like extinction curve. While we do not have host galaxy spectra available, several of our supernova spectra do show host emission lines, which we use to inform the model grid metallicities and extinction ranges. In particular, the host of SN\,2018avk has strong emission lines, where the ratio of H$\alpha$ to H$\beta$ constrains the extinction to be close to zero. Similarly, for the SN\,2018bym, SN\,2018avk and SN\,2018bgv host galaxies, we assume a model metallicity of 0.5~Z$_{\odot}$, while for the host of SN\,2018don we assume solar metallicity based on the relatively strong [\ion{N}{2}]$\lambda\lambda$6548,6584\AA\, host emission lines compared to H$\alpha$ seen in the spectra.

\begin{figure*}
    \centering
    \begin{tabular}{cc}
    \includegraphics[width=3.3in]{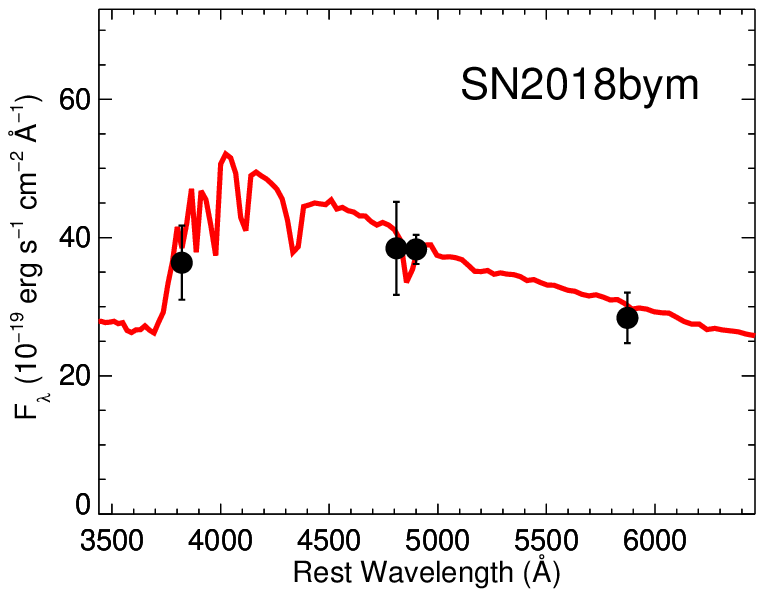} & \includegraphics[width=3.3in]{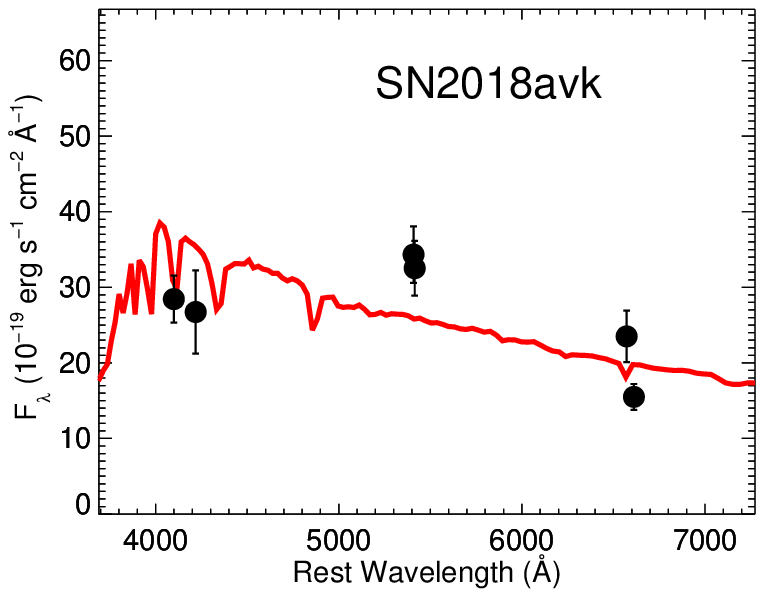} \\
    \includegraphics[width=3.3in]{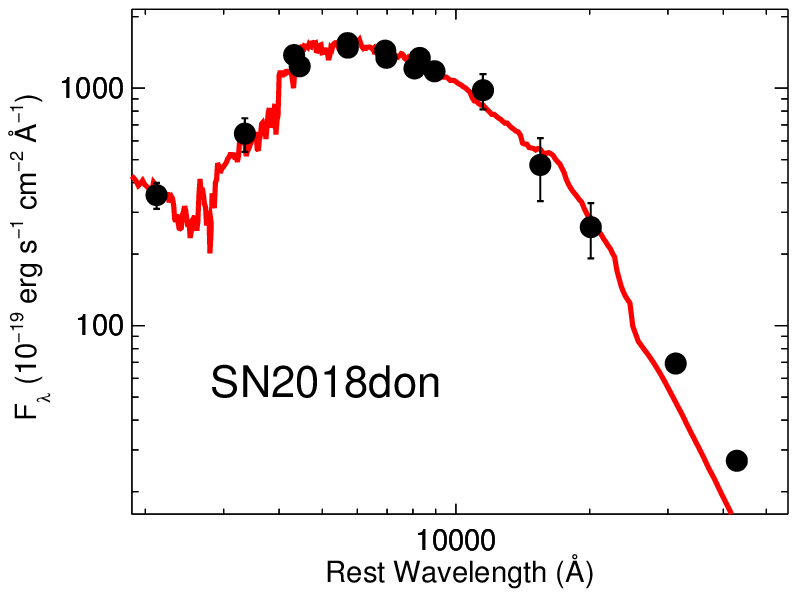} & \includegraphics[width=3.3in]{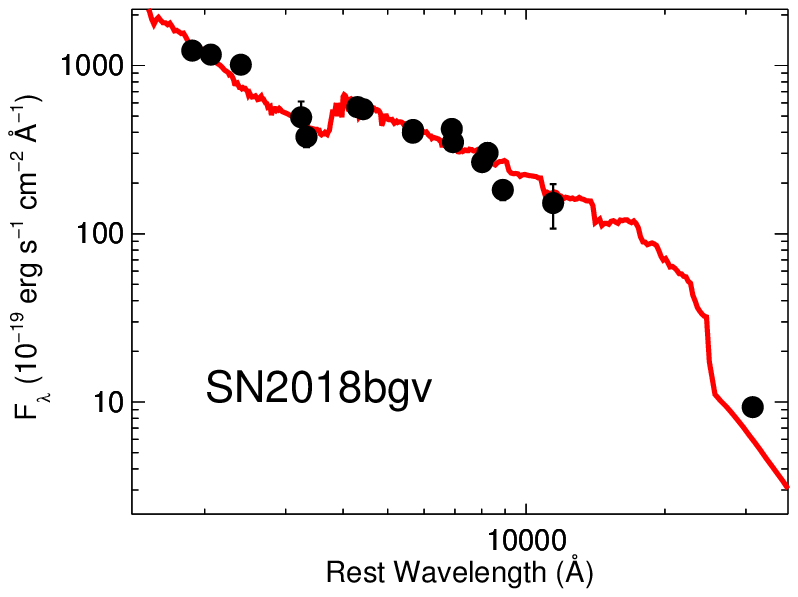}
    \end{tabular}
    \caption{Best-fit host galaxy SEDs, derived with the FAST stellar population synthesis code. Black points show the observed photometry, while the red lines show the model SED. SN\,2018don and SN\,2018bgv (bottom panels), which have a wider range of data, are shown on a log scale for clarity.
    \label{fig:host_sed}}
\end{figure*}

The derived host galaxy stellar masses, extinction, stellar population ages and star formation rates are listed in Table~\ref{tab:host_prop}, and the SED fits are shown in Figure~\ref{fig:host_sed}. The SEDs of the hosts of SN\,2018avk and especially SN\,2018bym are not particularly well constrained, and the latter lacks any UV data to constrain the star formation rate in the fit. The host of SN\,2018don has the best sampled SED, and is the only galaxy out of the four with derived stellar mass $>10^9~{\rm M}_{\odot}$. We note that while the best-fit model (with a reduced $\chi^2$ statistic of 0.6) has $A_V=0$, models with $A_V$ up to 1.8 are within the $1\sigma$ uncertainty range. If we assume $A_V \simeq 1.2$ as suggested by the supernova spectral comparisons, the resulting host model has a similar mass but younger stellar population age (as expected from the age-extinction degeneracy), and the fit has a reduced $\chi^2$ of 1.06. The galaxy photometry is therefore consistent with the possibility that the supernova is significantly reddened.

\begin{figure}
    \centering
    \includegraphics[width=3.5in]{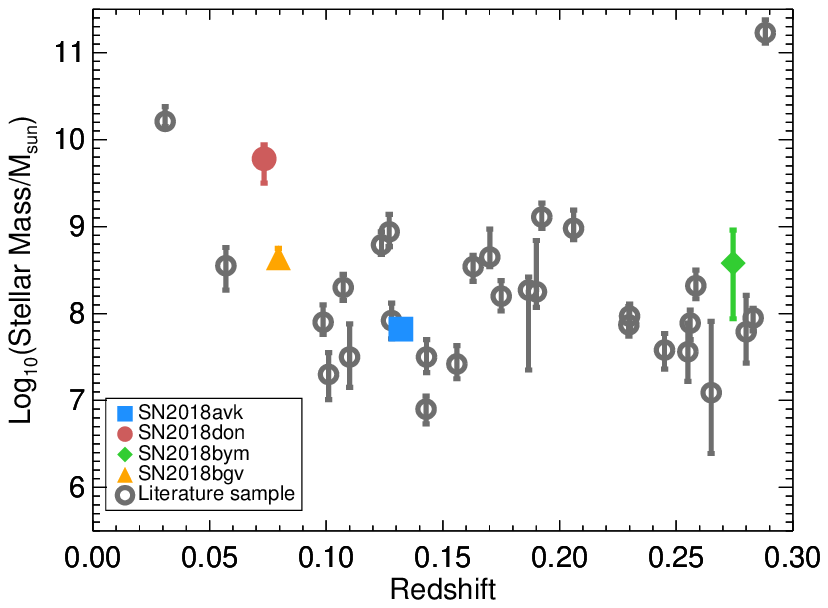}
    \caption{Host galaxy stellar mass plotted as a function of redshift for the four SLSN-I. Literature data for other low-redshift SLSN-I host galaxies are plotted as gray circles.}
    \label{fig:host_z_mass}
\end{figure}

To put the host galaxies in context, we plot the galaxy stellar masses in Figure~\ref{fig:host_z_mass}, compared to literature data over the same redshift range \citep{pqy+16,skl+18,bdp+18}. The hosts of SN\,2018avk, SN\,2018bym and SN\,2018bgv are consistent with the bulk of SLSN host galaxies at this redshift, which are dominated by dwarf galaxies. The host of SN\,2018don stands out by being one of the most massive SLSN-I host galaxies found to date at low redshift -- it is not unique, however: the host of PTF10uhf has a derived stellar mass $>10^{11}~{\rm M}_{\odot}$ \citep{pqy+16}, and the host of SN\,2017egm similarly has a stellar mass of several times $10^{10}~{\rm M}_{\odot}$ \citep{bdp+18,nbm+17}. Thus, SN\,2018don adds to the small but growing sample of SLSN-I that occur in high-mass, solar metallicity galaxies. We caution, however, that the global properties of the host galaxy do not necessarily correspond to the conditions at the explosion site, as the study of the very nearby SN\,2017egm has shown \citep{csx+17,itg+18, Yan2018}.

\section{Discussion}
\label{sec:disc}

\subsection{Diversity within the SLSN population}

While this paper only presents a few objects, the first SLSNe discovered by ZTF hints at the large diversity present in this population. It is worth noting that out of the four objects discussed here, only SN\,2018bym meets the ``original'' definition of a SLSN (peak absolute magnitude brighter than $-21~{\rm mag}$; \citealt{gal12}) as well as the definition more recently proposed by \citet{ipg+18}. This is consistent with results from recent compilation studies from untargeted surveys, however: the SLSN samples from Pan-STARRS1 \citep{lcb+18}, (i)PTF \citep{dgr+18} and DES \citep{ass+19} all find a significant population of objects with spectra consistent with SLSNe but luminosities extending at least down to $-20$~mag (and in the case of DES, down towards $-19$~mag). 

Additionally, the vast majority of SLSN-I discovered to date display rise timescales significantly slower than the 10~days observed in SN\,2018bgv \citep{lcb+18,dgr+18}. It is not clear whether this reflects the true underlying distribution of timescales in the population, or whether shorter timescale events are underrepresented because an easy way to discriminate most known SLSNe from more common transients like SN~Ia is to filter on longer light curve rise times. Complete spectroscopic surveys, such as the ZTF Bright Transient Survey \citep{fsk+18,kpm18}, will play an important role in mapping out the true diversity. 

\subsubsection{Powering the Fast Light Curve of SN\,2018bgv}
\label{sec:bgv_disc}
With a rest-frame $g$-band rise time of just 10 days as measured from the inferred explosion date (and 7 days as measured from time of half peak flux), SN\,2018bgv is starting to approach the timescales of so-called ``rapidly evolving transients'' or ``fast, blue, luminous transients'' \citep{dcs+14,awh+16,pcs+18,rgk+18}. There does not exist a unified definition of such objects, but is generally applied to transients with luminosities comparable to normal supernovae but significantly faster timescales, ruling out a $^{56}$Ni powering mechanism. We note that SN\,2018bgv would have met the criteria of \citet{awh+16}, who looked at objects from the Supernova Legacy Survey with a rise time of $\approx 10$~days and peak magnitudes between (what was then considered) ``typical'' SLSNe and SNe (i.e., $M_{\rm peak}\sim -20~{\rm mag}$). 
The link to rapidly evolving transients is also interesting because two objects in this category, detected and followed up in real time by iPTF and ZTF respectively, showed late-time spectra classifying them as broad-lined Type~Ic SNe (iPTF16asu; \citealt{wlk+17}, and ZTF18abukavn/SN\,2018gep; \citealt{hgs+19}). SN\,2018gep is particularly interesting in this context, as its spectrum taken $\sim 4$~days past explosion showed features similar to the hot photospheric phase of SLSN-I, including the characteristic \ion{O}{2} feature. At significantly higher redshift ($z = 0.677$), SN\,2011kl associated with the ultra-long gamma ray burst GRB111209A also showed comparable timescales and peak luminosity to SN\,2018bgv, and has been argued to show spectroscopic similarities to SLSN-I \citep{gmk+15,kso+19,lmb+17}.
From an empirical point of view, then, there seems to be a continuous transition between the properties of some stripped-envelope supernovae that have been considered fast, luminous transients, at least one with direct evidence of a central engine, and others that are considered SLSN-I. 

In Section~\ref{sec:magfits} and Table~\ref{tab:magfits}, we present a magnetar fit to the light curve of SN\,2018bgv derived with {\tt MOSFiT}. Compared to the SLSN sample analyzed with {\tt MOSFiT} in \citet{ngb17}, SN\,2018bgv is an outlier in terms of the magnetar properties (Fig.~\ref{fig:magfits}): it has both a lower ejecta mass, higher magnetic field and slower initial spin than any object in their literature sample. 
This is not in itself an argument against SN\,2018bgv being magnetar-powered, however, but rather another manifestation that the light curve properties of SN\,2018bgv are unusual among SLSNe observed to date. The low ejecta mass (and relatively high velocities) result in a short diffusion time, necessary to produce the fast rise. We note that the derived ejecta mass is still within the range found for SN~Ic in the literature, if on the low side \citep[e.g.][]{dsg+11,tsb+18,paj+19}. Thus, we find that a fast-rising object like SN\,2018bgv can still plausibly be powered by a magnetar, but requires slightly different parameters than the population of SLSNe observed to date.

A slightly different engine model also invoked for explaining fast and luminous transients is the birth of a pulsar in a binary neutron star system. Motivated by the properties of PSR J0737-3039A/B, \citet{Hotokezaka2017} show that the emission from the birth of the second pulsar could give rise to fast and luminous optical transients. This model resembles the magnetar model discussed above, i.e. internal heating by a pulsar wind nebula, but with some key differences. The magnetic field is generally lower ($\sim 10^{12}~{\rm G}$ rather than $\sim 10^{14}~{\rm G}$), so the pulsar spin-down luminosity is taken to be constant over the timescale of the transient. Additionally, the ejecta masses are low ($\sim 0.1~{\rm M}_{\odot}$), leading to a short diffusion time and the need to consider heating sources such as Compton scattering and bound-free absorption after the initial shock heating dominated phase. \citet{Hotokezaka2017} show reasonable fits to the four fast and luminous transients studied by \citet{awh+16}; given that SN\,2018bgv shows similar physical properties to these objects (namely a rise time of $\sim 10$~days and a peak luminosity of $\sim 10^{44}~{\rm erg~s}^{-1}$), it is likely its light curve can also be reproduced by a binary neutron star model. We consider a detailed exploration of the best engine-driven model for a transient such as SN\,2018bgv outside of the scope of this paper, but note that its well-sampled light curve offers an excellent starting point for future studies.   

The fast rise, high luminosities and blue colors of such rapidly evolving transients is often explained by either circumstellar interaction or shock cooling in an extended circumstellar medium \citep{dcs+14,rgk+18,wlk+17,hgs+19}. Having a naturally truncated energy input, such models can more easily explain short-duration light curves; recently \citet{ct19} found that CSM interaction models also did a better job of explaining SLSN-I light curves exhibiting either short-duration or symmetric light curves. Indeed, comparing the location of SN\,2018bgv in luminosity-duration phase space to the theoretical models explored in \citet{vbm+17}, the two models covering this location are the magnetar spin-down and CSM interaction. A detailed exploration of a CSM model generally requires hydrodynamical simulations out of the scope of this paper; for an estimate of the kind of parameters required, we again turn to {\tt MOSFiT} which implements a simplified CSM model based on the semianalytic relations derived in \citet{cwv12}. We largely use the default parameters for this model, but leave the opacity $\kappa$ as a free parameter allowed to vary between $0.05-0.2~{\rm cm}^2{\rm g}^{-1}$ like in the magnetar model (appropriate for hydrogen-free material), rather than fixing it at the default $0.34~{\rm cm}^2{\rm g}^{-1}$ (which is appropriate for H-rich CSM). We find the following best-fit parameters for SN\,2018bgv: ejecta mass $M_{\rm ej} = 2.1^{+1.8}_{-1.1}~{\rm M}_{\odot}$, CSM mass $M_{\rm CSM} = 1.8^{+0.8}_{-0.5}~{\rm M}_{\odot}$, and density $\rho = 3.0^{+2.7}_{-1.6} \times 10^{-12}~{\rm g~cm}^{-3}$ (assuming a constant density, i.e. a shell-like CSM). The fit is about equally good as the magnetar model in explaining the data (i.e., the distribution of the $\sigma$-parameter is nearly identical; we get $\sigma = 0.15 \pm 0.02$ in either case). The CSM model has fewer free parameters though, which is reflected in a better WAIC score (89.9 for the CSM model, vs. 82.6 for the magnetar model; MOSFiT does not calculate the error on the WAIC score which precludes a quantitative interpretation though). We conclude that CSM interaction can also explain the light curve properties of SN\,2018bgv; a more detailed exploration including actual hydrodynamical simulations, whether CSM interaction can also explain the spectra, and how realistic such a progenitor/mass-loss scenario is, we defer to future work.

\subsubsection{Understanding the Bumpy Light Curve of SN\,2018don}
\label{sec:18cue_disc}

Like SN\,2018bgv, SN\,2018don stands out from ``typical'' previously observed SLSN-I in multiple ways. To begin with, our classification of this object as a SLSN-I relies on its spectroscopic similarity to SN\,2007bi (Figure~\ref{fig:07bi_vs_cue}) and the corresponding assumption that the supernova is experiencing significant host galaxy extinction. This interpretation is supported by the magnetar fits presented in Section~\ref{sec:magfits}, which show that the derived parameters for SN\,2018don are comparable with the rest of the SLSN-I population if allowing for host extinction of $A_{\rm V} \simeq 2~{\rm mag}$. We also note that the light curve of SN\,2018don is difficult to reconcile with a standard, $^{56}$Ni powered SN~Ic model. To quantify this, we also run {\tt MOSFiT} with the {\tt default} i.e. Ni-powered model. We find that the best fit parameters require both a high ejecta mass ($M_{\rm ej} = 15.5_{-4.7}^{+11.4}~{\rm M}_{\odot}$) and an appreciable fraction of $^{56}$Ni ($F_{\rm Ni} = 0.17_{-0.08}^{+0.09}$), i.e. a $^{56}$Ni mass in the range $1-7$~M$_{\odot}$. Both the ejecta and nickel masses are significantly higher than is seen in the normal population of SNe~Ic (e.g., \citealt{tsb+18}). Additionally, the fit still requires some host extinction ($A_{\rm V} = 0.67_{-0.24}^{+0.17}~{\rm mag}$), and is a significantly worse fit to the data than the magnetar model despite fewer free parameters ($\sigma = 0.13 \pm 0.01~{\rm mag}$ and WAIC score of 162, vs. $\sigma = 0.08 \pm 0.01~{\rm mag}$  and WAIC of 222 for the magnetar model). Thus, we favor the interpretation of SN\,2018don as a SLSN with significant host extinction over that of a SN~Ic with an unusual light curve. 

\begin{figure*}
    \centering
    \begin{tabular}{cc}
    \includegraphics[width=3.4in]{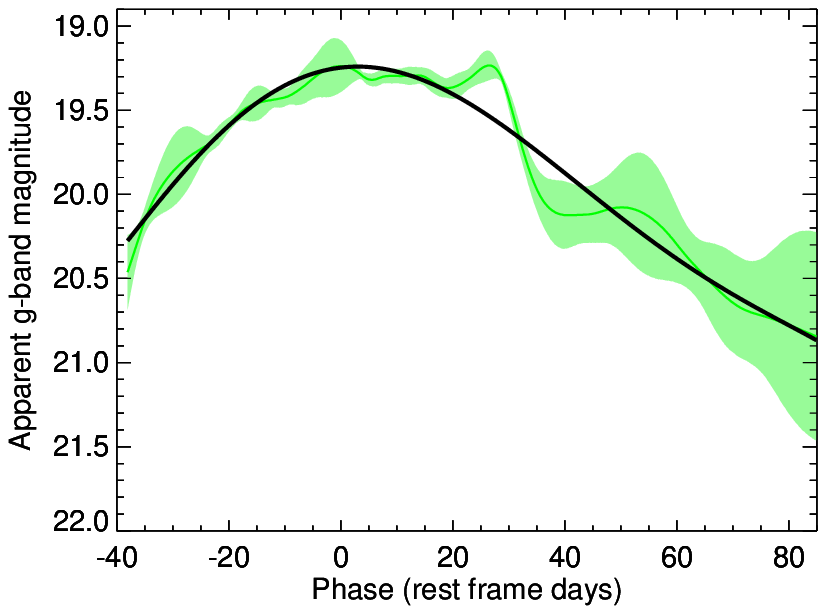} & \includegraphics[width=3.4in]{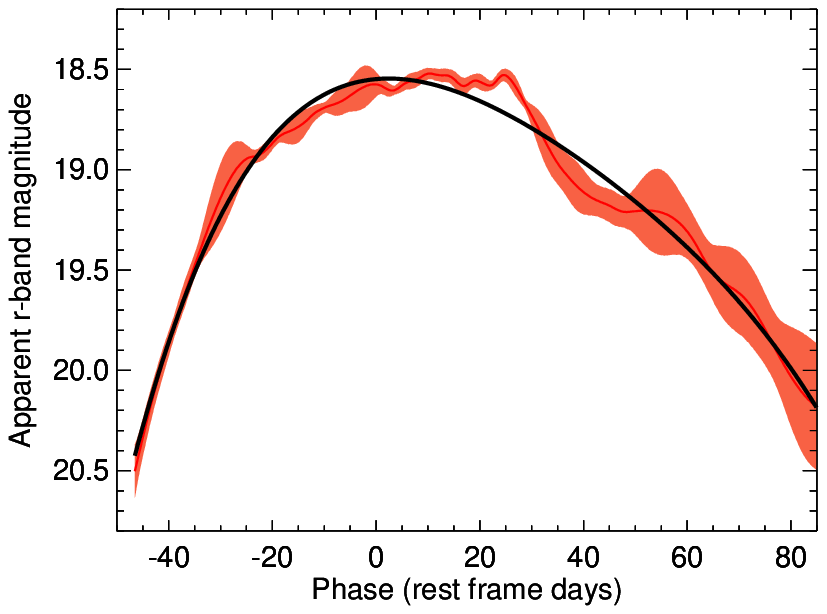}
    \end{tabular}
    \caption{Gaussian Process-smoothed light curve of SN\,2018don in $g$-band (left) and $r$-band (right). The shaded region shows the 3$\sigma$ error interval. Overplotted in black is a 5th-order polynomial fit to the light curve, demonstrating how the undulation/drop in the light curve at 30~days deviates significantly from a smooth decline. This, combined with that the feature is seen independently and at the same time in both $g$- and $r$-band, argues strongly for it being a real, physical feature of the light curve and not a statistical fluctuation. }
    \label{fig:18don_bumps}
\end{figure*}

While there does not exist an accepted definition of ``bumpiness'' in supernova light curves, there is clear evidence that the light curve of SN\,2018don deviates significantly from a smooth decline. The fact that the undulations are seen in both $g$- and $r$-band, at the same time, argues strongly that they are a real feature and not a statistical fluctuation. While individual data points are noisy, taken together they do constrain the light curve shape strongly in this region, which can be seen in the Gaussian Process-smoothed light curve. Figure~\ref{fig:18don_bumps} shows the result of the Gaussian Process interpolation, together with the 3$\sigma$ error interval, compared with a smoothly declining function (a 5th-order polynomial), demonstrating that the drop in the light curve at 30~days is significant at the $>3\sigma$ level in both filters.

Going by the assumption that SN\,2018don is best described as a SN\,2007bi-like SLSN-I, it joins a small but growing sample of such objects showing significant light curve undulations on the decline. Such undulations have been seen in a number of objects, including SN\,2007bi itself \citep{gmo+09} and PS1-14bj \citep{lcb+16}, though one of the most striking examples is SN\,2015bn \citep{nbs+16}, which had densely sampled multiband photometry and spectroscopy out to $\gtrsim 100$~days past peak light. Other examples of prominent light curve undulations in SLSN-I include PS1-12cil \citep{lcb+18}, iPTF15esb \citep{ylp+17}, iPTF13dcc \citep{vlg+17} and PS16aqv \citep{bnb+18}; as was also seen in SN\,2015bn and iPTF15esb, the undulations in SN\,2018don are more prominent in the bluer bands.

Such light curve undulations are not easily explained by powering mechanism such as radioactive decay or magnetar spin-down, which produce smoothly decaying light curves, at least under the simplifying assumption of complete energy trapping and constant opacity. Thus, explanations for light curve bumps and undulations generally involve either changes in the opacity, or an alternative power source. For example, the model of \citet{mvh+14} predicts a UV breakout when the \ion{O}{2} layer reaches the ejecta surface, leading to a rapid drop in UV opacity. This explanation was considered by \citet{nbs+16} for the ``knee'' in the light curve of SN\,2015bn around 30 days, which is also a similar timescale as the temporary flattening seen in SN\,2018don. In SN\,2015bn this change was also associated with a change in the spectrum, which transitioned from the hot photospheric phase showing \ion{O}{2} features prior to the light curve ``knee'', and displayed features typical of the cool photospheric phase afterwards. Given that SN\,2018don shows features typical of the cool photospheric phase also before the light curve drop and plateau, it is less clear that the same explanation will apply here, however.

Alternatively, it has been suggested that such light curve undulations result from circumstellar interaction, either in a light curve entirely powered by interaction, or as a modulation of the light curve on top of another power source such as radioactive decay or magnetar spin-down. Some SN~IIn (whose narrow hydrogen lines in the spectra are a tell-tale sign of CSM interaction) have been seen to exhibit such light curve bumps (e.g. SN\,2006jd; \citealt{stf+12}, SN\,2013Z; \citealt{nst+17}; see also the recent compilation in \citealt{nst+19}). In an interaction scenario, such light curve changes are naturally explained as the ejecta encounters a change in the CSM density; light curves with multiple peaks/bumps therefore require a CSM structure that deviates from a smooth profile (such as the $\rho \propto r^{-2}$ profile expected from a stellar wind). Light curves of several SLSN-I showing bumps have been successfully modelled as powered by interaction with multiple shells of circumstellar material \citep{lww+18}.
Such a CSM structure could arise from an eruptive mass-loss history; SLSNe in particular have been linked with stars encountering the pulsational pair-instability in their final stages of evolution \citep{wbh07,cw12b,woo17}, leading to the ejection of multiple shells of material. Several SLSN-I, including iPTF15esb, have also shown late-time H$\alpha$ emission, interpreted as arising from the supernova ejecta interacting with a H-rich CSM shell \citep{yqo+15,ylp+17}. Our spectra of SN\,2018don only extend to 44~days past peak, and do not show any H$\alpha$ emission. Recently, a CSM shell around a SLSN was also detected directly through a resonance scattering light echo \citep{lfv+18}. Thus, although SLSN-I spectra (including SN\,2018don) generally do not show classical signatures of CSM interaction in the form of narrow emission lines, there is increasing evidence that some SLSN-I progenitors experience significant, eruptive mass loss prior to explosion. As such, we consider CSM interaction a plausible explanation for the light curve variations seen in SN\,2018don, although a detailed exploration of the density profile required is out of the scope of this paper.

\subsection{Strategies for selecting SLSNe from large data sets}
The objects discussed in this paper, while few, serve to illustrate several challenges in identifying SLSNe in large, photometric data sets, particularly in an unbiased fashion and in real-time. This problem is important for several reasons: deriving the true luminosity function is necessary for both discerning the power source(s), and for devising search strategies for discovering SLSNe in deeper surveys such as LSST. Additionally, it serves to quantify the diversity in the population, including identifying potential sub-populations, and whether these are best explained as variations on the same underlying physical phenomenon, or representative of different underlying mechanisms.

As previous compilation studies have found that SLSNe generally exhibit slower timescales than SN~Ibc and SN~Ia \citep{nsj+15,lcb+18,dgr+18}, filtering on long observed rise times is one obvious strategy for reducing contamination from more common SNe. Such a strategy would have missed an object like SN\,2018bgv, and insofar as none of the large surveys presenting samples on SLSNe to date have been spectroscopically complete, it is tricky to ascertain whether such short-timescale SLSNe are intrinsically rare or systematically missed. While SN\,2018bgv, being relatively nearby for a SLSN ($z=0.079$) and having a peak observed magnitude of $\sim 17~{\rm mag}$ probably would have been picked up for classification regardless of whether it was recognized as a likely SLSN, it is unclear whether the same had been true of a similar object at $z\sim 0.2-0.3$ (which was the typical redshift of SLSNe found in PTF and iPTF, \citealt{dgr+18}). With correspondingly noisier photometry and slightly longer observed timescales due to time dilation, a higher-redshift analogue of SN\,2018bgv would be harder to distinguish against the background of SN~Ia. 

However, a striking property distinguishing SN\,2018bgv from SN~Ia is its color, showing $g-r \simeq -0.4~{\rm mag}$ at peak, while SN~Ia typically show $g-r \simeq 0~{\rm mag}$ \citep{Miller2017}. We note that the objects with the fastest rise timescales in the PS1 SLSN-I sample are also the ones with the highest measured blackbody temperatures \citep{lcb+18}, suggesting blue colors may be a way to distinguish the faster-evolving SLSN-I. This is opposite of the trend described in \citet{ipg+18}, however, who found that faster light curve timescales were weakly correlated with \textit{redder} colors at peak. (Time scale in their work was measured by the light curve decline 30 days past peak in the 400~nm filter, though other works have found that rise- and decline timescales are tightly correlated in SLSNe; \citealt{nsj+15,dgr+18}.) The full SLSN sample from ZTF, providing well-sampled (1-3 day cadence) light curves in two filters, will be in a good position to explore this question. 

Beyond light curve properties, other contextual information can also be used to pick out SLSNe. If a host galaxy redshift (photometric or spectroscopic) is available, this can be used to pick out intrinsically luminous transients; such a strategy has been successfully employed to find likely SLSNe at $z \gtrsim 2$ \citep{csg+12,mty+19,ccm+19}. Studies assessing the discovery rate and detectability of SLSNe in LSST have also assumed that at least photometric redshifts will be available \citep{vnb+18,vbm+19}. Challenges with this strategy include that many SLSNe (including 3 out of the 4 discussed in this paper) are found in dwarf galaxies, requiring correspondingly deeper photometry for reliable photometric redshifts; conversely requiring host galaxies with well-determined redshifts could bias the sample towards objects in brighter host galaxies. Additionally, objects like SN\,2018don shows that there can be SLSN-I with significant host galaxy reddening, which puts the absolute magnitude (as observed) outside of the range of typical SLSN-I. Similarly, it has also been suggested to exploit the fact that many SLSN-I are found in dwarf galaxies to use either host galaxy brightness/colors or the contrast between supernova and host galaxy brightness as a filter (e.g., \citealt{msr+14}). Again, while such a filter might have picked up the three objects in more ``typical'' SLSN-I host galaxies, it would have missed SN\,2018don, which is neither brighter than its host galaxy nor located in a particularly faint or blue host galaxy. The diversity displayed in the first SLSNe discovered by ZTF thus illustrates that no single parameter cut is sufficient to distinguish SLSNe from other types, and we face several challenges in selecting SLSNe from a large data stream; the strategy adopted by the ZTF SLSN group will be presented in Perley et al. (2019, in preparation).

\section{Conclusions}
\label{sec:conc}
We have presented light curves and spectra of four luminous supernovae found during the first months of the ZTF survey, and argue that they can all be classified spectroscopically as SLSN-I. Of the four, only SN\,2018bym resembles a ``classical'' SLSN-I as defined by an absolute magnitude $< -21~{\rm mag}$ \citep{gal12}, or as by falling into the ``4OPS'' parameter space defined statistically by \citet{ipg+18}. This variety echoes recent results from previous untargeted transient surveys, which have shown that the diversity in spectroscopically classified SLSN-I is larger than was found in the earlier literature samples \citep{lcb+18,dgr+18,ass+19}. 

All four SLSN-I discussed here have at least nightly (weather-permitting) light curve coverage in $g$ and $r$ from ZTF, allowing for excellent coverage of the rising phase of the light curve including color information, and enabling estimation of explosion times. We note in particular that SN\,2018bgv shows a rise time (explosion to peak in the rest-frame $g$-band) of just 10 days, the fastest seen in a SLSN-I to date, and is starting to approach the area of timescale-luminosity parameter space occupied by so-called fast, blue, luminous transients. The properties of SN\,2018bgv can still be explained by a magnetar-powered model, but requires an unusually small ejecta mass ($\simeq 1~{\rm M}_{\odot}$) in order to reproduce the fast rise-time. 

Like SN\,2018bgv, SN\,2018don pushes the boundaries of properties of observed SLSN-I to date, with an observed peak color of $g-r \simeq 0.7~{\rm mag}$ and a peak observed absolute magnitude of $r \simeq -19.0~{\rm mag}$. Based on spectroscopic comparisons with SN\,2007bi, we still classify this object as a SLSN-I, but with significant host galaxy reddening ($E(B-V)_{\rm host} \gtrsim 0.4~{\rm mag}$). Under this assumption, SN\,2018don is compatible with previously observed SLSNe luminosity-wise, though at the faint and slowly-evolving end of the distribution. Its other striking property is significant light curve undulations post-peak, including a drop of 0.75~mag in $g$-band and 0.6~mag in $r$-band over a time period of $\sim 10$~days, followed by a flattening before the light curve decline continues. These undulations are of the more extreme seen in SLSN-I, with iPTF15esb (which later showed signs of CSM interaction through emerging H$\alpha$ emission; \citealt{ylp+17}) being the closest observed analogue. We speculate that interaction may also be at play in causing the undulations in SN\,2018don.

The four objects presented here illustrate both the promise and challenges of finding SLSNe in large-scale surveys like ZTF. The demonstrated variety in both light curve timescales, observed colors, and host galaxy properties show how filtering on any one attribute is likely to miss objects that nevertheless should be considered SLSN-I. At the same time, the large area coverage and regular sampling in two filters by ZTF shows great promise in further mapping out the properties of SLSN-I, in particular increasing the sample of SLSNe with coverage through both the rise and decline and with continuous color information. As the detection efficiency during science validation will be low compared to the full survey (due to both lack of reference coverage and to algorithms still being tuned), the discovery rate of SLSNe in ZTF over the lifetime of the survey is expected to be considerably higher than these four objects over a time period of two months would suggest.
Future work, including discussions on candidate selection, completeness, rates and the properties of the first SLSN-I sample from ZTF (Perley et al., in preparation; Yan et al., in preparation) will explore these questions in detail.

\acknowledgements
Based on observations obtained with the Samuel Oschin Telescope 48-inch and the 60-inch Telescope at the Palomar Observatory as part of the Zwicky Transient Facility project. ZTF is supported by the National Science Foundation under Grant No. AST-1440341 and a collaboration including Caltech, IPAC, the Weizmann Institute for Science, the Oskar Klein Center at Stockholm University, the University of Maryland, the University of Washington, Deutsches Elektronen-Synchrotron and Humboldt University, Los Alamos National Laboratories, the TANGO Consortium of Taiwan, the University of Wisconsin at Milwaukee, and Lawrence Berkeley National Laboratories. Operations are conducted by COO, IPAC, and UW.
The Liverpool Telescope is operated on the island of La Palma by Liverpool John Moores University in the Spanish Observatorio del Roque de los Muchachos of the Instituto de Astrofisica de Canarias with financial support from the UK Science and Technology Facilities Council.
Some data presented herein were obtained at the W.M. Keck Observatory, which is operated as a scientific partnership among the California Institute of Technology, the University of California and the National Aeronautics and Space Administration. The Observatory was made possible by the generous financial support of the W.M. Keck Foundation. Partially based on observations made with the Nordic Optical Telescope, operated at the Observatorio del Roque de los Muchachos, La Palma, Spain, of the Instituto de Astrofisica de Canarias. Some of the data presented here were obtained with ALFOSC, which is provided by the Instituto de Astrofisica de Andalucia (IAA) under a joint agreement with the University of Copenhagen and NOTSA.
This research has made use of data obtained through the High Energy Astrophysics Science Archive Research Center Online Service, provided by the NASA/Goddard Space Flight Center.
This research has made use of NASA's Astrophysics Data System.
The ZTF forced-photometry service was funded under the Heising-Simons Foundation grant \#12540303 (PI: Graham).
This work was supported by the GROWTH project funded by the National Science Foundation under Grant No 1545949.
AGY’s research is supported by the EU via ERC grant No. 725161, the ISF GW excellence center, an IMOS space infrastructure grant and the BSF Transformative program as well as The Benoziyo Endowment Fund for the Advancement of Science, the Deloro Institute for Advanced Research in Space and Optics, The Veronika A. Rabl Physics Discretionary Fund, Paul and Tina Gardner and the WIS-CIT joint research grant;  AGY is the recipient of the Helen and Martin Kimmel Award for Innovative Investigation.
R.L. is supported by a Marie Sk\l{}odowska-Curie Individual Fellowship within the Horizon 2020 European Union (EU) Framework Programme for Research and Innovation (H2020-MSCA-IF-2017-794467).

\vspace{5mm}
\facilities{PO:1.2m, PO:Hale, Liverpool:2m, NOT, Keck:I}

\software{{\tt george} \citep{afg+15},
          {\tt heasoft} (\url{https://heasarc.nasa.gov/lheasoft/}),
          {\tt LAMBDAR} \citep{Wright2016a},
          {\tt MOSFiT} \citep{gnv+18},
          {\tt PhotoFit} \citep{sgg+19}}

\begin{deluxetable}{lccccc}
\tablecaption{List of SLSNe \label{tab:slsn_list}}
\tablehead{
\colhead{ZTF name} &
\colhead{IAU name} &
\colhead{RA} &
\colhead{Dec} &
\colhead{Redshift} &
\colhead{E(B-V)\tablenotemark{a}} \\
\colhead{} &
\colhead{} &
\colhead{(J2000)} &
\colhead{(J2000)} &
\colhead{} &
\colhead{(mag)}
}
\startdata
ZTF18aaisyyp & SN\,2018avk & \ra{13}{11}{27.72} & \dec{+65}{38}{16.7} & 0.132 & 0.012 \\
ZTF18aapgrxo & SN\,2018bym & \ra{18}{43}{13.42} & \dec{+45}{12}{28.2} & 0.274 & 0.052 \\
ZTF18aavrmcg & SN\,2018bgv & \ra{11}{02}{30.29} & \dec{+55}{35}{55.8} & 0.079 & 0.008 \\
ZTF18aajqcue & SN\,2018don & \ra{13}{55}{08.65} & \dec{+58}{29}{42.0} & 0.073  & 0.009 
\enddata
\tablenotetext{a}{From \citet{sf11}}
\end{deluxetable}

\begin{deluxetable}{lccccc}
\tablewidth{0pt}
\tabletypesize{\scriptsize}
\tablecaption{Photometry\label{tab:phot}}
\tablehead{
\colhead{Object} &
\colhead{MJD} &
\colhead{Phase} &
\colhead{Filter} &
\colhead{AB mag} &
\colhead{Telescope+Instrument} \\
\colhead{} &
\colhead{(days)} &
\colhead{(rest-frame days)} &
\colhead{} &
\colhead{} &
\colhead{}
}
\startdata
SN\,2018bym & 58232.5 &   -37.7 & g & $20.82 \pm 0.13$ & P48+ZTF \\
SN\,2018bym & 58232.5 &   -37.7 & g & $20.52 \pm 0.11$ & P48+ZTF \\
SN\,2018bym & 58234.4 &   -36.2 & g & $>20.30$ & P48+ZTF \\
SN\,2018bym & 58234.4 &   -36.2 & g & $>20.30$ & P48+ZTF \\
SN\,2018bym & 58234.5 &   -36.2 & g & $20.43 \pm 0.14$ & P48+ZTF \\
SN\,2018bym & 58234.5 &   -36.2 & g & $20.61 \pm 0.18$ & P48+ZTF \\
SN\,2018bym & 58234.5 &   -36.1 & g & $20.60 \pm 0.18$ & P48+ZTF \\
SN\,2018bym & 58235.4 &   -35.4 & g & $20.42 \pm 0.25$ & P48+ZTF \\
SN\,2018bym & 58235.4 &   -35.4 & g & $20.19 \pm 0.20$ & P48+ZTF \\
SN\,2018bym & 58235.4 &   -35.4 & g & $20.17 \pm 0.22$ & P48+ZTF
\enddata
\tablecomments{The full table is available in electronic form. A portion is shown here for guidance.}
\end{deluxetable}

\begin{deluxetable}{lcccccc}
\tablewidth{0pt}
\tabletypesize{\scriptsize}
\tablecaption{Summary of Spectroscopic Observations \label{tab:spec}}
\tablehead{
\colhead{Object} &
\colhead{Observation Date} & 
\colhead{Phase} &
\colhead{Telescope+Instrument} &
\colhead{Grating\tablenotemark{a}} &
\colhead{Exp. time\tablenotemark{a}} &
\colhead{Airmass} \\
\colhead{} &
\colhead{(YYYY MM DD.D)}  & 
\colhead{(rest-frame days)} &
\colhead{} &
\colhead{} &
\colhead{(s)} &
\colhead{} 
}
\startdata
SN\,2018avk & 2018 May 04 & -11.66 & NOT+ALFOSC & Gr 4 & 1800 & 1.3 \\
SN\,2018avk & 2018 May 14.3 & -2.56 & KeckI+LRIS & 400/3400,400/8500 & 840,800 & 1.4 \\
SN\,2018bym & 2018 May 31 & -9.34 & NOT+ALFOSC & Gr 4 & 2400 & 1.4 \\
SN\,2018bym & 2018 Jun 12 & +0.08 & P200+DBSP & 600/4000,316/7500 & 600 & 1.13 \\
SN\,2018bym & 2018 Jun 17.5 & +4.4 & KeckI+LRIS & 400/3400,400/8500 & 600,570 & 1.4 \\
SN\,2018bym & 2018 Jul 12.6 & {+24.10} & KeckI+LRIS & 400/3400,400/8500 & 600,600 & 2.0 \\
SN\,2018bgv & 2018 May 20  & {+4.45} & LT+SPRAT & Wasatch600 & 300 & 1.6 \\
SN\,2018bgv & 2018 May 23  & {+7.23} & NOT+ALFOSC & Gr 4 & $2\times2400$ & 1.3 \\
SN\,2018bgv & 2018 May 31  & {+14.64} & NOT+ALFOSC & Gr 4 & 900 & 1.3 \\
SN\,2018bgv & 2018 Jun 04  & {+18.34} & NOT+ALFOSC & Gr 4 & 2x2400 & 1.4 \\
SN\,2018bgv & 2018 Jun 10  & {+23.90} & NOT+ALFOSC & Gr 4 & 900 & 1.3 \\
SN\,2018bgv & 2018 Jun 13  & {+26.68} & NOT+ALFOSC & Gr 4 & $2\times2400$ & 1.3 \\
SN\,2018bgv & 2018 Jul 17  & {+58.18} & P200+DBSP & 600/4000,316/7500 & 1200 & 1.7 \\
SN\,2018don & 2018 May 16.3  & {-12.95} & P200+DBSP & 600/4000,316/7500 & 900 & 1.2 \\
SN\,2018don & 2018 May 16.9  & {-12.39} & LT+SPRAT & Wasatch600 & 300 & 1.2 \\
SN\,2018don & 2018 May 30    & {-0.19} & NOT+ALFOSC & Gr 4 & $2\times2400$ & 1.1 \\
SN\,2018don & 2018 Jun 09.4  & {+9.50} & P200+DBSP & 600/4000,316/7500 & 900 & 1.6 \\
SN\,2018don & 2018 Jun 10.9  & {+10.92} & LT+SPRAT & Wasatch600 & 900 & 1.2 \\
SN\,2018don & 2018 Jun 25    & {+24.04} & NOT+ALFOSC & Gr 4 & $2\times2400$ & 1.2 \\
SN\,2018don & 2018 Jul 17.2  & {+44.72} & P200+DBSP & 600/4000,316/7500 & 600 & 1.4 
\enddata
\tablenotetext{a}{Comma-separated values indicate setup for blue and red arms, respectively.}
\end{deluxetable}

\begin{deluxetable}{lccc}
\tablewidth{0pt}
\tabletypesize{\scriptsize}
\tablecaption{Host Galaxy Photometry\label{tab:host_phot}}
\tablehead{
\colhead{Object} &
\colhead{Filter} &
\colhead{AB mag} &
\colhead{Telescope} 
}
\startdata
SN\,2018avk & $g$ & 23.26 $\pm$ 0.12 & SDSS \\
SN\,2018avk & $r$ & 22.45 $\pm$ 0.12 & SDSS \\
SN\,2018avk & $i$ & 22.44 $\pm$ 0.16 & SDSS \\
SN\,2018avk & $g_{\rm P1}$ & 23.26 $\pm$ 0.22  & PS1\\
SN\,2018avk & $r_{\rm P1}$ & 22.51 $\pm$ 0.12  & PS1 \\
SN\,2018avk & $i_{\rm P1}$ & 22.88 $\pm$ 0.12  & PS1 \\
SN\,2018bym & $g$ & 23.02 $\pm$ 0.16  & CFHT \\
SN\,2018bym & $r$ & 22.42 $\pm$ 0.06  & CFHT\\
SN\,2018bym & $r_{\rm P1}$ & 22.46 $\pm$ 0.19  & PS1 \\
SN\,2018bym & $i_{\rm P1}$ & 22.35 $\pm$ 0.14  & PS1 \\
SN\,2018don & NUV & 22.01 $\pm$ 0.14 & GALEX\\
SN\,2018don & $u$ & 20.37 $\pm$ 0.17 & SDSS \\
SN\,2018don & $g$ & 18.99 $\pm$ 0.03 & SDSS \\
SN\,2018don & $r$ & 18.26 $\pm$ 0.02 & SDSS \\
SN\,2018don & $i$ & 17.92 $\pm$ 0.02 & SDSS \\
SN\,2018don & $z$ & 17.60 $\pm$ 0.07 & SDSS \\
SN\,2018don & $g_{\rm P1}$ & 19.04 $\pm$ 0.04 & PS1 \\
SN\,2018don & $r_{\rm P1}$ & 18.31 $\pm$ 0.02 & PS1 \\
SN\,2018don & $i_{\rm P1}$ & 17.98 $\pm$ 0.04 & PS1 \\
SN\,2018don & $z_{\rm P1}$ & 17.77 $\pm$ 0.04 & PS1 \\
SN\,2018don & $y_{\rm P1}$ & 17.58 $\pm$ 0.09 & PS1 \\
SN\,2018don & $J$ & 17.23 $\pm$ 0.18 & 2MASS \\
SN\,2018don & $H$ & 17.38 $\pm$ 0.32 & 2MASS \\
SN\,2018don & $K$ & 17.46 $\pm$ 0.28 & 2MASS \\
SN\,2018don & $W1$ & 17.94 $\pm$ 0.05 & WISE \\
SN\,2018don & $W2$ & 18.28 $\pm$ 0.09 & WISE \\
SN\,2018bgv & UVW2 & 20.92 $\pm$ 0.07 & UVOT \\
SN\,2018bgv & UVM2 & 20.77 $\pm$ 0.06 & UVOT \\
SN\,2018bgv & UVW1 & 20.60 $\pm$ 0.12 & UVOT \\
SN\,2018bgv & $U$ & 20.72 $\pm$ 0.26 & UVOT \\
SN\,2018bgv & $u$ & 20.96 $\pm$ 0.14 & SDSS \\
SN\,2018bgv & $g$ & 19.96 $\pm$ 0.03 & SDSS \\
SN\,2018bgv & $r$ & 19.74 $\pm$ 0.07 & SDSS \\
SN\,2018bgv & $i$ & 19.26 $\pm$ 0.06 & SDSS \\
SN\,2018bgv & $z$ & 19.23 $\pm$ 0.16 & SDSS \\
SN\,2018bgv & $g_{\rm P1}$ & 19.93 $\pm$ 0.03 & PS1 \\
SN\,2018bgv & $r_{\rm P1}$ & 19.70 $\pm$ 0.06 & PS1 \\
SN\,2018bgv & $i_{\rm P1}$ & 19.44 $\pm$ 0.06 & PS1 \\
SN\,2018bgv & $z_{\rm P1}$ & 19.42 $\pm$ 0.09 & PS1 \\
SN\,2018bgv & $y_{\rm P1}$ & 19.61 $\pm$ 0.14 & PS1 \\
SN\,2018bgv & $J$ & 19.26 $\pm$ 0.32 & 2MASS \\
SN\,2018bgv & $W1$ & 20.12 $\pm$ 0.11 & WISE 
\enddata
\end{deluxetable}


\begin{deluxetable}{lccccc}
\tablewidth{0pt}
\tabletypesize{\scriptsize}
\tablecaption{Summary of Light Curve Properties \label{tab:lc_prop}}
\tablehead{
\colhead{Object} &
\colhead{$g$-band Peak Date} &
\colhead{$g$-band Peak Luminosity} &
\colhead{Explosion date\tablenotemark{a}} &
\colhead{$g$-band rise (t$_{1/2}$)} &
\colhead{$g$-band decline (t$_{1/2}$)} \\
\colhead{} &
\colhead{(MJD)} &
\colhead{(mag)} & 
\colhead{(MJD)} &
\colhead{(rest-frame days)} &
\colhead{(rest-frame days)}
}
\startdata
SN\,2018avk & 58255.2 & $-20.30 \pm 0.02$ & 58199.0 $\pm$ 0.3 & $35.3_{-0.4}^{+0.6}$ & $66.8_{-12.0}^{+2.8}$ \\
SN\,2018bym & 58280.9 & $-22.05 \pm 0.01$ & 58230.1 $\pm$ 0.5 & $25.1_{-0.3}^{+0.6}$ & $30.1_{-0.3}^{+0.2}$ \\
SN\,2018bgv & 58253.2 & $-20.68 \pm 0.09$ & 58242.5 $\pm$ 0.1 & $7.0_{-0.1}^{+0.1}$ & $16.7_{-0.3}^{+0.3}$ \\
SN\,2018don & 58268.2 & $-18.65 \pm 0.05$\tablenotemark{b} & 58213.6 $\pm$ 0.5 & $32.6_{-1.3}^{+0.7}$ & $35.7_{-0.7}^{+20.6}$
\enddata
\tablenotetext{a}{Assuming the rise in flux is described by a second-order polynomial. Quoted error bars include statistical uncertainty only, and do not account for e.g. systematic uncertainty comparing different functional forms for the rise.}
\tablenotetext{b}{If assuming E(B-V) = 0.4~mag, the reddening-corrected $g$-band peak magnitude would be $-20.09$~mag.}
\end{deluxetable}

\begin{deluxetable}{lccccc}
\tablewidth{0pt}
\tabletypesize{\scriptsize}
\tablecaption{Summary of Magnetar Model Fits \label{tab:magfits}}
\tablehead{
\colhead{Object} &
\colhead{$P$} &
\colhead{$B$} &
\colhead{$M_{\rm ej}$} &
\colhead{Host $A_{\rm V}$} &
\colhead{$\sigma$} \\
\colhead{} &
\colhead{(ms)} &
\colhead{($10^{14}$~G}) &
\colhead{(M$_{\odot}$)} &
\colhead{(mag)} &
\colhead{(mag)}
}
\startdata
SN\,2018bgv & $7.2_{-1.6}^{+1.3}$ & $4.8_{-1.6}^{+2.7}$ & $1.3_{-0.2}^{+0.3}$ & $0.00_{-0.00}^{+0.19}$ & $0.14_{-0.02}^{+0.01}$ \\
SN\,2018bym & $2.0_{-0.3}^{+0.2}$ & $3.3_{-1.9}^{+3.5}$ & $8.1_{-0.5}^{+0.6}$ & $0.00_{-0.00}^{+0.01}$ & $0.10_{-0.01}^{+0.01}$ \\
SN\,2018avk & $3.6_{-0.8}^{+0.7}$ & $1.1_{-0.4}^{+1.0}$ & $3.9_{-0.5}^{+0.5}$ & $0.00_{-0.00}^{+0.03}$ & $0.18_{-0.01}^{+0.01}$ \\
SN\,2018don & $3.0_{-0.6}^{+0.4}$ & $2.5_{-1.7}^{+4.7}$ & $9.5_{-2.0}^{+3.0}$ & $2.26_{-0.05}^{+0.05}$ & $0.08_{-0.01}^{+0.01}$
\enddata
\end{deluxetable}

\begin{deluxetable}{lcccc}
\tablewidth{0pt}
\tabletypesize{\scriptsize}
\tablecaption{Summary of Host Galaxy Properties \label{tab:host_prop}}
\tablehead{
\colhead{Object} &
\colhead{$\log_{10}$(Stellar Mass)} &
\colhead{$A_{\rm V}$} &
\colhead{$\log_{10}$(Age)} &
\colhead{SFR\tablenotemark{a}}\\
\colhead{} &
\colhead{(M$_{\odot}$)} &
\colhead{(mag)} &
\colhead{(yr)} &
\colhead{(M$_{\odot}~{\rm yr}^-1$)}
}
\startdata
SN\,2018avk & 7.82$^{+0.07}_{-0.08}$ & 0.03$^{+0.00}_{-0.03}$ & 8.40$^{+0.51}_{-0.15}$ & \nodata \\ 
SN\,2018don & 9.78$^{+0.16}_{-0.28}$ & 0.00$^{+1.81}_{-0.00}$ & 9.50$^{+0.29}_{-1.35}$ & 0.087$^{+6.222}_{-0.008}$ \\ 
SN\,2018bym & 8.58$^{+0.38}_{-0.64}$ & 0.03$^{+1.41}_{-0.03}$ & 8.20$^{+1.22}_{-1.38}$ & \nodata \\ 
SN\,2018bgv & 8.64$^{+0.11}_{-0.10}$ & 0.00$^{+0.14}_{-0.00}$ & 9.00$^{+0.31}_{-0.33}$ & 0.316$^{+0.120}_{-0.047}$
\enddata
\tablenotetext{a}{No data indicates that the SFR is not constrained by the fit to the available data.}
\end{deluxetable}


\begin{thebibliography}{}
\expandafter\ifx\csname natexlab\endcsname\relax\def\natexlab#1{#1}\fi
\providecommand{\url}[1]{\href{#1}{#1}}

\bibitem[{{Ambikasaran} {et~al.}(2015){Ambikasaran}, {Foreman-Mackey},
  {Greengard}, {Hogg}, \& {O'Neil}}]{afg+15}
{Ambikasaran}, S., {Foreman-Mackey}, D., {Greengard}, L., {Hogg}, D.~W., \&
  {O'Neil}, M. 2015, IEEE Transactions on Pattern Analysis and Machine
  Intelligence, 38, arXiv:1403.6015

\bibitem[{{Angus} {et~al.}(2016){Angus}, {Levan}, {Perley}, {Tanvir}, {Lyman},
  {Stanway}, \& {Fruchter}}]{alp+16}
{Angus}, C.~R., {Levan}, A.~J., {Perley}, D.~A., {et~al.} 2016, \mnras, 458, 84

\bibitem[{{Angus} {et~al.}(2019){Angus}, {Smith}, {Sullivan}, {Inserra},
  {Wiseman}, {D'Andrea}, {Thomas}, {Nichol}, {Galbany}, \&
  {Childress}}]{ass+19}
{Angus}, C.~R., {Smith}, M., {Sullivan}, M., {et~al.} 2019, \mnras, 487, 2215

\bibitem[{{Arcavi} {et~al.}(2016){Arcavi}, {Wolf}, {Howell}, {Bildsten},
  {Leloudas}, {Hardin}, {Prajs}, {Perley}, {Svirski}, {Gal-Yam}, {Katz},
  {McCully}, {Cenko}, {Lidman}, {Sullivan}, {Valenti}, {Astier}, {Balland},
  {Carlberg}, {Conley}, {Fouchez}, {Guy}, {Pain}, {Palanque-Delabrouille},
  {Perrett}, {Pritchet}, {Regnault}, {Rich}, \& {Ruhlmann-Kleider}}]{awh+16}
{Arcavi}, I., {Wolf}, W.~M., {Howell}, D.~A., {et~al.} 2016, \apj, 819, 35

\bibitem[{{Barkat} {et~al.}(1967){Barkat}, {Rakavy}, \& {Sack}}]{brs67}
{Barkat}, Z., {Rakavy}, G., \& {Sack}, N. 1967, Physical Review Letters, 18,
  379

\bibitem[{{Bellm} \& {Sesar}(2016)}]{Bellm2016}
{Bellm}, E.~C., \& {Sesar}, B. 2016, {pyraf-dbsp: Reduction pipeline for the
  Palomar Double Beam Spectrograph}, , , ascl:1602.002

\bibitem[{{Bellm} {et~al.}(2019{\natexlab{a}}){Bellm}, {Kulkarni}, {Graham},
  {Dekany}, {Smith}, {Riddle}, {Masci}, {Helou}, {Prince}, {Adams},
  {Barbarino}, {Barlow}, {Bauer}, {Beck}, {Belicki}, {Biswas}, {Blagorodnova},
  {Bodewits}, {Bolin}, {Brinnel}, {Brooke}, {Bue}, {Bulla}, {Burruss}, {Cenko},
  {Chang}, {Connolly}, {Coughlin}, {Cromer}, {Cunningham}, {De}, {Delacroix},
  {Desai}, {Duev}, {Eadie}, {Farnham}, {Feeney}, {Feindt}, {Flynn},
  {Franckowiak}, {Frederick}, {Fremling}, {Gal-Yam}, {Gezari}, {Giomi},
  {Goldstein}, {Golkhou}, {Goobar}, {Groom}, {Hacopians}, {Hale}, {Henning},
  {Ho}, {Hover}, {Howell}, {Hung}, {Huppenkothen}, {Imel}, {Ip}, {Ivezi{\'c}},
  {Jackson}, {Jones}, {Juric}, {Kasliwal}, {Kaspi}, {Kaye}, {Kelley},
  {Kowalski}, {Kramer}, {Kupfer}, {Landry}, {Laher}, {Lee}, {Lin}, {Lin},
  {Lunnan}, {Giomi}, {Mahabal}, {Mao}, {Miller}, {Monkewitz}, {Murphy},
  {Ngeow}, {Nordin}, {Nugent}, {Ofek}, {Patterson}, {Penprase}, {Porter},
  {Rauch}, {Rebbapragada}, {Reiley}, {Rigault}, {Rodriguez}, {van Roestel},
  {Rusholme}, {van Santen}, {Schulze}, {Shupe}, {Singer}, {Soumagnac}, {Stein},
  {Surace}, {Sollerman}, {Szkody}, {Taddia}, {Terek}, {Van Sistine}, {van
  Velzen}, {Vestrand}, {Walters}, {Ward}, {Ye}, {Yu}, {Yan}, \&
  {Zolkower}}]{bkg+19}
{Bellm}, E.~C., {Kulkarni}, S.~R., {Graham}, M.~J., {et~al.}
  2019{\natexlab{a}}, \pasp, 131, 018002

\bibitem[{{Bellm} {et~al.}(2019{\natexlab{b}}){Bellm}, {Kulkarni}, {Barlow},
  {Feindt}, {Graham}, {Goobar}, {Kupfer}, {Ngeow}, {Nugent}, {Ofek}, {Prince},
  {Riddle}, {Walters}, \& {Ye}}]{bkb+19}
{Bellm}, E.~C., {Kulkarni}, S.~R., {Barlow}, T., {et~al.} 2019{\natexlab{b}},
  \pasp, 131, 068003

\bibitem[{{Bhirombhakdi} {et~al.}(2018){Bhirombhakdi}, {Chornock}, {Margutti},
  {Nicholl}, {Metzger}, {Berger}, {Margalit}, \& {Milisavljevic}}]{bcm+18}
{Bhirombhakdi}, K., {Chornock}, R., {Margutti}, R., {et~al.} 2018, \apjl, 868,
  L32

\bibitem[{{Blanchard} {et~al.}(2018{\natexlab{a}}){Blanchard}, {Gomez},
  {Berger}, {Nicholl}, {Kattner}, \& {Weiner}}]{bgb+18}
{Blanchard}, P., {Gomez}, S., {Berger}, E., {et~al.} 2018{\natexlab{a}}, The
  Astronomer's Telegram, 11714

\bibitem[{{Blanchard} {et~al.}(2018{\natexlab{b}}){Blanchard}, {Nicholl},
  {Berger}, {Chornock}, {Margutti}, {Milisavljevic}, {Fong}, {MacLeod}, \&
  {Bhirombhakdi}}]{bnb+18}
{Blanchard}, P.~K., {Nicholl}, M., {Berger}, E., {et~al.} 2018{\natexlab{b}},
  \apj, 865, 9

\bibitem[{{Blanton} \& {Roweis}(2007)}]{br07}
{Blanton}, M.~R., \& {Roweis}, S. 2007, \aj, 133, 734

\bibitem[{{Blondin} \& {Tonry}(2007)}]{snid}
{Blondin}, S., \& {Tonry}, J.~L. 2007, \apj, 666, 1024

\bibitem[{{Bose} {et~al.}(2018){Bose}, {Dong}, {Pastorello}, {Filippenko},
  {Kochanek}, {Mauerhan}, {Romero-Ca{\~n}izales}, {Brink}, {Chen}, {Prieto},
  {Post}, {Ashall}, {Grupe}, {Tomasella}, {Benetti}, {Shappee}, {Stanek},
  {Cai}, {Falco}, {Lundqvist}, {Mattila}, {Mutel}, {Ochner}, {Pooley},
  {Stritzinger}, {Villanueva}, {Zheng}, {Beswick}, {Brown}, {Cappellaro},
  {Davis}, {Fraser}, {de Jaeger}, {Elias-Rosa}, {Gall}, {Gaudi}, {Herczeg},
  {Hestenes}, {Holoien}, {Hosseinzadeh}, {Hsiao}, {Hu}, {Jaejin}, {Jeffers},
  {Koff}, {Kumar}, {Kurtenkov}, {Lau}, {Prentice}, {Reynolds}, {Rudy},
  {Shahbandeh}, {Somero}, {Stassun}, {Thompson}, {Valenti}, {Woo}, \&
  {Yunus}}]{bdp+18}
{Bose}, S., {Dong}, S., {Pastorello}, A., {et~al.} 2018, \apj, 853, 57

\bibitem[{{Bourne} {et~al.}(2012){Bourne}, {Maddox}, {Dunne}, {Auld}, {Baes},
  {Baldry}, {Bonfield}, {Cooray}, {Croom}, {Dariush}, {de Zotti}, {Driver},
  {Dye}, {Eales}, {Gomez}, {Gonz{\'a}lez-Nuevo}, {Hopkins}, {Ibar}, {Jarvis},
  {Lapi}, {Madore}, {Micha{\l}owski}, {Pohlen}, {Popescu}, {Rigby}, {Seibert},
  {Smith}, {Tuffs}, {van der Werf}, {Brough}, {Buttiglione}, {Cava},
  {Clements}, {Conselice}, {Fritz}, {Hopwood}, {Ivison}, {Jones}, {Kelvin},
  {Liske}, {Loveday}, {Norberg}, {Robotham}, {Rodighiero}, \& {Temi}}]{bmd+12}
{Bourne}, N., {Maddox}, S.~J., {Dunne}, L., {et~al.} 2012, \mnras, 421, 3027

\bibitem[{{Branch} {et~al.}(2002){Branch}, {Benetti}, {Kasen}, {Baron},
  {Jeffery}, {Hatano}, {Stathakis}, {Filippenko}, {Matheson}, {Pastorello},
  {Altavilla}, {Cappellaro}, {Rizzi}, {Turatto}, {Li}, {Leonard}, \&
  {Shields}}]{bbk+02}
{Branch}, D., {Benetti}, S., {Kasen}, D., {et~al.} 2002, ApJ, 566, 1005

\bibitem[{{Chambers} {et~al.}(2016){Chambers}, {Magnier}, {Metcalfe},
  {Flewelling}, {Huber}, {Waters}, {Denneau}, {Draper}, {Farrow}, \&
  {Finkbeiner}}]{cmm+16}
{Chambers}, K.~C., {Magnier}, E.~A., {Metcalfe}, N., {et~al.} 2016, arXiv
  e-prints, arXiv:1612.05560

\bibitem[{{Chambers} {et~al.}(2018){Chambers}, {Huber}, {Flewelling},
  {Magnier}, {Schultz}, {Lowe}, {Bulger}, {Smartt}, {Smith}, \&
  {Tonry}}]{2018don}
{Chambers}, K.~C., {Huber}, M.~E., {Flewelling}, H., {et~al.} 2018, Transient
  Name Server Discovery Report, 2018-972, 1

\bibitem[{{Chatzopoulos} \& {Tuminello}(2019)}]{ct19}
{Chatzopoulos}, E., \& {Tuminello}, R. 2019, \apj, 874, 68

\bibitem[{{Chatzopoulos} \& {Wheeler}(2012)}]{cw12b}
{Chatzopoulos}, E., \& {Wheeler}, J.~C. 2012, ApJ, 760, 154

\bibitem[{{Chatzopoulos} {et~al.}(2012){Chatzopoulos}, {Wheeler}, \&
  {Vinko}}]{cwv12}
{Chatzopoulos}, E., {Wheeler}, J.~C., \& {Vinko}, J. 2012, ApJ, 746, 121

\bibitem[{{Chen} {et~al.}(2017{\natexlab{a}}){Chen}, {Smartt}, {Yates},
  {Nicholl}, {Kr{\"u}hler}, {Schady}, {Dennefeld}, \& {Inserra}}]{csy+17}
{Chen}, T.-W., {Smartt}, S.~J., {Yates}, R.~M., {et~al.} 2017{\natexlab{a}},
  \mnras, 470, 3566

\bibitem[{{Chen} {et~al.}(2017{\natexlab{b}}){Chen}, {Schady}, {Xiao},
  {Eldridge}, {Schweyer}, {Lee}, {Yu}, {Smartt}, \& {Inserra}}]{csx+17}
{Chen}, T.-W., {Schady}, P., {Xiao}, L., {et~al.} 2017{\natexlab{b}}, \apjl,
  849, L4

\bibitem[{{Chevalier} \& {Irwin}(2011)}]{ci11}
{Chevalier}, R.~A., \& {Irwin}, C.~M. 2011, ApJL, 729, L6

\bibitem[{{Cooke} {et~al.}(2012){Cooke}, {Sullivan}, {Gal-Yam}, {Barton},
  {Carlberg}, {Ryan-Weber}, {Horst}, {Omori}, \& {D{\'\i}az}}]{csg+12}
{Cooke}, J., {Sullivan}, M., {Gal-Yam}, A., {et~al.} 2012, \nat, 491, 228

\bibitem[{{Curtin} {et~al.}(2019){Curtin}, {Cooke}, {Moriya}, {Tanaka},
  {Quimby}, {Bernard}, {Galbany}, {Jiang}, {Lee}, {Maeda}, {Morokuma},
  {Nomoto}, {Pignata}, {Pritchard}, {Suzuki}, {Takahashi}, {Tanaka},
  {Tominaga}, {Yamaguchi}, \& {Yasuda}}]{ccm+19}
{Curtin}, C., {Cooke}, J., {Moriya}, T.~J., {et~al.} 2019, \apjs, 241, 17

\bibitem[{{Cutri} {et~al.}(2013){Cutri}, {Wright}, {Conrow}, {Fowler},
  {Eisenhardt}, {Grillmair}, {Kirkpatrick}, {Masci}, {McCallon}, {Wheelock},
  {Fajardo-Acosta}, {Yan}, {Benford}, {Harbut}, {Jarrett}, {Lake}, {Leisawitz},
  {Ressler}, {Stanford}, {Tsai}, {Liu}, {Helou}, {Mainzer}, {Gettings},
  {Gonzalez}, {Hoffman}, {Marsh}, {Padgett}, {Skrutskie}, {Beck}, {Papin}, \&
  {Wittman}}]{Cutri2013a}
{Cutri}, R.~M., {Wright}, E.~L., {Conrow}, T., {et~al.} 2013, {Explanatory
  Supplement to the AllWISE Data Release Products}, Tech. rep.

\bibitem[{{De Cia} {et~al.}(2018){De Cia}, {Gal-Yam}, {Rubin}, {Leloudas},
  {Vreeswijk}, {Perley}, {Quimby}, {Yan}, {Sullivan}, {Fl{\"o}rs}, {Sollerman},
  {Bersier}, {Cenko}, {Gal-Yam}, {Maguire}, {Ofek}, {Prentice}, {Schulze},
  {Spyromilio}, {Valenti}, {Arcavi}, {Corsi}, {Howell}, {Mazzali}, {Kasliwal},
  {Taddia}, \& {Yaron}}]{dgr+18}
{De Cia}, A., {Gal-Yam}, A., {Rubin}, A., {et~al.} 2018, \apj, 860, 100

\bibitem[{{Dekany} {et~al.}(2016){Dekany}, {Smith}, {Belicki}, {Delacroix},
  {Duggan}, {Feeney}, {Hale}, {Kaye}, {Milburn}, {Murphy}, {Porter}, {Reiley},
  {Riddle}, {Rodriguez}, \& {Bellm}}]{dsb+16}
{Dekany}, R., {Smith}, R.~M., {Belicki}, J., {et~al.} 2016, in Society of
  Photo-Optical Instrumentation Engineers (SPIE) Conference Series, Vol. 9908,
  Ground-based and Airborne Instrumentation for Astronomy VI, 99085M

\bibitem[{{Delgado} {et~al.}(2018{\natexlab{a}}){Delgado}, {Harrison},
  {Hodgkin}, {Leeuwen}, {Rixon}, \& {Yoldas}}]{2018avk}
{Delgado}, A., {Harrison}, D., {Hodgkin}, S., {et~al.} 2018{\natexlab{a}},
  Transient Name Server Discovery Report, 2018-498, 1

\bibitem[{{Delgado} {et~al.}(2018{\natexlab{b}}){Delgado}, {Harrison},
  {Hodgkin}, {Leeuwen}, {Rixon}, \& {Yoldas}}]{2018bgv}
---. 2018{\natexlab{b}}, Transient Name Server Discovery Report, 2018-599, 1

\bibitem[{{Dessart}(2019)}]{des19}
{Dessart}, L. 2019, \aap, 621, A141

\bibitem[{{Dessart} {et~al.}(2012){Dessart}, {Hillier}, {Waldman}, {Livne}, \&
  {Blondin}}]{dhw+12}
{Dessart}, L., {Hillier}, D.~J., {Waldman}, R., {Livne}, E., \& {Blondin}, S.
  2012, MNRAS, 426, L76

\bibitem[{{Dong} {et~al.}(2018){Dong}, {Bose}, {Siltala}, {Holmbo},
  {Stritzinger}, {Stanek}, {Bersier}, {Tucker}, {Shappee}, {Huber}, \&
  {Fras}}]{dbs+18}
{Dong}, S., {Bose}, S., {Siltala}, L., {et~al.} 2018, The Astronomer's
  Telegram, 11654

\bibitem[{{Drout} {et~al.}(2011){Drout}, {Soderberg}, {Gal-Yam}, {Cenko},
  {Fox}, {Leonard}, {Sand}, {Moon}, {Arcavi}, \& {Green}}]{dsg+11}
{Drout}, M.~R., {Soderberg}, A.~M., {Gal-Yam}, A., {et~al.} 2011, ApJ, 741, 97

\bibitem[{{Drout} {et~al.}(2014){Drout}, {Chornock}, {Soderberg}, {Sanders},
  {McKinnon}, {Rest}, {Foley}, {Milisavljevic}, {Margutti}, {Berger},
  {Calkins}, {Fong}, {Gezari}, {Huber}, {Kankare}, {Kirshner}, {Leibler},
  {Lunnan}, {Mattila}, {Marion}, {Narayan}, {Riess}, {Roth}, {Scolnic},
  {Smartt}, {Tonry}, {Burgett}, {Chambers}, {Hodapp}, {Jedicke}, {Kaiser},
  {Magnier}, {Metcalfe}, {Morgan}, {Price}, \& {Waters}}]{dcs+14}
{Drout}, M.~R., {Chornock}, R., {Soderberg}, A.~M., {et~al.} 2014, \apj, 794,
  23

\bibitem[{{Duev} {et~al.}(2019){Duev}, {Mahabal}, {Masci}, {Graham},
  {Rusholme}, {Walters}, {Karmarkar}, {Frederick}, {Kasliwal}, {Rebbapragada},
  \& {Ward}}]{dima2019}
{Duev}, D.~A., {Mahabal}, A., {Masci}, F.~J., {et~al.} 2019, \mnras, 2039

\bibitem[{{Eftekhari} {et~al.}(2019){Eftekhari}, {Berger}, {Margalit},
  {Blanchard}, {Patton}, {Demorest}, {Williams}, {Chatterjee}, {Cordes},
  {Lunnan}, {Metzger}, \& {Nicholl}}]{ebm+19}
{Eftekhari}, T., {Berger}, E., {Margalit}, B., {et~al.} 2019, \apjl, 876, L10

\bibitem[{{Flewelling} {et~al.}(2016){Flewelling}, {Magnier}, {Chambers},
  {Heasley}, {Holmberg}, {Huber}, {Sweeney}, {Waters}, {Calamida}, \&
  {Casertano}}]{fmc+16}
{Flewelling}, H.~A., {Magnier}, E.~A., {Chambers}, K.~C., {et~al.} 2016, arXiv
  e-prints, arXiv:1612.05243

\bibitem[{{Foreman-Mackey} {et~al.}(2013){Foreman-Mackey}, {Hogg}, {Lang}, \&
  {Goodman}}]{fhl+13}
{Foreman-Mackey}, D., {Hogg}, D.~W., {Lang}, D., \& {Goodman}, J. 2013, \pasp,
  125, 306

\bibitem[{{Fremling} {et~al.}(2018){Fremling}, {Sharma}, {Kulkarni}, {Miller},
  {Taggart}, {Perley}, \& {Gooba}}]{fsk+18}
{Fremling}, C., {Sharma}, Y., {Kulkarni}, S.~R., {et~al.} 2018, The
  Astronomer's Telegram, 11688

\bibitem[{{Fremling} {et~al.}(2016){Fremling}, {Sollerman}, {Taddia}, {Ergon},
  {Fraser}, {Karamehmetoglu}, {Valenti}, {Jerkstrand}, {Arcavi}, {Bufano},
  {Elias Rosa}, {Filippenko}, {Fox}, {Gal-Yam}, {Howell}, {Kotak}, {Mazzali},
  {Milisavljevic}, {Nugent}, {Nyholm}, {Pian}, \& {Smartt}}]{fst+16}
{Fremling}, C., {Sollerman}, J., {Taddia}, F., {et~al.} 2016, A\&A, 593, A68

\bibitem[{{Gaia Collaboration} {et~al.}(2016){Gaia Collaboration}, {Prusti},
  {de Bruijne}, {Brown}, {Vallenari}, {Babusiaux}, {Bailer-Jones}, {Bastian},
  {Biermann}, {Evans}, \& et~al.}]{gaia16}
{Gaia Collaboration}, {Prusti}, T., {de Bruijne}, J.~H.~J., {et~al.} 2016,
  \aap, 595, A1

\bibitem[{{Gal-Yam}(2012)}]{gal12}
{Gal-Yam}, A. 2012, Science, 337, 927

\bibitem[{{Gal-Yam}(2019)}]{gal19}
---. 2019, \araa, 57, 305

\bibitem[{{Gal-Yam} {et~al.}(2009){Gal-Yam}, {Mazzali}, {Ofek}, {Nugent},
  {Kulkarni}, {Kasliwal}, {Quimby}, {Filippenko}, {Cenko}, {Chornock},
  {Waldman}, {Kasen}, {Sullivan}, {Beshore}, {Drake}, {Thomas}, {Bloom},
  {Poznanski}, {Miller}, {Foley}, {Silverman}, {Arcavi}, {Ellis}, \&
  {Deng}}]{gmo+09}
{Gal-Yam}, A., {Mazzali}, P., {Ofek}, E.~O., {et~al.} 2009, Nature, 462, 624

\bibitem[{{Gehrels} {et~al.}(2004){Gehrels}, {Chincarini}, {Giommi}, {Mason},
  {Nousek}, {Wells}, {White}, {Barthelmy}, {Burrows}, {Cominsky}, {Hurley},
  {Marshall}, {M{\'e}sz{\'a}ros}, {Roming}, {Angelini}, {Barbier}, {Belloni},
  {Campana}, {Caraveo}, {Chester}, {Citterio}, {Cline}, {Cropper}, {Cummings},
  {Dean}, {Feigelson}, {Fenimore}, {Frail}, {Fruchter}, {Garmire}, {Gendreau},
  {Ghisellini}, {Greiner}, {Hill}, {Hunsberger}, {Krimm}, {Kulkarni}, {Kumar},
  {Lebrun}, {Lloyd-Ronning}, {Markwardt}, {Mattson}, {Mushotzky}, {Norris},
  {Osborne}, {Paczynski}, {Palmer}, {Park}, {Parsons}, {Paul}, {Rees},
  {Reynolds}, {Rhoads}, {Sasseen}, {Schaefer}, {Short}, {Smale}, {Smith},
  {Stella}, {Tagliaferri}, {Takahashi}, {Tashiro}, {Townsley}, {Tueller},
  {Turner}, {Vietri}, {Voges}, {Ward}, {Willingale}, {Zerbi}, \&
  {Zhang}}]{gcg+04}
{Gehrels}, N., {Chincarini}, G., {Giommi}, P., {et~al.} 2004, \apj, 611, 1005

\bibitem[{{Gezari} {et~al.}(2009){Gezari}, {Halpern}, {Grupe}, {Yuan},
  {Quimby}, {McKay}, {Chamarro}, {Sisson}, {Akerlof}, {Wheeler}, {Brown},
  {Cenko}, {Rau}, {Djordjevic}, \& {Terndrup}}]{ghg+09}
{Gezari}, S., {Halpern}, J.~P., {Grupe}, D., {et~al.} 2009, ApJ, 690, 1313

\bibitem[{{Gomez} {et~al.}(2019){Gomez}, {Berger}, {Nicholl}, {Blanchard},
  {Villar}, {Patton}, {Chornock}, {Leja}, {Hosseinzadeh}, \&
  {Cowperthwaite}}]{gbn+19}
{Gomez}, S., {Berger}, E., {Nicholl}, M., {et~al.} 2019, \apj, 881, 87

\bibitem[{{Graham} {et~al.}(2019){Graham}, {Kulkarni}, {Bellm}, {Adams},
  {Barbarino}, {Blagorodnova}, {Bodewits}, {Bolin}, {Brady}, {Cenko}, {Chang},
  {Coughlin}, {De}, {Eadie}, {Farnham}, {Feindt}, {Franckowiak}, {Fremling},
  {Gezari}, {Ghosh}, {Goldstein}, {Golkhou}, {Goobar}, {Ho}, {Huppenkothen},
  {Ivezi{\'c}}, {Jones}, {Juric}, {Kaplan}, {Kasliwal}, {Kelley}, {Kupfer},
  {Lee}, {Lin}, {Lunnan}, {Mahabal}, {Miller}, {Ngeow}, {Nugent}, {Ofek},
  {Prince}, {Rauch}, {van Roestel}, {Schulze}, {Singer}, {Sollerman}, {Taddia},
  {Yan}, {Ye}, {Yu}, {Barlow}, {Bauer}, {Beck}, {Belicki}, {Biswas}, {Brinnel},
  {Brooke}, {Bue}, {Bulla}, {Burruss}, {Connolly}, {Cromer}, {Cunningham},
  {Dekany}, {Delacroix}, {Desai}, {Duev}, {Feeney}, {Flynn}, {Frederick},
  {Gal-Yam}, {Giomi}, {Groom}, {Hacopians}, {Hale}, {Helou}, {Henning},
  {Hover}, {Hillenbrand}, {Howell}, {Hung}, {Imel}, {Ip}, {Jackson}, {Kaspi},
  {Kaye}, {Kowalski}, {Kramer}, {Kuhn}, {Landry}, {Laher}, {Mao}, {Masci},
  {Monkewitz}, {Murphy}, {Nordin}, {Patterson}, {Penprase}, {Porter},
  {Rebbapragada}, {Reiley}, {Riddle}, {Rigault}, {Rodriguez}, {Rusholme}, {van
  Santen}, {Shupe}, {Smith}, {Soumagnac}, {Stein}, {Surace}, {Szkody}, {Terek},
  {Van Sistine}, {van Velzen}, {Vestrand}, {Walters}, {Ward}, {Zhang}, \&
  {Zolkower}}]{gkb+19}
{Graham}, M.~J., {Kulkarni}, S.~R., {Bellm}, E.~C., {et~al.} 2019, \pasp, 131,
  078001

\bibitem[{{Greiner} {et~al.}(2015){Greiner}, {Mazzali}, {Kann}, {Kr{\"u}hler},
  {Pian}, {Prentice}, {Olivares E.}, {Rossi}, {Klose}, {Taubenberger}, {Knust},
  {Afonso}, {Ashall}, {Bolmer}, {Delvaux}, {Diehl}, {Elliott}, {Filgas},
  {Fynbo}, {Graham}, {Guelbenzu}, {Kobayashi}, {Leloudas}, {Savaglio},
  {Schady}, {Schmidl}, {Schweyer}, {Sudilovsky}, {Tanga}, {Updike}, {van
  Eerten}, \& {Varela}}]{gmk+15}
{Greiner}, J., {Mazzali}, P.~A., {Kann}, D.~A., {et~al.} 2015, \nat, 523, 189

\bibitem[{{Guillochon} {et~al.}(2018){Guillochon}, {Nicholl}, {Villar},
  {Mockler}, {Narayan}, {Mandel}, {Berger}, \& {Williams}}]{gnv+18}
{Guillochon}, J., {Nicholl}, M., {Villar}, V.~A., {et~al.} 2018, \apjs, 236, 6

\bibitem[{{Guillochon} {et~al.}(2017){Guillochon}, {Parrent}, {Kelley}, \&
  {Margutti}}]{gpk+17}
{Guillochon}, J., {Parrent}, J., {Kelley}, L.~Z., \& {Margutti}, R. 2017, \apj,
  835, 64

\bibitem[{{Heger} \& {Woosley}(2002)}]{hw02}
{Heger}, A., \& {Woosley}, S.~E. 2002, ApJ, 567, 532

\bibitem[{{Ho} {et~al.}(2019){Ho}, {Goldstein}, {Schulze}, {Khatami}, {Perley},
  {Ergon}, {Gal-Yam}, {Corsi}, {Andreoni}, {Barbarino}, {Bellm},
  {Blagorodnova}, {Bright}, {Burns}, {Cenko}, {Cunningham}, {De}, {Dekany},
  {Dugas}, {Fender}, {Fransson}, {Fremling}, {Goldstein}, {Graham}, {Hale},
  {Horesh}, {Hung}, {Kasliwal}, {Kuin}, {Kulkarni}, {Kupfer}, {Lunnan},
  {Masci}, {Ngeow}, {Nugent}, {Ofek}, {Patterson}, {Petitpas}, {Rusholme},
  {Sai}, {Sfaradi}, {Shupe}, {Sollerman}, {Soumagnac}, {Tachibana}, {Taddia},
  {Walters}, {Wang}, {Yao}, \& {Zhang}}]{hgs+19}
{Ho}, A. Y.~Q., {Goldstein}, D.~A., {Schulze}, S., {et~al.} 2019, \apj, 887,
  169

\bibitem[{{Hotokezaka} {et~al.}(2017){Hotokezaka}, {Kashiyama}, \&
  {Murase}}]{Hotokezaka2017}
{Hotokezaka}, K., {Kashiyama}, K., \& {Murase}, K. 2017, \apj, 850, 18

\bibitem[{{Howell} {et~al.}(2005){Howell}, {Sullivan}, {Perrett}, {Bronder},
  {Hook}, {Astier}, {Aubourg}, {Balam}, {Basa}, {Carlberg}, {Fabbro},
  {Fouchez}, {Guy}, {Lafoux}, {Neill}, {Pain}, {Palanque-Delabrouille},
  {Pritchet}, {Regnault}, {Rich}, {Taillet}, {Knop}, {McMahon}, {Perlmutter},
  \& {Walton}}]{hsp+05}
{Howell}, D.~A., {Sullivan}, M., {Perrett}, K., {et~al.} 2005, \apj, 634, 1190

\bibitem[{{Inserra} {et~al.}(2018{\natexlab{a}}){Inserra}, {Prajs},
  {Gutierrez}, {Angus}, {Smith}, \& {Sullivan}}]{ipg+18}
{Inserra}, C., {Prajs}, S., {Gutierrez}, C.~P., {et~al.} 2018{\natexlab{a}},
  \apj, 854, 175

\bibitem[{{Inserra} {et~al.}(2013){Inserra}, {Smartt}, {Jerkstrand}, {Valenti},
  {Fraser}, {Wright}, {Smith}, {Chen}, {Kotak}, {Pastorello}, {Nicholl},
  {Bresolin}, {Kudritzki}, {Benetti}, {Botticella}, {Burgett}, {Chambers},
  {Ergon}, {Flewelling}, {Fynbo}, {Geier}, {Hodapp}, {Howell}, {Huber},
  {Kaiser}, {Leloudas}, {Magill}, {Magnier}, {McCrum}, {Metcalfe}, {Price},
  {Rest}, {Sollerman}, {Sweeney}, {Taddia}, {Taubenberger}, {Tonry},
  {Wainscoat}, {Waters}, \& {Young}}]{isj+13}
{Inserra}, C., {Smartt}, S.~J., {Jerkstrand}, A., {et~al.} 2013, ApJ, 770, 128

\bibitem[{{Inserra} {et~al.}(2018{\natexlab{b}}){Inserra}, {Smartt}, {Gall},
  {Leloudas}, {Chen}, {Schulze}, {Jerkstrand}, {Nicholl}, {Anderson}, {Arcavi},
  {Benetti}, {Cartier}, {Childress}, {Della Valle}, {Flewelling}, {Fraser},
  {Gal-Yam}, {Guti{\'e}rrez}, {Hosseinzadeh}, {Howell}, {Huber}, {Kankare},
  {Kr{\"u}hler}, {Magnier}, {Maguire}, {McCully}, {Prajs}, {Primak}, {Scalzo},
  {Schmidt}, {Smith}, {Smith}, {Tucker}, {Valenti}, {Wilman}, {Young}, \&
  {Yuan}}]{isg+18}
{Inserra}, C., {Smartt}, S.~J., {Gall}, E.~E.~E., {et~al.} 2018{\natexlab{b}},
  MNRAS, 475, 1046

\bibitem[{{Izzo} {et~al.}(2018){Izzo}, {Th{\"o}ne}, {Garc{\'\i}a-Benito}, {de
  Ugarte Postigo}, {Cano}, {Kann}, {Bensch}, {Della Valle},
  {Galad{\'\i}-Enr{\'\i}quez}, \& {Hedrosa}}]{itg+18}
{Izzo}, L., {Th{\"o}ne}, C.~C., {Garc{\'\i}a-Benito}, R., {et~al.} 2018, \aap,
  610, A11

\bibitem[{{Jerkstrand} {et~al.}(2016){Jerkstrand}, {Smartt}, \&
  {Heger}}]{jsh16}
{Jerkstrand}, A., {Smartt}, S.~J., \& {Heger}, A. 2016, MNRAS, 455, 3207

\bibitem[{{Kann} {et~al.}(2019){Kann}, {Schady}, {Olivares E.}, {Klose},
  {Rossi}, {Perley}, {Kr{\"u}hler}, {Greiner}, {Nicuesa Guelbenzu}, {Elliott},
  {Knust}, {Filgas}, {Pian}, {Mazzali}, {Fynbo}, {Leloudas}, {Afonso},
  {Delvaux}, {Graham}, {Rau}, {Schmidl}, {Schulze}, {Tanga}, {Updike}, \&
  {Varela}}]{kso+19}
{Kann}, D.~A., {Schady}, P., {Olivares E.}, F., {et~al.} 2019, \aap, 624, A143

\bibitem[{{Kasen} \& {Bildsten}(2010)}]{kb10}
{Kasen}, D., \& {Bildsten}, L. 2010, ApJ, 717, 245

\bibitem[{{Kasliwal} {et~al.}(2019){Kasliwal}, {Cannella}, {Bagdasaryan},
  {Hung}, {Feindt}, {Singer}, {Coughlin}, {Fremling}, {Walters}, {Duev},
  {Itoh}, \& {Quimby}}]{kcb+19}
{Kasliwal}, M.~M., {Cannella}, C., {Bagdasaryan}, A., {et~al.} 2019, \pasp,
  131, 038003

\bibitem[{{Komatsu} {et~al.}(2011){Komatsu}, {Smith}, {Dunkley}, {Bennett},
  {Gold}, {Hinshaw}, {Jarosik}, {Larson}, {Nolta}, {Page}, {Spergel},
  {Halpern}, {Hill}, {Kogut}, {Limon}, {Meyer}, {Odegard}, {Tucker}, {Weiland},
  {Wollack}, \& {Wright}}]{ksd+11}
{Komatsu}, E., {Smith}, K.~M., {Dunkley}, J., {et~al.} 2011, ApJs, 192, 18

\bibitem[{{Kriek} {et~al.}(2009){Kriek}, {van Dokkum}, {Labb{\'e}}, {Franx},
  {Illingworth}, {Marchesini}, \& {Quadri}}]{kvl+09}
{Kriek}, M., {van Dokkum}, P.~G., {Labb{\'e}}, I., {et~al.} 2009, ApJ, 700, 221

\bibitem[{{Kulkarni} {et~al.}(2018){Kulkarni}, {Perley}, \& {Miller}}]{kpm18}
{Kulkarni}, S.~R., {Perley}, D.~A., \& {Miller}, A.~A. 2018, \apj, 860, 22

\bibitem[{{Lang}(2014)}]{Lang2014a}
{Lang}, D. 2014, \aj, 147, 108

\bibitem[{{Leloudas} {et~al.}(2015){Leloudas}, {Schulze}, {Kr{\"u}hler},
  {Gorosabel}, {Christensen}, {Mehner}, {de Ugarte Postigo}, {Amor{\'{\i}}n},
  {Th{\"o}ne}, {Anderson}, {Bauer}, {Gallazzi}, {He{\l}miniak}, {Hjorth},
  {Ibar}, {Malesani}, {Morell}, {Vinko}, \& {Wheeler}}]{lsk+14}
{Leloudas}, G., {Schulze}, S., {Kr{\"u}hler}, T., {et~al.} 2015, MNRAS, 449,
  917

\bibitem[{{Liu} {et~al.}(2018){Liu}, {Wang}, {Wang}, \& {Dai}}]{lww+18}
{Liu}, L.-D., {Wang}, L.-J., {Wang}, S.-Q., \& {Dai}, Z.-G. 2018, \apj, 856, 59

\bibitem[{{Liu} {et~al.}(2017{\natexlab{a}}){Liu}, {Wang}, {Wang}, {Dai}, {Yu},
  \& {Peng}}]{lww+17}
{Liu}, L.-D., {Wang}, S.-Q., {Wang}, L.-J., {et~al.} 2017{\natexlab{a}}, \apj,
  842, 26

\bibitem[{{Liu} {et~al.}(2017{\natexlab{b}}){Liu}, {Modjaz}, \&
  {Bianco}}]{lmb+17}
{Liu}, Y.-Q., {Modjaz}, M., \& {Bianco}, F.~B. 2017{\natexlab{b}}, \apj, 845,
  85

\bibitem[{{Lunnan} {et~al.}(2013){Lunnan}, {Chornock}, {Berger},
  {Milisavljevic}, {Drout}, {Sanders}, {Challis}, {Czekala}, {Foley}, {Fong},
  {Huber}, {Kirshner}, {Leibler}, {Marion}, {McCrum}, {Narayan}, {Rest},
  {Roth}, {Scolnic}, {Smartt}, {Smith}, {Soderberg}, {Stubbs}, {Tonry},
  {Burgett}, {Chambers}, {Kudritzki}, {Magnier}, \& {Price}}]{lcb+13}
{Lunnan}, R., {Chornock}, R., {Berger}, E., {et~al.} 2013, ApJ, 771, 97

\bibitem[{{Lunnan} {et~al.}(2014){Lunnan}, {Chornock}, {Berger}, {Laskar},
  {Fong}, {Rest}, {Sanders}, {Challis}, {Drout}, {Foley}, {Huber}, {Kirshner},
  {Leibler}, {Marion}, {McCrum}, {Milisavljevic}, {Narayan}, {Scolnic},
  {Smartt}, {Smith}, {Soderberg}, {Tonry}, {Burgett}, {Chambers}, {Flewelling},
  {Hodapp}, {Kaiser}, {Magnier}, {Price}, \& {Wainscoat}}]{lcb+14}
---. 2014, ApJ, 787, 138

\bibitem[{{Lunnan} {et~al.}(2015){Lunnan}, {Chornock}, {Berger}, {Rest},
  {Fong}, {Scolnic}, {Jones}, {Soderberg}, {Challis}, {Drout}, {Foley},
  {Huber}, {Kirshner}, {Leibler}, {Marion}, {McCrum}, {Milisavljevic},
  {Narayan}, {Sanders}, {Smartt}, {Smith}, {Tonry}, {Burgett}, {Chambers},
  {Flewelling}, {Kudritzki}, {Wainscoat}, \& {Waters}}]{lcb+15}
---. 2015, ApJ, 804, 90

\bibitem[{{Lunnan} {et~al.}(2016){Lunnan}, {Chornock}, {Berger},
  {Milisavljevic}, {Jones}, {Rest}, {Fong}, {Fransson}, {Margutti}, {Drout},
  {Blanchard}, {Challis}, {Cowperthwaite}, {Foley}, {Kirshner}, {Morrell},
  {Riess}, {Roth}, {Scolnic}, {Smartt}, {Smith}, {Villar}, {Chambers},
  {Draper}, {Huber}, {Kaiser}, {Kudritzki}, {Magnier}, {Metcalfe}, \&
  {Waters}}]{lcb+16}
---. 2016, ApJ, 831, 144

\bibitem[{{Lunnan} {et~al.}(2018{\natexlab{a}}){Lunnan}, {Chornock}, {Berger},
  {Jones}, {Rest}, {Czekala}, {Dittmann}, {Drout}, {Foley}, {Fong}, {Kirshner},
  {Laskar}, {Leibler}, {Margutti}, {Milisavljevic}, {Narayan}, {Pan}, {Riess},
  {Roth}, {Sanders}, {Scolnic}, {Smartt}, {Smith}, {Chambers}, {Draper},
  {Flewelling}, {Huber}, {Kaiser}, {Kudritzki}, {Magnier}, {Metcalfe},
  {Wainscoat}, {Waters}, \& {Willman}}]{lcb+18}
---. 2018{\natexlab{a}}, ApJ, 852, 81

\bibitem[{{Lunnan} {et~al.}(2018{\natexlab{b}}){Lunnan}, {Fransson},
  {Vreeswijk}, {Woosley}, {Leloudas}, {Perley}, {Quimby}, {Yan},
  {Blagorodnova}, {Bue}, {Cenko}, {De Cia}, {Cook}, {Fremling}, {Gatkine},
  {Gal-Yam}, {Kasliwal}, {Kulkarni}, {Masci}, {Nugent}, {Nyholm}, {Rubin},
  {Suzuki}, \& {Wozniak}}]{lfv+18}
{Lunnan}, R., {Fransson}, C., {Vreeswijk}, P.~M., {et~al.} 2018{\natexlab{b}},
  Nature Astronomy, 2, 887

\bibitem[{{Mahabal} {et~al.}(2019){Mahabal}, {Rebbapragada}, {Walters},
  {Masci}, {Blagorodnova}, {van Roestel}, {Ye}, {Biswas}, {Burdge}, {Chang},
  {Duev}, {Golkhou}, {Miller}, {Nordin}, {Ward}, {Adams}, {Bellm}, {Branton},
  {Bue}, {Cannella}, {Connolly}, {Dekany}, {Feindt}, {Hung}, {Fortson},
  {Frederick}, {Fremling}, {Gezari}, {Graham}, {Groom}, {Kasliwal}, {Kulkarni},
  {Kupfer}, {Lin}, {Lintott}, {Lunnan}, {Parejko}, {Prince}, {Riddle},
  {Rusholme}, {Saunders}, {Sedaghat}, {Shupe}, {Singer}, {Soumagnac}, {Szkody},
  {Tachibana}, {Tirumala}, {van Velzen}, \& {Wright}}]{mrw+19}
{Mahabal}, A., {Rebbapragada}, U., {Walters}, R., {et~al.} 2019, \pasp, 131,
  038002

\bibitem[{{Maraston}(2005)}]{mar05}
{Maraston}, C. 2005, MNRAS, 362, 799

\bibitem[{{Margutti} {et~al.}(2018){Margutti}, {Chornock}, {Metzger},
  {Coppejans}, {Guidorzi}, {Migliori}, {Milisavljevic}, {Berger}, {Nicholl},
  {Zauderer}, {Lunnan}, {Kamble}, {Drout}, \& {Modjaz}}]{mcm+18}
{Margutti}, R., {Chornock}, R., {Metzger}, B.~D., {et~al.} 2018, \apj, 864, 45

\bibitem[{{Martin} {et~al.}(2005){Martin}, {Fanson}, {Schiminovich},
  {Morrissey}, {Friedman}, {Barlow}, {Conrow}, {Grange}, {Jelinsky},
  {Milliard}, {Siegmund}, {Bianchi}, {Byun}, {Donas}, {Forster}, {Heckman},
  {Lee}, {Madore}, {Malina}, {Neff}, {Rich}, {Small}, {Surber}, {Szalay},
  {Welsh}, \& {Wyder}}]{galex}
{Martin}, D.~C., {Fanson}, J., {Schiminovich}, D., {et~al.} 2005, \apjl, 619,
  L1

\bibitem[{{Masci} {et~al.}(2019){Masci}, {Laher}, {Rusholme}, {Shupe}, {Groom},
  {Surace}, {Jackson}, {Monkewitz}, {Beck}, {Flynn}, {Terek}, {Landry},
  {Hacopians}, {Desai}, {Howell}, {Brooke}, {Imel}, {Wachter}, {Ye}, {Lin},
  {Cenko}, {Cunningham}, {Rebbapragada}, {Bue}, {Miller}, {Mahabal}, {Bellm},
  {Patterson}, {Juri{\'c}}, {Golkhou}, {Ofek}, {Walters}, {Graham}, {Kasliwal},
  {Dekany}, {Kupfer}, {Burdge}, {Cannella}, {Barlow}, {Van Sistine}, {Giomi},
  {Fremling}, {Blagorodnova}, {Levitan}, {Riddle}, {Smith}, {Helou}, {Prince},
  \& {Kulkarni}}]{mlr+19}
{Masci}, F.~J., {Laher}, R.~R., {Rusholme}, B., {et~al.} 2019, \pasp, 131,
  018003

\bibitem[{{Mazzali} {et~al.}(2016){Mazzali}, {Sullivan}, {Pian}, {Greiner}, \&
  {Kann}}]{msp+16}
{Mazzali}, P.~A., {Sullivan}, M., {Pian}, E., {Greiner}, J., \& {Kann}, D.~A.
  2016, MNRAS, 458, 3455

\bibitem[{{McCrum} {et~al.}(2015){McCrum}, {Smartt}, {Rest}, {Smith}, {Kotak},
  {Rodney}, {Young}, {Chornock}, {Berger}, {Foley}, {Fraser}, {Wright},
  {Scolnic}, {Tonry}, {Urata}, {Huang}, {Pastorello}, {Botticella}, {Valenti},
  {Mattila}, {Kankare}, {Farrow}, {Huber}, {Stubbs}, {Kirshner}, {Bresolin},
  {Burgett}, {Chambers}, {Draper}, {Flewelling}, {Jedicke}, {Kaiser},
  {Magnier}, {Metcalfe}, {Morgan}, {Price}, {Sweeney}, {Wainscoat}, \&
  {Waters}}]{msr+14}
{McCrum}, M., {Smartt}, S.~J., {Rest}, A., {et~al.} 2015, MNRAS, 448, 1206

\bibitem[{{Meisner} {et~al.}(2017){Meisner}, {Lang}, \&
  {Schlegel}}]{Meisner2017a}
{Meisner}, A.~M., {Lang}, D., \& {Schlegel}, D.~J. 2017, \aj, 153, 38

\bibitem[{{Metzger} {et~al.}(2014){Metzger}, {Vurm}, {Hasco{\"e}t}, \&
  {Beloborodov}}]{mvh+14}
{Metzger}, B.~D., {Vurm}, I., {Hasco{\"e}t}, R., \& {Beloborodov}, A.~M. 2014,
  MNRAS, 437, 703

\bibitem[{{Miller} {et~al.}(2009){Miller}, {Chornock}, {Perley},
  {Ganeshalingam}, {Li}, {Butler}, {Bloom}, {Smith}, {Modjaz}, {Poznanski},
  {Filippenko}, {Griffith}, {Shiode}, \& {Silverman}}]{mcp+09}
{Miller}, A.~A., {Chornock}, R., {Perley}, D.~A., {et~al.} 2009, ApJ, 690, 1303

\bibitem[{{Miller} {et~al.}(2017){Miller}, {Kasliwal}, {Cao}, {Adams},
  {Goobar}, {Kne{\v{z}}evi{\'c}}, {Laher}, {Lunnan}, {Masci}, {Nugent},
  {Perley}, {Petrushevska}, {Quimby}, {Rebbapragada}, {Sollerman}, {Taddia}, \&
  {Kulkarni}}]{Miller2017}
{Miller}, A.~A., {Kasliwal}, M.~M., {Cao}, Y., {et~al.} 2017, The Astrophysical
  Journal, 848, 59

\bibitem[{{Modjaz} {et~al.}(2016){Modjaz}, {Liu}, {Bianco}, \&
  {Graur}}]{mlb+16}
{Modjaz}, M., {Liu}, Y.~Q., {Bianco}, F.~B., \& {Graur}, O. 2016, \apj, 832,
  108

\bibitem[{{Moriya} {et~al.}(2013){Moriya}, {Blinnikov}, {Tominaga}, {Yoshida},
  {Tanaka}, {Maeda}, \& {Nomoto}}]{mbt+13}
{Moriya}, T.~J., {Blinnikov}, S.~I., {Tominaga}, N., {et~al.} 2013, MNRAS, 428,
  1020

\bibitem[{{Moriya} {et~al.}(2019){Moriya}, {Tanaka}, {Yasuda}, {Jiang}, {Lee},
  {Maeda}, {Morokuma}, {Nomoto}, {Quimby}, {Suzuki}, {Takahashi}, {Tanaka},
  {Tominaga}, {Yamaguchi}, {Bernard}, {Cooke}, {Curtin}, {Galbany},
  {Gonz{\'a}lez-Gait{\'a}n}, {Pignata}, {Pritchard}, {Komiyama}, \&
  {Lupton}}]{mty+19}
{Moriya}, T.~J., {Tanaka}, M., {Yasuda}, N., {et~al.} 2019, \apjs, 241, 16

\bibitem[{{Morrissey} {et~al.}(2007){Morrissey}, {Conrow}, {Barlow}, {Small},
  {Seibert}, {Wyder}, {Budav{\'a}ri}, {Arnouts}, {Friedman}, {Forster},
  {Martin}, {Neff}, {Schiminovich}, {Bianchi}, {Donas}, {Heckman}, {Lee},
  {Madore}, {Milliard}, {Rich}, {Szalay}, {Welsh}, \& {Yi}}]{mcb+07}
{Morrissey}, P., {Conrow}, T., {Barlow}, T.~A., {et~al.} 2007, \apjs, 173, 682

\bibitem[{{Nicholl} {et~al.}(2017{\natexlab{a}}){Nicholl}, {Berger},
  {Margutti}, {Blanchard}, {Guillochon}, {Leja}, \& {Chornock}}]{nbm+17}
{Nicholl}, M., {Berger}, E., {Margutti}, R., {et~al.} 2017{\natexlab{a}},
  \apjl, 845, L8

\bibitem[{{Nicholl} {et~al.}(2018{\natexlab{a}}){Nicholl}, {Gomez},
  {Blanchard}, \& {Berger}}]{ngb+18}
{Nicholl}, M., {Gomez}, S., {Blanchard}, P., \& {Berger}, E.
  2018{\natexlab{a}}, The Astronomer's Telegram, 11648

\bibitem[{{Nicholl} {et~al.}(2017{\natexlab{b}}){Nicholl}, {Guillochon}, \&
  {Berger}}]{ngb17}
{Nicholl}, M., {Guillochon}, J., \& {Berger}, E. 2017{\natexlab{b}}, \apj, 850,
  55

\bibitem[{{Nicholl} \& {Smartt}(2016)}]{ns16}
{Nicholl}, M., \& {Smartt}, S.~J. 2016, MNRAS, 457, L79

\bibitem[{{Nicholl} {et~al.}(2013){Nicholl}, {Smartt}, {Jerkstrand}, {Inserra},
  {McCrum}, {Kotak}, {Fraser}, {Wright}, {Chen}, {Smith}, {Young}, {Sim},
  {Valenti}, {Howell}, {Bresolin}, {Kudritzki}, {Tonry}, {Huber}, {Rest},
  {Pastorello}, {Tomasella}, {Cappellaro}, {Benetti}, {Mattila}, {Kankare},
  {Kangas}, {Leloudas}, {Sollerman}, {Taddia}, {Berger}, {Chornock}, {Narayan},
  {Stubbs}, {Foley}, {Lunnan}, {Soderberg}, {Sanders}, {Milisavljevic},
  {Margutti}, {Kirshner}, {Elias-Rosa}, {Morales-Garoffolo}, {Taubenberger},
  {Botticella}, {Gezari}, {Urata}, {Rodney}, {Riess}, {Scolnic}, {Wood-Vasey},
  {Burgett}, {Chambers}, {Flewelling}, {Magnier}, {Kaiser}, {Metcalfe},
  {Morgan}, {Price}, {Sweeney}, \& {Waters}}]{nsj+13}
{Nicholl}, M., {Smartt}, S.~J., {Jerkstrand}, A., {et~al.} 2013, Nature, 502,
  346

\bibitem[{{Nicholl} {et~al.}(2015){Nicholl}, {Smartt}, {Jerkstrand}, {Inserra},
  {Sim}, {Chen}, {Benetti}, {Fraser}, {Gal-Yam}, {Kankare}, {Maguire}, {Smith},
  {Sullivan}, {Valenti}, {Young}, {Baltay}, {Bauer}, {Baumont}, {Bersier},
  {Botticella}, {Childress}, {Dennefeld}, {Della Valle}, {Elias-Rosa},
  {Feindt}, {Galbany}, {Hadjiyska}, {Le Guillou}, {Leloudas}, {Mazzali},
  {McKinnon}, {Polshaw}, {Rabinowitz}, {Rostami}, {Scalzo}, {Schmidt},
  {Schulze}, {Sollerman}, {Taddia}, \& {Yuan}}]{nsj+15}
---. 2015, MNRAS, 452, 3869

\bibitem[{{Nicholl} {et~al.}(2016){Nicholl}, {Berger}, {Smartt}, {Margutti},
  {Kamble}, {Alexander}, {Chen}, {Inserra}, {Arcavi}, {Blanchard}, {Cartier},
  {Chambers}, {Childress}, {Chornock}, {Cowperthwaite}, {Drout}, {Flewelling},
  {Fraser}, {Gal-Yam}, {Galbany}, {Harmanen}, {Holoien}, {Hosseinzadeh},
  {Howell}, {Huber}, {Jerkstrand}, {Kankare}, {Kochanek}, {Lin}, {Lunnan},
  {Magnier}, {Maguire}, {McCully}, {McDonald}, {Metzger}, {Milisavljevic},
  {Mitra}, {Reynolds}, {Saario}, {Shappee}, {Smith}, {Valenti}, {Villar},
  {Waters}, \& {Young}}]{nbs+16}
{Nicholl}, M., {Berger}, E., {Smartt}, S.~J., {et~al.} 2016, ApJ, 826, 39

\bibitem[{{Nicholl} {et~al.}(2018{\natexlab{b}}){Nicholl}, {Blanchard},
  {Berger}, {Alexander}, {Metzger}, {Bhirombhakdi}, {Chornock}, {Coppejans},
  {Gomez}, {Margalit}, {Margutti}, \& {Terreran}}]{nbb+18}
{Nicholl}, M., {Blanchard}, P.~K., {Berger}, E., {et~al.} 2018{\natexlab{b}},
  \apjl, 866, L24

\bibitem[{{Nyholm} {et~al.}(2017){Nyholm}, {Sollerman}, {Taddia}, {Fremling},
  {Moriya}, {Ofek}, {Gal-Yam}, {De Cia}, {Roy}, {Kasliwal}, {Cao}, {Nugent}, \&
  {Masci}}]{nst+17}
{Nyholm}, A., {Sollerman}, J., {Taddia}, F., {et~al.} 2017, \aap, 605, A6

\bibitem[{{Nyholm} {et~al.}(2020){Nyholm}, {Sollerman}, {Tartaglia}, {Taddia},
  {Fremling}, {Blagorodnova}, {Filippenko}, {Gal-Yam}, {Howell},
  {Karamehmetoglu}, {Kulkarni}, {Laher}, {Leloudas}, {Masci}, {Kasliwal},
  {Mor{\r{a}}}, {Moriya}, {Ofek}, {Papadogiannakis}, {Quimby}, {Rebbapragada},
  \& {Schulze}}]{nst+19}
{Nyholm}, A., {Sollerman}, J., {Tartaglia}, L., {et~al.} 2020, \aap, 637, A73

\bibitem[{{Ofek}(2014)}]{Ofek2014a}
{Ofek}, E.~O. 2014, {MATLAB package for astronomy and astrophysics},
  Astrophysics Source Code Library, , , ascl:1407.005

\bibitem[{{Oke} \& {Gunn}(1982)}]{og82}
{Oke}, J.~B., \& {Gunn}, J.~E. 1982, PASP, 94, 586

\bibitem[{{Oke} {et~al.}(1995){Oke}, {Cohen}, {Carr}, {Cromer}, {Dingizian},
  {Harris}, {Labrecque}, {Lucinio}, {Schaal}, {Epps}, \& {Miller}}]{occ+95}
{Oke}, J.~B., {Cohen}, J.~G., {Carr}, M., {et~al.} 1995, PASP, 107, 375

\bibitem[{{Pastorello} {et~al.}(2010){Pastorello}, {Smartt}, {Botticella},
  {Maguire}, {Fraser}, {Smith}, {Kotak}, {Magill}, {Valenti}, {Young},
  {Gezari}, {Bresolin}, {Kudritzki}, {Howell}, {Rest}, {Metcalfe}, {Mattila},
  {Kankare}, {Huang}, {Urata}, {Burgett}, {Chambers}, {Dombeck}, {Flewelling},
  {Grav}, {Heasley}, {Hodapp}, {Kaiser}, {Luppino}, {Lupton}, {Magnier},
  {Monet}, {Morgan}, {Onaka}, {Price}, {Rhoads}, {Siegmund}, {Stubbs},
  {Sweeney}, {Tonry}, {Wainscoat}, {Waterson}, {Waters}, \&
  {Wynn-Williams}}]{psb+10}
{Pastorello}, A., {Smartt}, S.~J., {Botticella}, M.~T., {et~al.} 2010, ApJL,
  724, L16

\bibitem[{{Patterson} {et~al.}(2019){Patterson}, {Bellm}, {Rusholme}, {Masci},
  {Juric}, {Krughoff}, {Golkhou}, {Graham}, {Kulkarni}, {Helou}, \& {Zwicky
  Transient Facility Collaboration}}]{pbr+19}
{Patterson}, M.~T., {Bellm}, E.~C., {Rusholme}, B., {et~al.} 2019, \pasp, 131,
  018001

\bibitem[{{Perley}(2019)}]{Perley2019}
{Perley}, D.~A. 2019, \pasp, 131, 084503

\bibitem[{{Perley} {et~al.}(2016){Perley}, {Quimby}, {Yan}, {Vreeswijk}, {De
  Cia}, {Lunnan}, {Gal-Yam}, {Yaron}, {Filippenko}, {Graham}, {Laher}, \&
  {Nugent}}]{pqy+16}
{Perley}, D.~A., {Quimby}, R.~M., {Yan}, L., {et~al.} 2016, ApJ, 830, 13

\bibitem[{{Piascik} {et~al.}(2014){Piascik}, {Steele}, {Bates}, {Mottram},
  {Smith}, {Barnsley}, \& {Bolton}}]{pia+14}
{Piascik}, A.~S., {Steele}, I.~A., {Bates}, S.~D., {et~al.} 2014, in Society of
  Photo-Optical Instrumentation Engineers (SPIE) Conference Series, Vol. 9147,
  Ground-based and Airborne Instrumentation for Astronomy V, 91478H

\bibitem[{{Prentice} {et~al.}(2019){Prentice}, {Ashall}, {James}, {Short},
  {Mazzali}, {Bersier}, {Crowther}, {Barbarino}, {Chen}, {Copperwheat},
  {Darnley}, {Denneau}, {Elias-Rosa}, {Fraser}, {Galbany}, {Gal-Yam},
  {Harmanen}, {Howell}, {Hosseinzadeh}, {Inserra}, {Kankare}, {Karamehmetoglu},
  {Lamb}, {Limongi}, {Maguire}, {McCully}, {Olivares E}, {Piascik}, {Pignata},
  {Reichart}, {Rest}, {Reynolds}, {Rodr{\'\i}guez}, {Saario}, {Schulze},
  {Smartt}, {Smith}, {Sollerman}, {Stalder}, {Sullivan}, {Taddia}, {Valenti},
  {Vergani}, {Williams}, \& {Young}}]{paj+19}
{Prentice}, S.~J., {Ashall}, C., {James}, P.~A., {et~al.} 2019, \mnras, 485,
  1559

\bibitem[{{Pursiainen} {et~al.}(2018){Pursiainen}, {Childress}, {Smith},
  {Prajs}, {Sullivan}, {Davis}, {Foley}, {Asorey}, {Calcino}, {Carollo},
  {Curtin}, {D'Andrea}, {Glazebrook}, {Gutierrez}, {Hinton}, {Hoormann},
  {Inserra}, {Kessler}, {King}, {Kuehn}, {Lewis}, {Lidman}, {Macaulay},
  {M{\"o}ller}, {Nichol}, {Sako}, {Sommer}, {Swann}, {Tucker}, {Uddin},
  {Wiseman}, {Zhang}, {Abbott}, {Abdalla}, {Allam}, {Annis}, {Avila}, {Brooks},
  {Buckley-Geer}, {Burke}, {Carnero Rosell}, {Carrasco Kind}, {Carretero},
  {Castander}, {Cunha}, {Davis}, {De Vicente}, {Diehl}, {Doel}, {Eifler},
  {Flaugher}, {Fosalba}, {Frieman}, {Garc{\'{\i}}a-Bellido}, {Gruen},
  {Gruendl}, {Gutierrez}, {Hartley}, {Hollowood}, {Honscheid}, {James},
  {Jeltema}, {Kuropatkin}, {Li}, {Lima}, {Maia}, {Martini}, {Menanteau},
  {Ogando}, {Plazas}, {Roodman}, {Sanchez}, {Scarpine}, {Schindler}, {Smith},
  {Soares-Santos}, {Sobreira}, {Suchyta}, {Swanson}, {Tarle}, {Tucker}, \&
  {Walker}}]{pcs+18}
{Pursiainen}, M., {Childress}, M., {Smith}, M., {et~al.} 2018, \mnras, 481, 894

\bibitem[{{Quimby} {et~al.}(2011){Quimby}, {Kulkarni}, {Kasliwal}, {Gal-Yam},
  {Arcavi}, {Sullivan}, {Nugent}, {Thomas}, {Howell}, {Nakar}, {Bildsten},
  {Theissen}, {Law}, {Dekany}, {Rahmer}, {Hale}, {Smith}, {Ofek}, {Zolkower},
  {Velur}, {Walters}, {Henning}, {Bui}, {McKenna}, {Poznanski}, {Cenko}, \&
  {Levitan}}]{qkk+11}
{Quimby}, R.~M., {Kulkarni}, S.~R., {Kasliwal}, M.~M., {et~al.} 2011, Nature,
  474, 487

\bibitem[{{Quimby} {et~al.}(2018){Quimby}, {De Cia}, {Gal-Yam}, {Leloudas},
  {Lunnan}, {Perley}, {Vreeswijk}, {Yan}, {Bloom}, {Cenko}, {Cooke}, {Ellis},
  {Filippenko}, {Kasliwal}, {Kleiser}, {Kulkarni}, {Matheson}, {Nugent}, {Pan},
  {Silverman}, {Sternberg}, {Sullivan}, \& {Yaron}}]{qdg+18}
{Quimby}, R.~M., {De Cia}, A., {Gal-Yam}, A., {et~al.} 2018, \apj, 855, 2

\bibitem[{{Rau} {et~al.}(2009){Rau}, {Kulkarni}, {Law}, {Bloom}, {Ciardi},
  {Djorgovski}, {Fox}, {Gal-Yam}, {Grillmair}, {Kasliwal}, {Nugent}, {Ofek},
  {Quimby}, {Reach}, {Shara}, {Bildsten}, {Cenko}, {Drake}, {Filippenko},
  {Helfand}, {Helou}, {Howell}, {Poznanski}, \& {Sullivan}}]{rkl+09}
{Rau}, A., {Kulkarni}, S.~R., {Law}, N.~M., {et~al.} 2009, \pasp, 121, 1334

\bibitem[{{Rest} {et~al.}(2018){Rest}, {Garnavich}, {Khatami}, {Kasen},
  {Tucker}, {Shaya}, {Olling}, {Mushotzky}, {Zenteno}, {Margheim},
  {Strampelli}, {James}, {Smith}, {F{\"o}rster}, \& {Villar}}]{rgk+18}
{Rest}, A., {Garnavich}, P.~M., {Khatami}, D., {et~al.} 2018, Nature Astronomy,
  2, 307

\bibitem[{{Roming} {et~al.}(2005){Roming}, {Kennedy}, {Mason}, {Nousek}, {Ahr},
  {Bingham}, {Broos}, {Carter}, {Hancock}, {Huckle}, {Hunsberger}, {Kawakami},
  {Killough}, {Koch}, {McLelland}, {Smith}, {Smith}, {Soto}, {Boyd},
  {Breeveld}, {Holland}, {Ivanushkina}, {Pryzby}, {Still}, \& {Stock}}]{rkm+05}
{Roming}, P.~W.~A., {Kennedy}, T.~E., {Mason}, K.~O., {et~al.} 2005, \ssr, 120,
  95

\bibitem[{{Schlafly} \& {Finkbeiner}(2011)}]{sf11}
{Schlafly}, E.~F., \& {Finkbeiner}, D.~P. 2011, ApJ, 737, 103

\bibitem[{{Schulze} {et~al.}(2018){Schulze}, {Kr{\"u}hler}, {Leloudas},
  {Gorosabel}, {Mehner}, {Buchner}, {Kim}, {Ibar}, {Amor{\'{\i}}n},
  {Herrero-Illana}, {Anderson}, {Bauer}, {Christensen}, {de Pasquale}, {de
  Ugarte Postigo}, {Gallazzi}, {Hjorth}, {Morrell}, {Malesani}, {Sparre},
  {Stalder}, {Stark}, {Th{\"o}ne}, \& {Wheeler}}]{skl+18}
{Schulze}, S., {Kr{\"u}hler}, T., {Leloudas}, G., {et~al.} 2018, \mnras, 473,
  1258

\bibitem[{{Shappee} {et~al.}(2014){Shappee}, {Prieto}, {Grupe}, {Kochanek},
  {Stanek}, {De Rosa}, {Mathur}, {Zu}, {Peterson}, {Pogge}, {Komossa}, {Im},
  {Jencson}, {Holoien}, {Basu}, {Beacom}, {Szczygie{\l}}, {Brimacombe},
  {Adams}, {Campillay}, {Choi}, {Contreras}, {Dietrich}, {Dubberley},
  {Elphick}, {Foale}, {Giustini}, {Gonzalez}, {Hawkins}, {Howell}, {Hsiao},
  {Koss}, {Leighly}, {Morrell}, {Mudd}, {Mullins}, {Nugent}, {Parrent},
  {Phillips}, {Pojmanski}, {Rosing}, {Ross}, {Sand}, {Terndrup}, {Valenti},
  {Walker}, \& {Yoon}}]{spg+14}
{Shappee}, B.~J., {Prieto}, J.~L., {Grupe}, D., {et~al.} 2014, \apj, 788, 48

\bibitem[{{Skrutskie} {et~al.}(2006){Skrutskie}, {Cutri}, {Stiening},
  {Weinberg}, {Schneider}, {Carpenter}, {Beichman}, {Capps}, {Chester},
  {Elias}, {Huchra}, {Liebert}, {Lonsdale}, {Monet}, {Price}, {Seitzer},
  {Jarrett}, {Kirkpatrick}, {Gizis}, {Howard}, {Evans}, {Fowler}, {Fullmer},
  {Hurt}, {Light}, {Kopan}, {Marsh}, {McCallon}, {Tam}, {Van Dyk}, \&
  {Wheelock}}]{2mass}
{Skrutskie}, M.~F., {Cutri}, R.~M., {Stiening}, R., {et~al.} 2006, \aj, 131,
  1163

\bibitem[{{Sorokina} {et~al.}(2016){Sorokina}, {Blinnikov}, {Nomoto}, {Quimby},
  \& {Tolstov}}]{sbn16}
{Sorokina}, E., {Blinnikov}, S., {Nomoto}, K., {Quimby}, R., \& {Tolstov}, A.
  2016, ApJ, 829, 17

\bibitem[{{Soumagnac} {et~al.}(2019){Soumagnac}, {Ganot}, {Gal-yam}, {Ofek},
  {Yaron}, {Waxman}, {Schulze}, {Yang}, {Rubin}, {Cenko}, {Sollerman},
  {Perley}, {Fremling}, {Nugent}, {Neill}, {Karamehmetoglu}, {Bellm}, {Bruch},
  {Burruss}, {Cunningham}, {Dekany}, {Golkhou}, {Irani}, {Kasliwal},
  {Konidaris}, {Kulkarni}, {Kupfer}, {Laher}, {Masci}, {Morag}, {Riddle},
  {Rigault}, {Rusholme}, {van Roestel}, \& {Zackay}}]{sgg+19}
{Soumagnac}, M.~T., {Ganot}, N., {Gal-yam}, A., {et~al.} 2019, arXiv e-prints,
  arXiv:1907.11252

\bibitem[{{Steele} {et~al.}(2004){Steele}, {Smith}, {Rees}, {Baker}, {Bates},
  {Bode}, {Bowman}, {Carter}, {Etherton}, {Ford}, {Fraser}, {Gomboc}, {Lett},
  {Mansfield}, {Marchant}, {Medrano-Cerda}, {Mottram}, {Raback}, {Scott},
  {Tomlinson}, \& {Zamanov}}]{ssr+04}
{Steele}, I.~A., {Smith}, R.~J., {Rees}, P.~C., {et~al.} 2004, in \procspie,
  Vol. 5489, Ground-based Telescopes, ed. J.~M. {Oschmann}, Jr., 679--692

\bibitem[{{Stritzinger} {et~al.}(2012){Stritzinger}, {Taddia}, {Fransson},
  {Fox}, {Morrell}, {Phillips}, {Sollerman}, {Anderson}, {Boldt}, {Brown},
  {Campillay}, {Castellon}, {Contreras}, {Folatelli}, {Habergham}, {Hamuy},
  {Hjorth}, {James}, {Krzeminski}, {Mattila}, {Persson}, \& {Roth}}]{stf+12}
{Stritzinger}, M., {Taddia}, F., {Fransson}, C., {et~al.} 2012, ApJ, 756, 173

\bibitem[{{Tachibana} \& {Miller}(2018)}]{Tachibana2018}
{Tachibana}, Y., \& {Miller}, A.~A. 2018, Publications of the Astronomical
  Society of the Pacific, 130, 128001

\bibitem[{{Taddia} {et~al.}(2018){Taddia}, {Stritzinger}, {Bersten}, {Baron},
  {Burns}, {Contreras}, {Holmbo}, {Hsiao}, {Morrell}, {Phillips}, {Sollerman},
  \& {Suntzeff}}]{tsb+18}
{Taddia}, F., {Stritzinger}, M.~D., {Bersten}, M., {et~al.} 2018, \aap, 609,
  A136

\bibitem[{{Tonry} {et~al.}(2018{\natexlab{a}}){Tonry}, {Stalder}, {Denneau},
  {Heinze}, {Weiland}, {Rest}, {Smith}, {Smartt}, {Young}, \&
  {Fulton}}]{2018bym}
{Tonry}, J., {Stalder}, B., {Denneau}, L., {et~al.} 2018{\natexlab{a}},
  Transient Name Server Discovery Report, 2018-707, 1

\bibitem[{{Tonry} {et~al.}(2018{\natexlab{b}}){Tonry}, {Denneau}, {Heinze},
  {Stalder}, {Smith}, {Smartt}, {Stubbs}, {Weiland}, \& {Rest}}]{tdh+18}
{Tonry}, J.~L., {Denneau}, L., {Heinze}, A.~N., {et~al.} 2018{\natexlab{b}},
  \pasp, 130, 064505

\bibitem[{{Villar} {et~al.}(2017){Villar}, {Berger}, {Metzger}, \&
  {Guillochon}}]{vbm+17}
{Villar}, V.~A., {Berger}, E., {Metzger}, B.~D., \& {Guillochon}, J. 2017,
  \apj, 849, 70

\bibitem[{{Villar} {et~al.}(2018){Villar}, {Nicholl}, \& {Berger}}]{vnb+18}
{Villar}, V.~A., {Nicholl}, M., \& {Berger}, E. 2018, \apj, 869, 166

\bibitem[{{Villar} {et~al.}(2019){Villar}, {Berger}, {Miller}, {Chornock},
  {Rest}, {Jones}, {Drout}, {Foley}, {Kirshner}, {Lunnan}, {Magnier},
  {Milisavljevic}, {Sanders}, \& {Scolnic}}]{vbm+19}
{Villar}, V.~A., {Berger}, E., {Miller}, G., {et~al.} 2019, \apj, 884, 83

\bibitem[{{Vreeswijk} {et~al.}(2017){Vreeswijk}, {Leloudas}, {Gal-Yam}, {De
  Cia}, {Perley}, {Quimby}, {Waldman}, {Sullivan}, {Yan}, {Ofek}, {Fremling},
  {Taddia}, {Sollerman}, {Valenti}, {Arcavi}, {Howell}, {Filippenko}, {Cenko},
  {Yaron}, {Kasliwal}, {Cao}, {Ben-Ami}, {Horesh}, {Rubin}, {Lunnan}, {Nugent},
  {Laher}, {Rebbapragada}, {Wo{\'z}niak}, \& {Kulkarni}}]{vlg+17}
{Vreeswijk}, P.~M., {Leloudas}, G., {Gal-Yam}, A., {et~al.} 2017, \apj, 835, 58

\bibitem[{{Wheeler} {et~al.}(2017){Wheeler}, {Chatzopoulos}, {Vink{\'o}}, \&
  {Tuminello}}]{wcv+17}
{Wheeler}, J.~C., {Chatzopoulos}, E., {Vink{\'o}}, J., \& {Tuminello}, R. 2017,
  \apjl, 851, L14

\bibitem[{{Whitesides} {et~al.}(2017){Whitesides}, {Lunnan}, {Kasliwal},
  {Perley}, {Corsi}, {Cenko}, {Blagorodnova}, {Cao}, {Cook}, {Doran},
  {Frederiks}, {Fremling}, {Hurley}, {Karamehmetoglu}, {Kulkarni}, {Leloudas},
  {Masci}, {Nugent}, {Ritter}, {Rubin}, {Savchenko}, {Sollerman}, {Svinkin},
  {Taddia}, {Vreeswijk}, \& {Wozniak}}]{wlk+17}
{Whitesides}, L., {Lunnan}, R., {Kasliwal}, M.~M., {et~al.} 2017, \apj, 851,
  107

\bibitem[{{Woosley}(2010)}]{woo10}
{Woosley}, S.~E. 2010, ApJL, 719, L204

\bibitem[{{Woosley}(2017)}]{woo17}
---. 2017, ApJ, 836, 244

\bibitem[{{Woosley} {et~al.}(2007){Woosley}, {Blinnikov}, \& {Heger}}]{wbh07}
{Woosley}, S.~E., {Blinnikov}, S., \& {Heger}, A. 2007, Nature, 450, 390

\bibitem[{{Wright} {et~al.}(2016){Wright}, {Robotham}, {Bourne}, {Driver},
  {Dunne}, {Maddox}, {Alpaslan}, {Andrews}, {Bauer}, {Bland-Hawthorn},
  {Brough}, {Brown}, {Clarke}, {Cluver}, {Davies}, {Grootes}, {Holwerda},
  {Hopkins}, {Jarrett}, {Kafle}, {Lange}, {Liske}, {Loveday}, {Moffett},
  {Norberg}, {Popescu}, {Smith}, {Taylor}, {Tuffs}, {Wang}, \&
  {Wilkins}}]{Wright2016a}
{Wright}, A.~H., {Robotham}, A.~S.~G., {Bourne}, N., {et~al.} 2016, \mnras,
  460, 765

\bibitem[{{Yan} {et~al.}(2018){Yan}, {Perley}, {De Cia}, {Quimby}, {Lunnan},
  {Rubin}, \& {Brown}}]{Yan2018}
{Yan}, L., {Perley}, D.~A., {De Cia}, A., {et~al.} 2018, The Astrophysical
  Journal, 858, 91

\bibitem[{{Yan} {et~al.}(2015){Yan}, {Quimby}, {Ofek}, {Gal-Yam}, {Mazzali},
  {Perley}, {Vreeswijk}, {Leloudas}, {de Cia}, {Masci}, {Cenko}, {Cao},
  {Kulkarni}, {Nugent}, {Rebbapragada}, {Wo{\'z}niak}, \& {Yaron}}]{yqo+15}
{Yan}, L., {Quimby}, R., {Ofek}, E., {et~al.} 2015, ApJ, 814, 108

\bibitem[{{Yan} {et~al.}(2017){Yan}, {Lunnan}, {Perley}, {Gal-Yam}, {Yaron},
  {Roy}, {Quimby}, {Sollerman}, {Fremling}, {Leloudas}, {Cenko}, {Vreeswijk},
  {Graham}, {Howell}, {De Cia}, {Ofek}, {Nugent}, {Kulkarni}, {Hosseinzadeh},
  {Masci}, {McCully}, {Rebbapragada}, \& {Wo{\'z}niak}}]{ylp+17}
{Yan}, L., {Lunnan}, R., {Perley}, D.~A., {et~al.} 2017, ApJ, 848, 6

\bibitem[{{York} {et~al.}(2000){York}, {Adelman}, {Anderson}, {Anderson},
  {Annis}, {Bahcall}, {Bakken}, {Barkhouser}, {Bastian}, {Berman}, {Boroski},
  {Bracker}, {Briegel}, {Briggs}, {Brinkmann}, {Brunner}, {Burles}, {Carey},
  {Carr}, {Castander}, {Chen}, {Colestock}, {Connolly}, {Crocker}, {Csabai},
  {Czarapata}, {Davis}, {Doi}, {Dombeck}, {Eisenstein}, {Ellman}, {Elms},
  {Evans}, {Fan}, {Federwitz}, {Fiscelli}, {Friedman}, {Frieman}, {Fukugita},
  {Gillespie}, {Gunn}, {Gurbani}, {de Haas}, {Haldeman}, {Harris}, {Hayes},
  {Heckman}, {Hennessy}, {Hindsley}, {Holm}, {Holmgren}, {Huang}, {Hull},
  {Husby}, {Ichikawa}, {Ichikawa}, {Ivezi{\'c}}, {Kent}, {Kim}, {Kinney},
  {Klaene}, {Kleinman}, {Kleinman}, {Knapp}, {Korienek}, {Kron}, {Kunszt},
  {Lamb}, {Lee}, {Leger}, {Limmongkol}, {Lindenmeyer}, {Long}, {Loomis},
  {Loveday}, {Lucinio}, {Lupton}, {MacKinnon}, {Mannery}, {Mantsch}, {Margon},
  {McGehee}, {McKay}, {Meiksin}, {Merelli}, {Monet}, {Munn}, {Narayanan},
  {Nash}, {Neilsen}, {Neswold}, {Newberg}, {Nichol}, {Nicinski}, {Nonino},
  {Okada}, {Okamura}, {Ostriker}, {Owen}, {Pauls}, {Peoples}, {Peterson},
  {Petravick}, {Pier}, {Pope}, {Pordes}, {Prosapio}, {Rechenmacher}, {Quinn},
  {Richards}, {Richmond}, {Rivetta}, {Rockosi}, {Ruthmansdorfer}, {Sand ford},
  {Schlegel}, {Schneider}, {Sekiguchi}, {Sergey}, {Shimasaku}, {Siegmund},
  {Smee}, {Smith}, {Snedden}, {Stone}, {Stoughton}, {Strauss}, {Stubbs},
  {SubbaRao}, {Szalay}, {Szapudi}, {Szokoly}, {Thakar}, {Tremonti}, {Tucker},
  {Uomoto}, {Vanden Berk}, {Vogeley}, {Waddell}, {Wang}, {Watanabe},
  {Weinberg}, {Yanny}, {Yasuda}, \& {SDSS Collaboration}}]{yaa+00}
{York}, D.~G., {Adelman}, J., {Anderson}, John~E., J., {et~al.} 2000, \aj, 120,
  1579

\bibitem[{{Zackay} {et~al.}(2016){Zackay}, {Ofek}, \& {Gal-Yam}}]{zog16}
{Zackay}, B., {Ofek}, E.~O., \& {Gal-Yam}, A. 2016, \apj, 830, 27

\end{thebibliography}
\end{document}